\documentclass[amsmath,amssymb,article,onecolumn]{revtex4}

\usepackage[english]{babel}
\usepackage{natbib}
\usepackage{multirow}

\usepackage{rotating} % per ruotare tabelle con \begin{sidewaystable}

\usepackage{color}

\begin{document}

\title[Properties of the Water Molecule from QMC calculations ]
{Molecular properties by Quantum Monte Carlo: an investigation on the role of the wave function ansatz and the basis set in the water molecule \footnote{Reprinted and adapted with permission from J. Chem. Theory Comput., DOI: 10.1021/ct400382m. Copyright 2013 American Chemical Society.}}

\author{Andrea Zen}
\email{andrea.zen@uniroma1.it}
\affiliation{Dipartimento di Fisica, La Sapienza - Universit\`a di Roma , Piazzale Aldo Moro 2, 00185 Rome, Italy}

\author{Ye Luo}
\affiliation{Scuola Internazionale Superiore di Studi Avanzati (SISSA), via Bonomea 265, 34136 Trieste, Italy}

\author{Sandro Sorella}
\email{sorella@sissa.it}
\affiliation{Scuola Internazionale Superiore di Studi Avanzati (SISSA) and Democritos National Simulation Center, Istituto Officina dei Materiali del CNR, via Bonomea 265, 34136 Trieste, Italy}

\author{Leonardo Guidoni}
\email{leonardo.guidoni@univaq.it}
\affiliation{Dipartimento di Scienze Fisiche e Chimiche, Universit\`a degli studi de L'Aquila, Via Vetoio, 67100 Coppito, L'Aquila, Italy}

%%%%%%%%%%%%%%%%%%%%%%%%%%%%%%%%%%%%%%%%%%%%%%%%%%%%%%%%%%%%%%%%%%%%%
\begin{abstract}
%%%%%%%%%%%%%%%%%%%%%%%%%%%%%%%%%%%%%%%%%%%%%%%%%%%%%%%%%%%%%%%%%%%%%

Quantum Monte Carlo methods are accurate and promising many body techniques for electronic structure calculations which, in the last years, are encountering a growing interest thanks to their favorable scaling with the system size and their efficient parallelization, particularly suited for the modern high performance computing facilities. 
The ansatz of the  wave function and its variational flexibility are crucial points for both  the accurate description of molecular properties and  the capabilities of the method to tackle large systems. 
In this paper, we extensively analyze, using different variational ansatzes, several properties of the water molecule, namely: the total energy, the dipole and quadrupole momenta, the ionization and atomization energies, the equilibrium configuration, and the harmonic and fundamental frequencies of vibration. 
The investigation mainly focuses on variational Monte Carlo calculations, although several lattice regularized diffusion Monte Carlo calculations are also reported.
Through a systematic study, we provide a useful guide to the choice of the wave function, the pseudo potential, and the basis set for QMC calculations. 
We also introduce a new strategy for the definition of the atomic orbitals involved in the Jastrow - Antisymmetrised Geminal power wave function, in order to drastically reduce  the number of variational parameters. 
This scheme significantly improves the efficiency of QMC energy minimization in case of large basis sets.

%%%%%%%%%%%%%%%%%%%%%%%%%%%%%%%%%%%%%%%%%%%%%%%%%%%%%%%%%%%%%%%%%%%%%
\end{abstract}
%%%%%%%%%%%%%%%%%%%%%%%%%%%%%%%%%%%%%%%%%%%%%%%%%%%%%%%%%%%%%%%%%%%%%

\maketitle

%%%%%%%%%%%%%%%%%%%%%%%%%%%%%%%%%%%%%%%%%%%%%%%%%%%%%%%%%%%%%%%%%%%%%
\section{Introduction}
%%%%%%%%%%%%%%%%%%%%%%%%%%%%%%%%%%%%%%%%%%%%%%%%%%%%%%%%%%%%%%%%%%%%%

Quantum Monte Carlo\cite{Austin:2012kt,Needs:2010p28123,Assaraf:2007gn,Foulkes:2001p19717} (QMC) stands for a number of alternative stochastic methods that are used for electronic structure calculations of solids and molecules. 
These techniques range from the simplest and computationally cheapest variational Monte Carlo (VMC) scheme, 
to the more sophisticated and computationally expensive projection methods, such as 
the diffusion Monte Carlo\cite{Reynolds:1982en,DePasquale:1988bc,UMRIGAR:1993p25301,Mitas:1991kr} (DMC), 
the Green function Monte Carlo\cite{Kalos:1962gq,Trivedi:1990bj,Buonaura:1998p25304,Sorella:2000p17651} (GFMC), 
the lattice regularized diffusion Monte Carlo\cite{Casula:2005p14138,Casula:2010p14082} (LRDMC), 
the auxiliary field quantum Monte Carlo\cite{AlSaidi:2006gj,Zhang:2003ed,Baer:1998jh} (AFQMC), 
the released node quantum Monte Carlo\cite{Chen:1995cf,Ceperley:1984ha},
the self-healing diffusion Monte Carlo \cite{Bajdich:2010iu} (SHDMC),
the reptation quantum Monte Carlo\cite{Baroni:1999p27757,Yuen:2009gv} (RQMC),
or the recent full configuration interation Quantum Monte carlo\cite{Booth:2009p27842} (FCI-QMC).
% recente interesse e caratteristiche
In recent years, the QMC methods are encountering a growing interest due to the favorable scaling of the algorithms with the system size\cite{Coccia:2012kz} (the computational cost scales with the number $N$ of electrons as ${N}^{m}$ with $m$ between 3 and 4), an accuracy comparable to those of other high-level correlated quantum chemistry methods  \cite{Barborini:2012iy,Filippi:2012hg,Kolorenc:2011hv,Maezono:2010ic,Valsson:2010p25419,Spanu:2009p12613,Zimmerman:2009hh,Sterpone:2008p12640,Sorella:2007p12646,Schautz:2004p25285,Caffarel:1993td}, and their readiness for the implementation in modern highly parallel supercomputer facilities\cite{Coccia:2012kz}.

% recent improvements
Despite QMC techniques have been known for more than three decades, their application has  been quite limited if compared to other methods like Density Functional Theory (DFT), or to traditional quantum chemistry methods, such as Coupled-cluster\cite{Bartlett:2007kv} (CC),
 Configuration interaction\cite{Sherrill:1999dw} (CI), 
 M{\o}eller-Plesset perturbation theory\cite{Moller:1934kf,HeadGordon:1988fd}, and
Complete Active Space Self-Consistent Field (CASSCF).
Due to the large computational cost,
the use of QMC has been often restricted to particularly challenging systems, especially those characterized by the presence of strong electron correlation. 
% errore stocastico
This is probably due to the underlying stochastic nature of QMC that, on one side, it is responsible for the favorable scaling with the number of electrons and the intrinsic parallelization but, on the other side, it yields expectation values $\left< {\cal O} \right>$ of any operator $\cal O$ affected by an associated stochastic error $\sigma_{\cal O}$, converging to zero quite slowly, 
namely  like the inverse square root of the computational time.
Since this scaling has usually a large prefactor,  the stochastic error affecting QMC calculations is typically much larger than the corresponding numerical errors affecting non-stochastic  computational methods.

% soluzione dei problemi: forze, errori stocastici, migliori ansatz
Strictly connected with such underlying stochastic error of QMC is the ``historical'' challenge to calculate reliable ionic forces. The straightforward employment of the finite difference methods is quite inefficient, due to the propagation of the stochastic errors when energy differences are considered. However, a number of technical improvements have significantly reduced these problems, making possible to realize calculations of increasing complexity with an affordable computational cost.
The introduction of the correlated sampling \cite{Filippi:2000p25406}  (CS) and the space warp coordinate transformation\cite{UMRIGAR:1989tq} (SWCT) lead to large improvements in calculating energy differences between two different wave functions, and then on the force evaluation. Anyway, the finite difference approaches have a computational cost proportional to 3 times the number of atoms, making it is prohibitive for large molecules. 
Concerning the analytical approaches for the calculation of the forces, 
large improvements have been achieved by the introduction of the reweighting methods for the stochastic sampling \cite{Assaraf:2000vr,Assaraf:2003gq,Attaccalite:2008p12639}, 
which allows to overcome the well known problem of the infinite variance of the force estimators. 
A further step in the direction of an efficient and accurate computation of the QMC forces has been recently accomplished by 
\citet{Sorella:2010p23644}, who proposed a combined use of the reweighting method, the CS and the SWCT techniques. Thanks to the use of the algorithmic adjoint differentiation (AAD), all the components of the ionic forces  are calculated in a computational time that is only about four times the one  of an ordinary energy calculation, in both the cases of all electrons  and  pseudo potential calculations.
% recenti risultati ottimizzazione geometria
Thanks to these improvements in the force evaluations, in the last years several optimizations of molecular geometries, based on QMC calculations, have been reported\cite{Chiesa:2005p28489,Wagner:2010p25393,Sorella:2010p23644,Barborini:2012iy,Barborini:2012it,Coccia:2012ex,Coccia:2012kz}, for molecules of growing size and complexity. 
Another important issue emerging form recent QMC literature, is the possibility to calculate ground state molecular properties beyond  energies and geometries, such as, for instance, the polarizability and the electronic density\cite{Assaraf:2007gn,Coccia:2012fi}. Recently, by using a procedure based on the multidimensional fitting of the potential energy surface (PES) of a molecule in proximity of its configurational minimum, it has also been  shown how it is possible to calculate the harmonic vibrational frequencies and the anharmonic corrections by QMC, despite the presence of stochastic errors\cite{Zen:2012br}.

% wf ansatz
Another area in which the QMC has recently undergone  a remarkable progress is the introduction and the characterization of several new wave function ansatzes. The wave function variational ansatz is, indeed, of fundamental importance for the accuracy and the reliability of both variational and diffusion Monte Carlo results, as it emerges from different works \cite{Umrigar:2007p12662,Clark:2011ie,Barborini:2012iy,Barborini:2012it,Coccia:2012fi,Morales:2012kk,Xu:2013fc}. As expected, the definition of the wave function is more important for the VMC technique rather than for the corresponding DMC projection method, since the latter only depends  on the nodal surface of the variational wave function.
% Remarkably, the bias of the results from the wave function ansatz is observed both if VMC is adopted, and if the computationally expensive projection methods are used within the fixed node (FN) approximation. 
%%% wave function ansatz
%CAS in determinantal \cite{Toulouse:2008p27527}
%Multideterminant QMC (Scusetia) \cite{Clark:2011ie} \cite{Morales:2012kk}
%Linear Scaling Generalized Valence Bond Approach (Filippi) \cite{Fracchia:2012el}
A typical QMC wave function is given by an antisymmetric determinantal part, aimed to describe static correlation effects, and a bosonic part, termed the Jastrow factor\cite{Jastrow:1955en}, which recovers most of the dynamical correlation effects.
Going beyond the simplest wave function where a single Slater determinant is correlated with a Jastrow term, among  recent wave function developments we can include: the Jastrow antisymmetrized geminal power (JAGP) wave function \cite{Casula:2003p12694},  the Pfaffian wave function\cite{Bajdich:2008p18507,Bajdich:2006p18510}, 
wave functions with backflow correction\cite{Holzmann:2003p28608},  and many others multi-determinant-Jastrow functions\cite{Toulouse:2008p27527,Clark:2011ie,Morales:2012kk,Fracchia:2012el}. 
%messaggio:
%JAGP meglio di JHF 
%LRDMC= J *, if J is good.
%Pseudo Filippi OK 
%la qualita' del Variazionale (accurate Variational .. non so).
%Vantaggio enorme sia dal punto di vista computazionale che sulle forze.
The JAGP is a particularly interesting and promising ansatz, due to its ability to represent a multiderminant wave function in an implicit and compact way. Moreover, the presence of the Jastrow factor allows to satisfy the size consistency that, as observed in ref.\cite{Sorella:2007p12646} and more recently by \citet{Neuscamman:2012hm}, is not fulfilled by the simple AGP ansatz.

In many papers, convergence studies have been carried out as a function of the basis set size \cite{Sorella:2007p12646,Petruzielo:2011p24345,Sterpone:2008p12640,Marchi:2009p12614,Coccia:2012fi}. 
An emerging trend is that the optimization of all wave function parameters, including the coefficients of the contractions and the exponents of the primitive gaussians, can accelerate the convergence of many observables. As expected, different observables (such as geometries, energies, polarizabilities and vibrational frequencies) have a different convergence behavior with respect to the size of the basis set and the number and kind of parameters to be optimized, as pointed out for instance by Coccia et al. in the case of the ethyne polarizability \cite{Coccia:2012fi}. 

In QMC the interplay between the variational ansatz and the basis set size is quite intricate and not yet completely understood. The number  of variational parameters is also a crucial issue for the QMC wave function optimization, since it grows both with the wave function complexity and the size of the basis set. Also the kind  of variational parameter (linear or nonlinear) can be important for the practical stability of the wave function optimization algorithms.

In order to investigate systematically the behavior of the different variational ansatzes, together with  the choice of the basis sets for the determinantal and the Jastrow part of the wave function, in the present work we propose an extensive study of the molecular properties of the water molecule, both with all electron wave functions and pseudo potentials.
The investigation mainly focuses on the VMC scheme, although several LRDMC calculations are also reported.
We have considered as a test case the water molecule, because it is a sufficiently small system to afford different calculations of several properties, but it still preserves  a certain degree of complexity,  allowing a meaningful application of the various approaches.
Moreover, the water molecule has been widely studied and characterized both experimentally\cite{Clough:1973bh,Verhoeven:1970jd,Benedict:1956id} and by using  highly accurate {\em ab initio} computational approaches\cite{Feller:1987dm,Partridge:1997p26051,Schwenke:2000hf,Lodi:2008ic,Csaszar:2005dl,KIM:1995p23441,Feller:2009p23440}, which provide useful benchmarks for our QMC calculations. 
The accuracy of the different approaches has been tested versus a number of different properties, namely the
energy, the dipole, the quadrupole, the ionization and atomization energies, the structural minimum, the harmonic and the fundamental frequencies of vibration.  
In addition to this systematic study we introduce in the present paper a new scheme of building the atomic orbitals involved in the wave functions, called hereafter  {\em atomic hybrid orbitals}. Due to the tight relationships between the variational parameters and the basis sets, the proposed scheme would be particularly suitable and computationally convenient in the treatments of large systems with large basis sets.

The paper includes a  self contained description of the used wave function ansatzes in  Section~\ref{sec.qmc.wf}, and of the QMC techniques  in Section~\ref{sec:QMCmethods}. In these sections some novel methodological improvements are also presented. 
Together with the new hybrid orbitals, we also provide an improvement for open non-periodic systems of the {\em reweighting method} proposed by  \citet{Attaccalite:2008p12639} . 
%- GUIDA AL LAVORO: OBIETTIVO, METODO, OUTLINE DI ARTICOLO
In Section~\ref{sec.comput.det} we provide some additional  details
about the computation that we have performed, 
that are reported and discussed in Section~\ref{sec.results}, followed by a conclusive discussion in Section~\ref{sec.conclusion} of the impact of the work and of the future perspectives.

%%%%%%%%%%%%%%%%%%%%%%%%%%%%%%%%%%%%%%%%%%%%%%%%%%%%%%%%%%%%%%%%%%%%%
%%%%%%%%%%%%%%%%%%%%%%%%%%%%%%%%%%%%%%%%%%%%%%%%%%%%%%%%%%%%%%%%%%%%%

%  \input{2.QMC-WF}
%%%%%%%%%%%%%%%%%%%%%%%%%%%%%%%%%%%%%%%%%%%%%%%%%%%%%%%%%%%%%%%%%%%%%
\section{Functional form of the QMC wave function}\label{sec.qmc.wf}
%%%%%%%%%%%%%%%%%%%%%%%%%%%%%%%%%%%%%%%%%%%%%%%%%%%%%%%%%%%%%%%%%%%%%
%%%%%%%%%%%%%%%%%%%%%%%%%%%%%%%%%%%%%%%%%%%%%%%%%%%%%%%%%%%%%%%%%%%%%

% QMC w.f.: determinantal part x Jastrow
The usual form of a QMC wave function\cite{Foulkes:2001p19717} is 
the product of an antisymmetric (fermionic) function $\Psi_A$, 
and a symmetric (bosonic) exponential function $J=e^U$:
\begin{equation}
\Psi_{QMC}\left( \bar{\textbf{x}} \right) = 
  \Psi_{A} \left( \bar{\textbf{x}}\right)
  {J\left( \bar{\textbf{x}} \right)} .
\label{equ:psiQMC}
\end{equation}
Both $\Psi_A$
and $J$, 
hereafter called respectively the {\em determinantal part} and the the {\em Jastrow factor} of the wave function, 
depend on the spatial $\textbf{r}_i$ and spin $\sigma_i$ coordinates 
of the $N$ electrons in the system, being
$\bar{\textbf{x}}=\{ \textbf{x}_i \}_{i=1,\ldots,N}$
and
$\textbf{x}_i=(\textbf{r}_i,\sigma_i)$.
The determinantal part $\Psi_A$,
sum of one or more Slater determinants,
completely defines the nodal surface of $\Psi_{QMC}$, 
%(the $3N-1$ dimensional surface where $\Psi_{QMC}(\bar{\textbf{x}}=0)$,
and it is responsible for the description of the static correlation.
The Jastrow factor, explicitly dependent on the inter-electronic distances, 
describes the dynamical correlation between the electrons and is used also to satisfy the cusp conditions\cite{Kato:1957jo,Foulkes:2001p19717}. 
% The inclusion of this  correlation factor determines a severe complication in the calculation of integrals involving the total wave function $\Psi_{QMC}$, because they are no longer factorizable as in standard Hartree-Fock theory. This is the main reason why the quantum Monte Carlo method is almost unavoidable for correlated wave functions because it allows us to deal  with  high dimensional integrals with a feasible computational effort. 

% static correlation
%It has to be noticed that in those systems where  the static correlation  is important (i.e., a single Slater determinant is unable to provide a reliable and accurate description of the wave function), the Jastrow term is not enough to significantly improve  the accuracy of the wave function, because this term, being positive, cannot modify the nodal surface of $\Psi_{QMC}$ .
%When projection methods are used, as for instance DMC or LRDMC, the fixed node approximation\cite{Foulkes:2001p19717}  has to be introduced, and the quality of the results depends on the accuracy in the wave function description of the nodal surface.
%For these reasons in QMC, as in all other Quantum Chemical approaches, it is important to use an antisymmetric function $\Psi_A$ that provides a reliable description of the static correlation of the system, i.e., of the nodal surface. 

$\Psi_{QMC}$, and its constituting determinantal and Jastrow parts, are functionally dependent on some parameters, that are optimized in order to minimize the corresponding variational energy.
The optimized wave function should provide the best description of the electronic properties, and of the static and dynamical correlation, within the limitations of the considered ansatz.
However, when the number of variational parameters of the wave function increases, their optimization can become very challenging. It is therefore crucial to adopt a parametric wave function that has a large variational flexibility but, at the same time, a limited number of tunable parameters.  

% DESCRIVERE DET UP DET DOWN
%An important simplification on the QMC wave function, that gives also computational advantages, is that it is not necessary that $\Psi_A$ is antisymmetric upon exchange of electrons with unlike spins; it is enough that $\Psi_A$ is antisymmetric for exchange of electrons with like spin.
%This can be done because it can be proven that this function gives the same expectation value for any spin-independent operator of the corresponding completely antisymmetric wave function, see Sec.IV.E of \citet{Foulkes:2001p19717}.
%For example, instead of considering a single large Slater determinant including all the spin up and down electrons, it is equivalent but computationally  convenient to consider the product of two Slater determinants, with spin up  and with spin down electrons\cite{Foulkes:2001p19717}, respectively.

In the next paragraph  we will provide a synthetic description of the atomic orbitals that are used in the determinantal and the Jastrow parts of the wave function.
Next we will review the different forms for the determinantal part $\Psi_A$  
that are considered in this work, namely the Antisymmetrized Geminal Power (AGP), the single Slater Determinant (SD), and the AGP with fixed number of molecular orbitals (AGPn*). 
Afterwards we will provide a description of the Jastrow factor.

\subsection{Atomic orbitals}\label{sec.orbitals}

The choice of the primitive atomic orbitals and the contractions is important to achieve a rapid basis set convergence (BSC) and balanced calculations, both for QMC as well as  for many  other electronic structure methods.
However  in QMC calculations, at variance with other techniques, all the basis set parameters (included the exponents and the contraction coefficients) are  often optimized during the minimization of the variational energy. An appropriate choice for the contraction scheme is particularly important in the AGP wave function, since for this wave function the atomic orbital contractions and the number of wave function parameters are closely related, as we will see in Section~\ref{sec.AGP}. 

A generic atomic orbital $\phi^a_{\mu_a}(\textbf{r}_{ia})$ of the atom $a$ is  written in terms of the radial vector 
$\textbf{r}_{ia} = \textbf{r}_{i} - \textbf{R}_{a} $
connecting the nucleus of the atom $a$ with the position $\textbf{r}_{i}$ of the electron $i$.
In this work we will consider three different types of atomic orbitals: 
{\em i.} the {\em uncontracted orbitals},
{\em ii.} the {\em contracted orbitals}, and
{\em iii.} the {\em contracted atomic hybrid orbitals}.

An uncontracted orbital 
$\phi_{l,m}$, 
having azimuthal quantum number $l$ and  magnetic quantum number $m$, 
is the product of an angular part, {\em i.e.} real spherical harmonic, and a radial part.
The latter may have several functional forms;
in this work we have considered only the two most used: 
the Slater type orbitals (STO)
\begin{equation}\label{eq.STO}
\phi_{l,m}^{STO}({\bf r};\zeta) \propto r^l e^{-\zeta r} Z_{l,m}(\Omega) ;
\end{equation}
and the Gaussian type orbitals (GTO)
\begin{equation}\label{eq.GTO}
\phi_{l,m}^{GTO}({\bf r};\zeta) \propto r^l e^{-\zeta r^2} Z_{l,m}(\Omega) ,
\end{equation}
where $Z_{l,m}(\Omega)$ is the real spherical harmonic and $r=\|{\bf r}\|$.
The proportionality constant is fixed by the normalization and depends on the parameter $\zeta$.
Other parametric forms for the atomic orbitals exist, see for instance \citet{Petruzielo:2011p24345}, but are not used in this work.

In our implementation,  the nuclear cusp condition is satisfied by an electron-nucleous interaction term that is included in the Jastrow factor. 
For this reason we need atomic orbitals with no cusps at the nuclei. 
This is automatically satisfied by all the GTO and STO orbitals in Eq.~(\ref{eq.STO}) and (\ref{eq.GTO}), with the exception of  the STO orbital $s$ (i.e., $l=m=0$).
For this reason, the latter orbital is replaced by the following:
\begin{equation}\label{eq.STO.s}
\phi_{0,0}^{STO}({\bf r};\zeta) \propto (1+\zeta r) e^{-\zeta r}.
\end{equation}
%We wish to underline that 
Each of the uncontracted orbitals described above depends parametrically only on the value of the $\zeta$ in the exponent.

The contracted orbitals \emph{$\phi_{l,m}^{K}$} are simple generalizations of the uncontracted orbitals, where the radial part is the summation of the radial parts of several uncontracted orbitals (GTOs, STOs, or mixed).
Therefore a contracted orbital is:
\begin{equation}\label{eq.orb_contr}
\phi_{l,m}^{K} (\textbf{r};\{\zeta_k,c_k\}) = 
  \sum_{k=1}^{K} c_k \phi^{X_k}_{l,m}(\textbf{r},\zeta_k) 
\end{equation}
where $X_k$ can be GTO or STO, and $K$ is the number of summed uncontracted orbitals.
The number of variational parameters in $\phi_{l,m}^{K}$ is $2K-1$, given by the $K$ exponents and the $K$ coefficients, minus one due to the overall normalization of the orbital. 

In this work we have introduced and tested another type of contracted orbital, hereafter indicated with the name of {\em atomic hybrid orbital}.
It represents a further ``drastic'' generalization of the contraction of an orbital that is rather similar to the well known expansion in natural hybrid  orbitals\cite{NYO}.
It is written in the following way:
\begin{equation}
\phi_{a}(\textbf{r};\{\zeta_{k,l}, c^k_{l,m}\}) = 
  \sum_{l=0}^{l_{MAX}} 
    \sum_{k=1}^{K_l} 
      \sum_{m=-l}^{+l} 
        c^k_{l,m} \phi^{X_{k,l}}_{l,m}(\textbf{r},\zeta_{k,l}) Z_{l,m}(\Omega) 
\end{equation}
The number of parameters here is given by the sum of the number of exponents and of the coefficients. 
The number of exponents $\{ \zeta_{k,l} \}$ is given by
$n_z = \sum_l^{l_{MAX}} n_z^l$,
being $n_z^l$ the number of exponents with angular momentum $l$.
% (as usual, given $l$, all the terms for $m=-l,\ldots,+l$ will have the same exponents).
The number of coefficients $\{ c^k_{l,m} \}$ is
$n_c = \sum_l^{l_{MAX}} (2l+1) * n_z^l - 1$, the minus one being introduced for the normalization.
An atomic hybrid orbital $\phi_{a}$, related to the atom $a$, is  written as the sum of all the uncontracted orbitals, of any azimuthal and magnetic quantum numbers, that we want to use to describe the atom.
For the description of an atom it is generally required to use more than one atomic hybrid orbital (the number of which will be in the following indicated between  brace parenthesis).
 Both the exponents and the coefficients have to be conveniently optimized and in principle they can be different (especially the coefficients)  even for  different atoms of the same type appearing  in the same molecule.

% Differenza con ANO
These atomic hybrid orbitals somehow remind the well known natural orbitals\cite{NYO}, but differently from natural orbitals,  our hybrid orbitals are not necessarily orthonormal and are obtained by straightforward optimization of the energy.

%%% determinantal part:
\subsection{The AGP wave function}\label{sec.AGP}
% - AGP \citenum{Casula:2004p12689,Casula:2003p12694}

The Antisymmetrized Geminal Power %(AGP) 
is a particular pairing wave function which describes the correlations between pairs of electrons by means of a two-particle geminal function.
Initially introduced to describe spin unpolarized systems\cite{Hurley:1953fo}, it has been generalized in order to describe also spin polarized system, i.e., systems with unpaired electrons\cite{Coleman:1972ds,Casula:2003p12694}.
Hereafter we limit our description to the case of spin unpolarized systems,
and we refer to the work of \citet{Casula:2003p12694} for the generalization to spin polarized systems.
%,{Marchi:2009p12614}.

A spin unpolarized system, with zero total spin, has  the number $N^{\uparrow}$ of electrons with spin up   equal to the number $N^{\downarrow}$ of  electrons with spin down and to one half of the total number of electrons $N$.
In this case the AGP wave function is:
\begin{equation}
\Psi_{AGP} \left( \bar{\textbf{x}}  \right) = \hat{\cal A} \left[
   G \left( \textbf{x}_{1};\textbf{x}_{2} \right)
   G \left( \textbf{x}_{3} ;\textbf{x}_{4}\right)
   \ldots
   G \left( \textbf{x}_{N-1} ;\textbf{x}_{N}\right)  \right] 
\label{equ:AGP}
\end{equation}
where 
%${\textbf x}_i = (\textbf{r}_{i},\sigma_i)$ denotes formally the cartesian coordinates $\textbf{r}_{i}$ of the electron $i$ and its spin $\sigma_i$,  
$\hat{\cal A}$ is the antisymmetric operator, and 
$G \left( \textbf{x}_{i};\textbf{x}_{j} \right)$ 
is the geminal function,  a  product of a spin singlet and a symmetric spatial wave function 
$g\left( \textbf{r}_{i},\textbf{r}_{j} \right)$: 
\begin{equation}
G ( \textbf{x}_{i} ;\textbf{x}_{j}) = 
   g\left( \textbf{r}_{i},\textbf{r}_{j} \right) 
     \frac{ \delta(\sigma_i,\uparrow)\delta(\sigma_j,\downarrow)-\delta(\sigma_i,\downarrow)\delta(\sigma_j,\uparrow)}{\sqrt{2}}
%   \frac{  \left| \uparrow\right\rangle_{i} \left| \downarrow\right\rangle_{j}-\left|\uparrow \right\rangle_{j}\left|\downarrow \right\rangle_{i} }{\sqrt{2}} 
\end{equation}
It can be shown\cite{Casula:2003p12694} that the spatial part of $\Psi_{AGP}$ can be written as the determinant of a matrix ${\bf M}^{AGP}$ of dimension $N/2 \times N/2$ whose elements are:
$ M^{AGP}_{ij} = g\left( \textbf{r}_{i},\textbf{r}_{N/2+j} \right) $, 
with $i,j=1,..,N/2$.

The spatial geminal function $g$ is written in terms of single electron atomic wave functions:
\begin{equation}
g\left( \textbf{r}_{i},\textbf{r}_{j} \right) =
  \sum_{a,b}^{M} \sum_{\mu_a}^{L_a} \sum_{\mu_b}^{L_b} 
    \lambda_{\mu_a, \mu_b}^{a,b}
    \phi^{a}_{\mu_a}\left(\textbf{r}_{ia}\right)
    \phi^{b}_{\mu_b}\left(\textbf{r}_{jb} \right)
\label{equ:phiG}
\end{equation} 
where $a$ and $b$ are the atom indexes, running from 1 to the number  $M$  of atoms in the system, and $\mu_a$ labels the $L_a$ local atomic orbitals $\phi^{a}_{\mu_a}$ used to describe the atom $a$.
The local orbital $\phi^{a}_{\mu_a}$ is a function of the difference
$\textbf{r}_{ia}=\textbf{r}_{i}-\textbf{R}_{a}$
between the position $\textbf{r}_i$ of the electron $i$ and the position $\textbf{R}_{a}$ of the nucleus $a$. % and that correspond to quantum numbers ($n,l,l_z$). 
The $\lambda^{a,b}_{\mu_a,\mu_b}$ coefficients in Eq.~(\ref{equ:phiG}) represent the weight of the superposition of different orbitals, analogously to the valence bond representation, or in other words the contribution of the atomic orbital $\mu_a$ of the atom $a$ and the atomic orbital $\mu_b$ of the atom $b$ to the formation of the chemical bond between $a$ and $b$.
The set of coefficients $\lambda^{a,b}_{\mu_a,\mu_b}$ defines the %symmetric 
square matrix $\Lambda$ of size $L \times L$, where $L=\sum_a^M L_a$ is the total number of atomic orbitals defining our basis set.
In order to ensure that the total spin is conserved,
 the condition  
$\lambda^{a,b}_{\mu_a,\mu_b}=\lambda^{b,a}_{\mu_b,\mu_a}$ is required, i.e., the  $\Lambda$ matrix is symmetric.
This implies that the number of independent parameters in the $\Lambda$ matrix is $L(L+1)/2$.
Moreover,
if molecular symmetries are present,
it is possible to introduce additional constraints on the elements of the $\Lambda$ matrix, that can significantly reduce the number of independent parameters of the wave function\cite{Casula:2004p12689}.

% diagonalize lambda matrix
In the following sections we will consider other functional forms for the determinantal part of the wave function.
The relation between the AGP and  those other wave functions can be easily understood by rewriting the pairing function $g( \textbf{r}_{i},\textbf{r}_{j} )$ in an equivalent way, where the $\Lambda$ matrix is diagonalized.
In order to diagonalize $\Lambda$, it is convenient to  take into account that the atomic orbitals are not necessarily orthogonal each other, namely the overlap matrix 
$ S^{a,b}_{\mu_a,\mu_b} \equiv \left< \phi^{a}_{\mu_a} \right| \left. \phi^{b}_{\mu_b} \right> \ne {\bf 1}$, 
and by using a standard generalized diagonalization:
\begin{equation} \label{gendiag}
\Lambda {\bf S P} = {\bf P} \bar \Lambda .
\end{equation} 
In \ref{gendiag} 
each column of the matrix $\bf P$ represents a generalized eigenvector of $\Lambda$,
and the  corresponding eigenvalues  $\bar\lambda_\alpha$ constitute the elements of 
the diagonal matrix ${\bar \Lambda} = diag(\bar\lambda_1,\ldots,\bar\lambda_L)$,
sorted in decreasing order of their absolute value:
$|\bar\lambda_1| \ge |\bar\lambda_2| \ge \ldots \ge |\bar\lambda_L |\ge 0$. 
Thus, from $ {\bf P}^T {\bf S P} = {\bf 1}$, by right multiplying both sides of \ref{gendiag}  for the 
matrix ${\bf P}^T= ({\bf SP})^{-1}$ we obtain   $\Lambda = {\bf P}  {\bar \Lambda}  {\bf P}^T$.
Then, by substituting it in Eq.~(\ref{equ:phiG}), we finally obtain that 
the pairing function is:
\begin{equation}
g\left( \textbf{r}_{i},\textbf{r}_{j} \right) =
  \sum_{\alpha=1}^{L}  
    \bar \lambda_{\alpha} 
    \Phi_{\alpha}\left(\textbf{r}_{i}\right)
    \Phi_{\alpha}\left(\textbf{r}_{j} \right)
\label{equ:phiG_2}
\end{equation} 
where we have defined the orthogonal single particle orbitals:
\begin{equation}
\Phi_\alpha(\textbf{r}_i) = \sum_a^M \sum_{\mu_a}^{L_a} P_\alpha^{a,\mu_a} \phi^a_{\mu_a}(\textbf{r}_{ia}) 
\end{equation}
which will be afterwards named molecular orbitals (MOs).
The complete basis set (CBS) for the pairing function in Eq.~(\ref{equ:phiG_2}) is reached in the limit $L \to \infty$, namely in the limit of considering an infinite number of MOs.

\subsection{The SD wave function}\label{sec.SD}
%\subsection{SD and JSD wave function}\label{sec.SD}
% - molecular orbitals \cite{Marchi:2009p12614}

It can be reasonably expected that the leading terms in the expansion of the pairing function $g$ in Eq.~(\ref{equ:phiG_2}) are provided by a limited set of MOs associated to the eigenvalues $\bar \lambda_\alpha$ 
largest in absolute value $| \bar \lambda_\alpha|$.  
Therefore, by considering a truncated pairing function, where only a subset $n\ll L$ of the MOs are used instead of all the $L$ orbitals appearing in Eq.~(\ref{equ:phiG_2}), 
we have that, if $n$ is large enough to provide the leading behavior of $g$,  the quality of the parametrical wave function is not significantly affected. This truncation substantially reduce the number of variational parameters (working with a $n\times L$ matrix instead of the larger $L\times L$ matrix $\Lambda$).

The lowest number of orbitals that we have to consider to describe an unpolarized system of $N$ electrons is exactly equal to the number of electron pairs $N/2$. %, because with a lower number of MOs it is not possible to write an antisymmetric function. 
Thus, within this minimal approach, the pairing function is:
\begin{equation}
g^{SD}\left( \textbf{r}_{i},\textbf{r}_{j} \right) =
  \sum_{\alpha=1}^{N/2}  
    \bar \lambda_{\alpha} 
    \Phi_{\alpha}\left(\textbf{r}_{i}\right)
    \Phi_{\alpha}\left(\textbf{r}_{j} \right)
\label{equ:phiG_SD}
\end{equation} 
It can be seen that the antisymmetrization operator in Eq.~(\ref{equ:AGP}), applied to the truncated pairing function in Eq.~(\ref{equ:phiG_SD}), singles out only one Slater determinant (SD), therefore hereafter this wave function will be referred as the SD function.
We also observe that the MO weights $\bar \lambda_{\alpha}$ affect only the overall pre factor of this Slater determinant, so that their actual values are irrelevant in this case.
This SD function is the equivalent of a restricted Hartree-Fock (RHF) function, in HF calculations, or of a restricted Kohn-Sham function, in DFT calculations.
However, within a QMC scheme, a Jastrow factor is always introduced in the wave function, in order to enhance the description of the dynamical correlations between the electrons. 
When a Jastrow factor, of the type that will be described in Section~\ref{sec.Jastrow}, is applied to a SD function, it will be  referred as Jastrow correlated single determinant (JSD) function.

It has been observed in several cases\cite{Foulkes:2001p19717,Casula:2003p12694,Marchi:2009p12614} that a JSD wave function is able to describe the atoms with an high level of accuracy.
However, for several molecular systems the JSD function is unable to provide an equally accurate and reliable description of several properties.
For these cases, the JAGP function results to provide a much more accurate description.
An important property to be considered for a reliable description of a molecular system is whether the wave function is size consistent.
The JAGP is size consistent\cite{Sorella:2007p12646,Neuscamman:2012hm} 
in all cases where the JSD is size consistent, namely when the spin/angular momentum  of the compound is the sum of the spin/angular momentum  of the fragments.
A remarkable exception is when the fragments are S=1/2 atoms, such as the $H_2$ and $F_2$,  where the JAGP is size consistent and the JSD is not. 
%This size inconsistency affects the values of the atomization energies, the bond energies, and the shape op the PES out of the minimum. 
%
In addition to this,  there are several other reasons to use the JAGP rather than a simpler JSD:
\begin{description}
\item{\em i.} It is more accurate at a similar computational cost.
\item{\em ii.} It is a more compact representation of the determinantal part within a localized atomic basis, thus  
it is simple to implement constraints which avoid to optimize variationally irrelevant parameters.
For instance the symmetries, such as the translation, can be simply implemented as constraints in the $\Lambda$ matrix.
\item{\em iii.} %Since geminals are written in terms of local atomic orbitals, it is simple to implement constraints which avoid to optimize variationally irrelevant parameters. For instance, 
For large systems, a big reduction of the variational freedom is possible by disregarding matrix elements of $\Lambda$ corresponding to localized orbitals very far in space.
\end{description}

\subsection{The AGPn* wave function}\label{sec.AGPn}

% - AGP n*
%
% size consistency of JAGP
% [23] S. Sorella, M. Casula, and D. Rocca, J. Chem. Phys. 127, 014 105 (2007).
%Since JAGP satisfies the size consistency of singlet fragments, provided that the Jastrow term is large enough\cite{Sorella:2007p12646,Neuscamman:2012hm}. 

In order to improve upon the simple JSD wave function for a more accurate 
description of molecules, we have to include in the pairing function $g$ a number $n$  of MOs larger than $N/2$, $n=N/2$ corresponding to the  JSD function. 
Since a JSD function provides an accurate description of the atoms, a natural criterium for the choice of the number of MOs is by requiring that, when the atoms are at large distances, we cannot obtain an energy below the sum of the JSD atomic energies.
The number $n^*$ of MOs defined in this way is  determined by the requirement that:
\begin{equation}
n^* \le \sum_a^M N^{\uparrow}(A_a) + m - 1
\end{equation}
where
$A_1,\ldots,A_M$ identify the $M$ atoms forming the system, 
$N^{\uparrow}(A_a)$ is the number of spin up electrons for a description of the atom $A_a$, and 
$m$ is equal to the minimum number of identical atoms in the system
(for further details and for a discussion of the case of polarized systems see \citet{Marchi:2009p12614}).

Therefore, the pairing function associated to $n^*$ is defined as:
\begin{equation}
g^{n^*}\left( \textbf{r}_{i},\textbf{r}_{j} \right) =
  \sum_{\alpha=1}^{n^*}  
    \bar \lambda_{\alpha} 
    \Phi_{\alpha}\left(\textbf{r}_{i}\right)
    \Phi_{\alpha}\left(\textbf{r}_{j} \right)
\label{equ:phiG_AGPn*}
\end{equation} 
The Jastrow correlated AGP function obtained by the antisymmetrization of the geminal $g^{n^*}$ will be hereafter indicated with JAGPn*.\cite{Marchi:2009p12614}

\subsection{ The Jastrow factor }\label{sec.Jastrow}
%short description of Jastrow part (1,2,3,4 body)

The bosonic Jastrow term, $J=e^U$, represents a compact and efficient way to introduce explicitly the electronic correlation in the wave function, because it depends directly on distances between electrons.
Several different implementations of the the Jastrow term are used in the QMC codes.  
The Jastrow that we have used in this work consists of several terms that account for the 2-body, 3-body and 4-body interaction between the electrons and the nuclei.
The exponent $U$ of the Jastrow factor can therefore  be conveniently written as the sum of three independent functions:
\begin{equation}
U = U_{en} + U_{ee} + U_{een[n]} .
\end{equation}

The leading contribution is given by 
$U_{ee}\left( \bar{\textbf{r}} \right)$, that is a homogeneous \textit{two electron} interaction term.
It depends only on the relative distance between pairs of electrons and it improves the electron-electron correlation, besides satisfying the electron-electron cusp condition for unlike spin. 
The cusp condition for like spin is not satisfied, as this would lead to spin 
%contamination\cite{Filippi:1996eb,Casula:2003p12694}, however this is a minor problem because the probability for like spin electrons to be close is very small, just for the Pauli principle.  
contamination\cite{Filippi:1996eb,Casula:2003p12694}. However this is a minor problem because the probability for like spin electrons to be close is very small, because of the Pauli principle.  
The functional form that we have used for $U_{ee}$ is:
% 2bJ
\begin{equation}
U_{ee}\left(\bar{\textbf{r}}\right) = 
  \sum_{i<j}^N u_2(r_{ij})
\label{equ:2BJas}
\end{equation}
where $r_{ij}=\|\textbf{r}_{i}-\textbf{r}_{j}\|$ is the distance between electrons $i$ and $j$, and 
$ u_2(x)= { 1-e^{- b_2 x} \over 2 b_2 } $
is a function of the variational parameter $b_2$.

The term $U_{en}$ is a \textit{one electron} interaction term which improves the electron-nucleus correlation and satisfies the nuclear cusp condition. 
%For this reason it is particularly useful with gaussian type basis set (GTO), for {\em all electron} calculations.
Its functional form is:
% 1bJ
\begin{equation}
U_{en}( \bar{\textbf{r}},\bar{\textbf{R}} ) =
  - \sum_{a}^M 
      \left[
        (2 Z_a)^{3\over4} 
          \sum_{i}^N  
            u_1\left( \sqrt[4]{2 Z_a} r_{ia} \right) 
      \right]
  + \sum_{a}^M \sum_{\nu_a}^{L^J_a}
      \left[ \sum_{i}^N 
        f^a_{\nu_a} \chi^a_{\nu_a}(\textbf{r}_{ia}) 
      \right]
\label{equ:1BJas}
\end{equation}
where the vector 
$\textbf{r}_{ia}=\textbf{r}_i-\textbf{R}_a$
is the difference between the position of the nucleus $a$ and the electron $i$,  
$r_{ia}=\|\textbf{r}_{ia}\|$ is their distance, 
$Z_a$ is the electronic charge of the nucleus $a$,
 $L^J_a$ is the number of atomic orbitals $\chi^a_{\nu_a}$ that are used to describe the atom $a$ (they are similar to the $\phi^a_{\mu_a}$ orbitals used for the determinantal part), 
$f^a_{\nu_a}$ are variational parameters and
the function 
$ u_1(x)= { 1-e^{- b_1 x} \over 2 b_1 } $
is used to satisfy the electron-nucleus cusp condition, and it depends parametrically on the value of $b_1$.

The term $U_{een[n]}$ is an inhomogeneous {\em two electron} interaction term, and it has the following form:
% 3bJ
\begin{equation}
U_{een[n]}\left( \bar{\textbf{r}}, \bar{\textbf{R}} \right) = 
  \sum_{i<j}^N \left[
    \sum_{a,b}^M 
      \sum_{\nu_a}^{L_a^J} \sum_{\nu_b}^{L_b^J}
        f^{a,b}_{\nu_a,\nu_b} 
          \chi^a_{\nu_a}\left(\textbf{r}_{ia}\right)
          \chi^b_{\nu_b}\left(\textbf{r}_{jb}\right)
  \right] ,
\label{equ:3BJas}
\end{equation}
where the $\chi^a_{\mu_a}$ are the same atomic orbitals that appear also in $U_{eu}$, second term in the right hand side of Eq.~(\ref{equ:1BJas}), and $f^{a,b}_{\nu_a,\nu_b}$ are variational parameters. 
In Eq.~(\ref{equ:3BJas}) are included both 
the three body e-e-n interactions and 
the four body e-e-n-n interactions, 
for $a=b$ and for $a\ne b$, respectively.

% spostata descrizione core il Computational Details

%%%%%%%%%%%%%%%%%%%%%%%%%%%%%%%%%%%%%%%%%%%%%%%%%%%%%%%%%%%%%%%%%%%%%
%%%%%%%%%%%%%%%%%%%%%%%%%%%%%%%%%%%%%%%%%%%%%%%%%%%%%%%%%%%%%%%%%%%%%

%  \input{3.QMC-Methods}
%%%%%%%%%%%%%%%%%%%%%%%%%%%%%%%%%%%%%%%%%%%%%%%%%%%%%%%%%%%%%%%%%%%%%
\section{Quantum Monte Carlo methods}\label{sec:QMCmethods}
%%%%%%%%%%%%%%%%%%%%%%%%%%%%%%%%%%%%%%%%%%%%%%%%%%%%%%%%%%%%%%%%%%%%%
%%%%%%%%%%%%%%%%%%%%%%%%%%%%%%%%%%%%%%%%%%%%%%%%%%%%%%%%%%%%%%%%%%%%%

% cite:
% Casula:2003p12694    minimization (atoms)
% Casula:2004p12689    minimization (molecules)
% Sorella:2005p14143    optimization VMC in comparison with other methods
% Casula:2005p14138   LRDMC 
%% Umrigar:2007p12662   opt.VMC (Jastrow+determinantal) -> small sign problem
% Sorella:2007p12646   opt.vmc + LRDMC a->0 + size consistency
% Attaccalite:2008p12639   forces regularization + dynamics
% Marchi:2009p12614   MOs
% Casula:2010p14082   LRDMC with nonlocal pseudopot.
% Azadi:2010p14081   DFT nel turbo
% Sorella:2010p23644   forze VMC con Algorithmic differentiation 
% Sorella:2011kd   linearized auxiliary fields MC
% Barborini:2012iy   ethylene: vmc, forze
% Mazzola:2012ch   dyn VMC finite T
% Coccia:2012fi    VMC con dipoli e quadrupoli e opt.

%\subsection{Variational Monte Carlo and Lattice Regularized Diffusion Monte Carlo}
%\subsection{Wave function optimization}

% VMC 

The expectation value of an observable $\cal O$, with corresponding quantum mechanical operator $\hat O$,
is evaluated as 
$\left< {\cal O} \right> \equiv {\left<\psi | {\hat O} |\psi \right> \over \left<\psi | \psi \right>}$, 
involving the computation of $3N$-dimensional integrals.
Differently from HF or post-HF approaches, with QMC wave functions these integrals do not factorize, 
due to the presence of the Jastrow factor.
In Section~\ref{sec.stochastic.eval} we  review some aspects about the stochastic approach adopted to evaluate these integrals within VMC.
In Section~\ref{sec.energy} we  discuss the specific case of the energy evaluation, in Section~\ref{sec.wf.opt} the variational optimization of the wave function parameters, in Section~\ref{sec.force} the force evaluation, and in Section~\ref{sec.reweight} the reweighting technique used to have a well behaved expectation value of the force for open systems (namely, having finite variance).
Next, in Section~\ref{sec:dipole.quadrupole} we discuss the dipole and quadrupole evaluations.
Finally, in Section~\ref{sec.FN}, we briefly review some aspects of the projection Monte Carlo approaches.

%In the wave-function-based methods, in order to evaluate the expectation value 
%$\left< {\cal O} \right> \equiv {\left<\psi | {\hat O} |\psi \right> \over \left<\psi | \psi \right>}$ 
%for an observable $\cal O$, as the energy or the force, where  $\hat O$ is the corresponding quantum mechanical operator, we have to compute a $3N$-dimensional integral, i.e., we have to integrate over the coordinates of the $N$ electrons.
%Using the typical wave functions adopted in HF or post-HF calculations,  this $3N$-dimensional integral can be factorized in lower dimensional integrals.
%However in our case the presence of the Jastrow term in the wave function  implies that the $3N$-dimensional integral is not factorable.
%For this reason the standard variational Monte Carlo\cite{Foulkes:2001p19717} technique has been used.

\subsection{Stochastic evaluation of the expectation value of an observable }\label{sec.stochastic.eval}

VMC is a stochastic method for the estimation of the expectation value $\left< {\cal O} \right>$ 
associated to a parametric wave function $\psi$.
The method is based on the fact that any expectation value $\left< {\cal O} \right>$ can be rewritten as:
\begin{equation}\label{eq.VMC}
\left< {\cal O} \right> = { \left< O_L(\bar {\bf x}) 
                              \psi(\bar {\bf x})^2 / W(\bar {\bf x}) 
                            \right>_{P(\bar {\bf x})} \over 
                            \left< \psi(\bar {\bf x})^2 / W(\bar {\bf x}) \right>_{P(\bar {\bf x})} }
\end{equation}
where 
$O_L(\bar {\bf x}) \equiv \psi(\bar {\bf x})^{-1} \hat O \psi(\bar {\bf x})$ 
is the so called {\em local value} of the operator $\hat O$ calculated in the specific electronic configuration $\bar {\bf x}$, $P(\bar {\bf x})$ is an appropriately chosen probability density distribution determined by a known positive weight $W(\bar {\bf x})$ up to a normalization constant, 
namely $ P(\bar{\bf x}) = W(\bar { \bf x } ) / \int  W(\bar {\bf x }^\prime) d\bar {\bf x }^\prime $   and 
$\left< f(\bar {\bf x})/W(\bar {\bf x}) \right>_{P(\bar {\bf x})}$ represents the expectation value $E[f]$ of a function $f(\bar {\bf x})$ that is calculated by sampling over a probability distribution $P(\bar {\bf x})$ the function $f(\bar {\bf x})/W(\bar {\bf x})$.
The most common and simple choice for the positive weight $W$ is 
%$\Pi = \psi^2/\left< \psi|\psi \right>$, 
$W(\bar {\bf x}) =  \psi(\bar {\bf x})^2$, 
in which case the denominator in Eq.~(\ref{eq.VMC}) is identically one and the expression for $\left< {\cal O} \right>$ simplifies in:
\begin{equation}\label{eq.VMC.standard}
\left< {\cal O} \right> = \left< O_L(\bar {\bf x}) \right>_{\Pi(\bar {\bf x})}
\end{equation}
which is usually referred as {\em standard sampling}.
Notice that in quantum Monte Carlo it is not necessary to know the rather involved 
normalization constant 
$ \int  W( \bar {\bf x }^\prime) d\bar {\bf x }^\prime $ to generate configurations according to the probability distribution 
$P (\bar {\bf x }) $. Only weight ratios $W(\bar {\bf x }^\prime)/W (\bar {\bf x })$  between different
 cofigurations are necessary. This makes 
the variational quantum Monte Carlo computationally feasable, as long as the weight $W$ is known and easy to compute.

% stochastic - Metropolis
Within VMC, the expectation values appearing in the right hand side of Eqs.~(\ref{eq.VMC}) or  (\ref{eq.VMC.standard}) are estimated statistically.
In particular, in Eq.~(\ref{eq.VMC})
the desired expectation value $ \left< {\cal O} \right> $
is calculated as $ { E(O) \over E(D) } $,
being the nominator 
$E[O]=\left<O_L(\bar {\bf x}) \psi(\bar {\bf x})^2  /W(\bar {\bf x})\right>_{P(\bar {\bf x})}$ 
and the denominator 
$E(D)= \left<\psi(\bar {\bf x})^2/W(\bar {\bf x})\right>_{P(\bar {\bf x})}$. 
%In the standard sampling case $E(D)=1$.
Both the numerator and the denominator can be computed 
by generating a finite set of $S$ independent points 
$\{ \bar {\bf x}_s \}_{s=1,\ldots,S}$, 
%${\bar {\bf x}}_s$, 
distributed according to the probability density distribution $P(\bar {\bf x})$ and typically generated with the Metropolis algorithm. 
Then one can compute $E(O)$ and $E(D)$ by standard averaging an appropriate function:
\begin{equation}\label{eq.estimate.MC}
E(f) \equiv  {1 \over S} 
\sum_{s=1}^{S} {f(\bar {\bf x}_s) \psi(\bar {\bf x}_s)^2\over W(\bar {\bf x}_s)} .
\end{equation}
%where $f=O_L$ ($f=1$) for the numerator (denominator) in Eq.~(\ref{eq.VMC}).
For a large but finite  sampling $S$, the estimates $A_S[f,P]$ for the numerator and the denominator 
are affected by very correlated stochastic errors $\sigma_S[f,P]$, 
therefore special techniques are required to evaluate how this  error affects the uncertainty in their ratio,  whenever a non trivial reweighting technique is employed.
The standard deviation $\sigma_S[f,P]$ is defined as the square root of the variance of 
the estimate $A_S[f,P]$.
If we assume the applicability of the central limit theorem, which in particular requires that the second moment of the probability distribution of $f(\bar {\bf x}_s) \psi(\bar {\bf x}_s)^2/W(\bar {\bf x}_s)$ exists, we have that the probability distribution for the estimate $A_S[f,P]$ is normally distributed with mean $E[f]$ and  standard deviation: 
\begin{equation}\label{eq.sigma.MC}
\sigma_S[f,P] \equiv \sqrt{ VAR\{ A_S[f,P] \} } 
              = \sqrt{ {1\over S} VAR\left\{ {f(\bar {\bf x}_s) \psi(\bar {\bf x}_s)^2\over W(\bar {\bf x}_s)}\right\} } \;.
\end{equation}
% validity CLT
For the sake of completeness it should be mentioned that the applicability of the central limit theorem depends on some properties of the probability distribution of 
$f(\bar {\bf x}_s) \psi(\bar {\bf x}_s)^2 /W(\bar {\bf x}_s)$. The fact that the second moment exists only ensures the applicability of the theorem in its most general form, where the normality of the distribution is reached in the limit of infinite sampling $S\to \infty$.
For a finite sampling $S<\infty$ 
the normal distribution is not generally satisfied, as it was indeed  observed by J.R.~Trail\cite{Trail:2008ft} in the form of heavy tails.

% uncertainty 1/sqrt(s)
Observe in Eq.~(\ref{eq.sigma.MC}) that the margin of uncertainty for the estimate of $E[f]$ using $A_S[f]$ goes to zero in the limit of infinite sampling $S\to \infty$.
Moreover, if $VAR\left\{ {f(\bar {\bf x}_s)\psi(\bar {\bf x}_s)^2/W(\bar {\bf x}_s)}\right\}$ is finite, we have that the uncertainty $\sigma_S[f,P]$ on $A_S[f]$ converges to zero as $1/\sqrt{S}$, which is a very favorable scaling considering that there is no dependence on the dimensionality of the space ($3N$) where the sample points $\{\bar {\bf x}_s\}$ are defined.

% important sampling - zero variance
\ref{eq.sigma.MC} also sheds lights on the importance of the probability density function $P$. 
A bad choice of $W$ leads to a variance  $VAR\left\{ {f(\bar {\bf x}_s)\psi(\bar {\bf x}_s)^2/W(\bar {\bf x}_s)}\right\}$ that is not even finite, whereas a good $W$ yields a finite value of $VAR\left\{ {f(\bar {\bf x}_s)\psi(\bar {\bf x}_s)^2/W(\bar {\bf x}_s)}\right\}$, as we will see in the following sections.
Moreover, whenever the estimator $O_L(\bar {\bf x}) =\bar O$ 
is independent of $\bar {\bf x}$ we see an 
important property in the calculation of physical expectation values, namely that, for any 
choice of the weight $W$, the evaluation of the ratio:
\begin{equation}
\left< {\cal O} \right>  \equiv {  
\sum_{s=1}^{S} {O_L(\bar {\bf x}_s) \psi(\bar {\bf x}_s)^2\over W(\bar {\bf x}_s)} \over 
\sum_{s=1}^{S} {\psi(\bar {\bf x}_s)^2\over W(\bar {\bf x}_s)}
}
\end{equation}
yields always the same value $\bar O$, namely has zero variance, regardless of the fact that both the numerator and the denominator may have finite variances.  
This highlights once more the fact that a method like bootstrap or jackknife\cite{KUNSCH:1989ws,Wolff:2004cu} is necessary to exploit 
the correlation between the numerator and the denominator in the evaluation of the standard deviation 
corresponding to the physical average $\left< {\cal O} \right>$.
 
%In particular, there exists the limit case where $P$ could be chosen in order to make the value ${f(\bar {\bf x}_s)/W(\bar {\bf x}_s)}$ constant independently on the point $\bar {\bf x}_s$ where it is calculated.
%This would lead to a variance that is zero and to an estimation in Eq.~(\ref{eq.estimate.MC}) which is exact independently to the sampling size $S$.
%If this happens, we say that our estimation of $\left< \cal O \right>$ fulfills the so called {\em zero-variance principle}.

In order to simplify the notation, in the following sections 
the functional dependence of the wave function $\psi$, the local operator $O_L$ and the density probability distributions $P$ and $\Pi$ on $\bar {\bf x}$ , 
will be left implicit.

% energy evaluation
\subsection{Energy evaluation}\label{sec.energy}
The most important quantity that is evaluated in VMC is the energy. %, {\em i.e.}, the expectation value of the Hamiltonian  operator $\hat H$.
Considering Eq.~(\ref{eq.VMC}) and (\ref{eq.VMC.standard}), the energy evaluation is determined by the values of the {\em local energy} 
$H_L ( \bar {\bf r} ) \equiv \left( \psi^{-1} \hat H \psi \right)_{\bar{\bf r}}$,
being  $\hat H$ the Hamiltonian  operator.
For instance, using the standard sampling technique, Eq.~(\ref{eq.VMC.standard}), the VMC evaluation of the energy ${\cal E}[\psi]$ for the wave function $\psi$, involves the calculation of 
\begin{equation}\label{eq.E_VMC}
{\cal E}[\psi] = { \int H_L ( \bar { \bf r } ) \psi^2 d\bar {\bf r} \over  \int \psi^2 d\bar {\bf r}},
\end{equation}
where the integration is over the $3N$ Cartesian coordinates $\bar {\bf r}$ of the electrons.

% eigenfunction -> zero variance
If we consider a wave function $\psi_i$, which  is an eigenfunction of the Hamiltonian with eigenvalue $E_i$, the corresponding local energy is $H_L[\psi_i]=E_i$ independently of the point $\bar {\bf x}_s$ where it is evaluated.
This is true in particular for the ground state (GS) of the system, that is typically the target of electronic structure calculations.
%SSS
%{\bf  NO. The zero variance is always satisfied if you consider the correlation between numerator  and the denominator, you will get always the same exact number even by reweighting with a random  W. Please correct.} 
This shows that, in case we are using an exact eigenfunction of the system and  the  standard sampling,  the zero-variance principle is satisfied. Moreover, it can be seen that also by sampling with a general weight W, as in Eq.~(\ref{eq.VMC}), an exact eigenfunction always fulfills the zero-variance principle.

% nodal surface
However, in proximity of the {\em nodal surface} 
%(i.e., the ipersurface in the electronic configurational space defined by the condition $\psi(\bar {\bf x})=0$) 
the local energy $H_L$ is divergent, unless we are sampling an exact eigenfunction of the Hamiltonian.
% with standard sampling
Indeed, 
if an electron in the system is close, say at a distance $\delta \ll 1$, to the nodal surface, 
we have that the wave function $\psi$ vanishes linearly with this distance, i.e., $\psi \propto \delta$, 
but for a generic $\psi$ that is not an exact eigenfunction of $H$, we have that $\hat H \psi \propto 1$, 
therefore the local energy diverges as $H_L \propto \delta^{-1}$. 
The application of the standard sampling, Eq.~(\ref{eq.VMC.standard}), leads to the  integral $\int H_L \psi^2 d\bar {\bf r}$, 
that in the proximity of this divergence is $\propto \int_0^1 \delta d\delta$, therefore it is well behaved.
In order to have a stochastic error on $\left<\cal H \right>$ that converges to zero as $1/\sqrt{S}$,
% being $S$ the sampling size, 
it is also necessary that the variance is well behaved.
The calculation of the variance for the standard sampling leads to the integral $\int {H_L}^2 \psi^2 d\bar {\bf r}$, 
that in the proximity of the nodal surface is $\propto \int_0^1 1 d\delta$, therefore the variance of the energy is also finite.

%  with reweight
However the standard sampling approach is problematic for the estimation of the nuclear forces, as it will be shown in Section~\ref{sec.force}, because its variance is not finite due to the divergences in proximity of the nodal surface. 
In order to overcome this problem, we have sampled both the energies and the forces using Eq.~(\ref{eq.VMC}), with a density probability distribution $P$ that is proportional to $\psi^2$ everywhere except in proximity of the nodal surface, where its value is a non zero constant.
The details of this sampling function $P$ will be given in Section~\ref{sec.reweight}, and the method is called {\em reweighting sampling}.
By  using the reweighting sampling a stochastic  evaluation  of the nuclear forces as well as of the energy remains well behaved.

It is worth mentioning that other divergences can exist in the local energy,  besides the one in proximity of the nodal surface, namely in the following cases: 
(i) the electron-nucleus coalescence, 
(ii) the electron-electron coalescence,
and for open systems also (iii) for electrons approaching infinity.
However, for the wave function we have considered in this work, the first two cases are already managed by the Jastrow factor, 
through the terms in Eq.~(\ref{equ:1BJas}) and (\ref{equ:2BJas}) that satisfy respectively the nuclear-electron and the electron-electron cusp conditions.
The divergence (iii) will be discussed in Section~\ref{sec.reweight}.

\subsection{Wave function optimization}\label{sec.wf.opt}

% SR: \cite{Sorella:2007p12646,Umrigar:2007p12662}
% SRH: \cite{Toulouse:2007p27522,Toulouse:2008p27527}
% with MOs: \cite{Marchi:2009p12614} projection on a rank-r geminal

% variational principle
According to the variational principle, the exact ground state energy $E_{GS}$ represents the lowest bound for any variational wave function, including the parametrized wave functions that are considered in VMC calculations.
The set of parameters $\bar \alpha$ of the variational wave function are therefore optimized in order to minimize the corresponding variational energy ${\cal E}[\psi_{\bar \alpha}]$. As a consequence of the fact that the wave function is approaching to an eigenstate, also the variance of the energy decreases and approaches zero.

% optimization w.f. - parameters
In order to optimize the variational parameters $\bar \alpha$ we use in this work  the stochastic reconfiguration\cite{Sorella:2007p12646,Umrigar:2007p12662} method (SR) and the more recent linear method\cite{Sorella:2005p14143,Toulouse:2007p27522,Toulouse:2008p27527} based on an 
efficient estimate of the Hessian matrix  (SRH).
Both SR and SRH (for the details we refer to the cited references) are iterative methods where the variational parameters are evolved by incremental changes 
$\bar \alpha \to \bar \alpha'=\bar \alpha + \Delta \bar \alpha$  
using the generalized force 
$\bar f \equiv -{ \partial {\cal E}[\psi_{\bar \alpha}] \over \partial \bar \alpha }$ 
acting on the parameters,
and the matrix $\bf S$, 
whose elements are 
$S_{kl} \equiv \left< 
 {\partial \over \partial \alpha_k } {\psi_{\bar \alpha} \over \|\psi_{\bar \alpha}\|} \right| \left.
 {\partial \over \partial \alpha_l } {\psi_{\bar \alpha} \over \|\psi_{\bar \alpha}\|} \right> $,
that takes into account the correlation between the parameters in the wave function. 
In SRH also partial information of the energy second  derivatives is taken into account, 
and the method is generally faster and more efficient.

In particular, within SR, a generic parameter $\alpha_k$ is changed at each iteration by 
$ \Delta \alpha_k = \Delta t {S^{-1}}_{kl} f_l $, being $\Delta t$ an appropriate small number.
In case $\bf S$ is the identity matrix, the SR optimization would correspond to a simple {\em steepest descent} optimization of the wave function.
The computational advantage of SR over a simple steepest descent, in terms of velocity of convergence, has been observed in several cases\cite{Sorella:2013bz} %{Sorella_book2013}
and it is roughly proportional to the condition number of the matrix $\bf S$. Since in a correlated 
wave function the non linear coupling between different variational parameters makes 
this matrix necessarily very ill conditioned (with high condition number), the gain 
in the optimization may be often drastic, that is 
certainly true for large number of variational parameters.
%{\bf  S. Sorella book Springer 2013 eds. F. Mancini } 
%
A recent work\cite{Mazzola:2012ch}
provides a simple geometrical interpretation of the advantage of the SR optimization over the steepest descent.
Indeed, \citet{Mazzola:2012ch} have shown  that the matrix $\bf S$ is actually the {\em metric}, to be intended in a differential geometry sense, 
where the parametrized normalized wave function 
$\psi_{\bar \alpha} \over \|\psi_{\bar \alpha}\|$ 
lives.
According to this point of view, it follows that SR can be interpreted as a steepest descent in this curved space, where the parameters are moved in the direction of the force 
along locally orthogonal and independent directions.
%but avoiding to change the normalization of $\psi_{\bar \alpha}$, that would be an useless change that do not minimize the variational energy.

\subsection{Force evaluation}\label{sec.force}

If we assume the Born-Oppenheimer approximation and a classical description of the nuclei, the 3-dimensional force acting on atom $a$ is, by definition:   
\begin{equation}\label{eq.force}
{\bf F}_a \equiv - 
  {\nabla_a {\cal E}[\psi]},
%  {d {\cal E}[\psi] \over d{\bf R}_a},
\end{equation}
where 
$\nabla_a \equiv {d \over d{\bf R}_a}$ 
%${d \over d{\bf R}_a}$ 
is the gradient relative to the cartesian coordinates ${\bf R}_a$ of the nucleus $a$, and 
${\cal E}[\psi]$ is the variational energy, as written in Eq.~(\ref{eq.E_VMC}), associated to the electronic wave function $\psi$ for a configuration $\bar {\bf R}$ of the atoms.
The terms in ${\cal E}[\psi]$ that are functionally dependent on the atomic coordinates are:
the Hamiltonian $\hat H \equiv \hat H_{\bar {\bf R}}$, 
and the wave function $\psi\equiv \psi_{\bar \alpha,\bar {\bf R}}$, 
which has an implicit dependence on ${\bar {\bf R}}$ in the $p$ parameters $\bar \alpha=\{ \alpha_1,\ldots,\alpha_p \} \equiv \bar \alpha_{\bar {\bf R}}$, which have to be  optimized for each $\bar {\bf R}$ in order to minimize the variational energy, 
and also an explicit dependence, if $\psi$ is  defined using a local basis set, as in our work.
Therefore the local energy 
$ H_L 
 %\equiv H_L^{\bar \alpha,\bar {\bf R}}
 \equiv { \hat H_{\bar {\bf R}} \psi_{\bar \alpha,\bar {\bf R}} \over \psi_{\bar \alpha,\bar {\bf R}}}$ 
that appears in ${\cal E}[\psi]$
depends on $\bar {\bf R}$ both through the wave function and the Hamiltonian.

By substitution of Eq.~(\ref{eq.E_VMC}) into (\ref{eq.force}), it is straightforward to obtain the following {\em analytical} expression for the force:
% espressione: HF + PULAY + parameters
\begin{eqnarray} \label{eq.force.VMC}
{\bf F}_a      &=& {\bf F}_a^{HF} + {\bf F}_a^{P} + {\bf F}_a^{\bar \alpha}  \\
{\bf F}_a^{HF} &=& - { \int {\partial H_L \over 
                             \partial {\bf R}_a} 
                               \psi^2 d\bar r \over 
                       \int \psi^2 d\bar r}  \nonumber \\ 
{\bf F}_a^{P}  &=&  -2 { \int (H_L - {\cal E}[\psi]) 
                           {\partial \log|\psi| \over 
                            \partial {\bf R}_a} 
                              \psi^2 d\bar r \over 
                         \int \psi^2 d\bar r} \nonumber \\
{\bf F}_a^{\bar \alpha} 
               &=& - \sum_{k=1}^{p} {\partial {\cal E}[\psi] \over 
                                     \partial \alpha_k} 
                              \cdot {\partial \alpha_k \over 
                                     \partial {\bf R}_a} \nonumber                
\end{eqnarray}
where the three terms that constitute the total force: ${\bf F}_a^{HF}$, ${\bf F}_a^{P}$ and ${\bf F}_a^{\bar \alpha}$  are respectively given by the explicit dependence on ${\bf R}_a$ of the local energy and of the wave function, and the implicit dependence on ${\bf R}_a$ of the parameters of the wave function.

% term derivative parameters
The term ${\bf F}_a^{\bar \alpha}$ is, in principle, the most complicated  to be evaluated, because of this implicit dependence % of $\bar \alpha_{\bar {\bf R}}$ on $\bar {\bf R}$ 
which makes the derivative 
${\partial \alpha_k \over \partial {\bf R}_a}$ difficult to evaluate.
Fortunately, if the values of $\bar \alpha_{\bar {\bf R}}$ correspond to a minimum for the energy ${\cal E}[\psi]$, 
then ${\partial {\cal E}[\psi] \over \partial \alpha_k} = 0$ for the Euler condition, and ${\bf F}_a^{\bar \alpha}=0$. For this reason the term ${\bf F}_a^{\bar \alpha}$ has been neglected in our calculations.

% terms: HF and P
The other two terms, 
${\bf F}_a^{HF}$ and ${\bf F}_a^{P}$, are usually referred to as the {\em Hellmann-Feynman} term and the {\em Pulay} term, respectively.
Actually the Hellmann-Feynman term ${\bf F}_a^{HF}$ resembles the term 
$\int (\nabla_a \hat H) \psi^2 d\bar {\bf r}$ 
that comes from the application of the Hellmann-Feynman theorem, although it is not exactly the same because in general $\nabla_a H_L \ne \nabla_a \hat H$.
Moreover, in VMC calculations, the Hellmann-Feynman theorem is not even applicable, because $\psi$ is neither normalized nor an eigenstate of $\hat H$.
But in the limit case where $\psi$ is an eigenstate of $\hat H$, and consequently $H_L=E[\psi]$, % and $\nabla_a H_L = \nabla_a E[\psi]$, 
the Pulay term ${\bf F}_a^{P}$ is zero and the only contribution to the force comes from ${\bf F}_a^{HF}$.
As a consequence of this, it is expected that the more $\psi$ approaches an eigenstate of $\hat H$, the lower is the ${\bf F}_a^{P}$ component of the force.

% Space Warp Transformation Coordinate
The analytical expression of the force in Eq.~(\ref{eq.force.VMC})
is correct and 
is significantly more accurate and efficient than the corresponding expression based only on the 
Hellmann-Feynman contribution, as observed by \citet{Sorella:2010p23644}. 
The efficiency is defined as  the inverse of the computational time to reach the required stochastic precision, and in the specific case of the water dimer studied in ref. \cite{Sorella:2010p23644}, an improvement of two orders of magnitude was obtained. 
However, \citet{Sorella:2010p23644} showed that a further improvement of about one order of magnitude is possible using the analytical expression with  the differential 
{\em space warp coordinate transformation} (SWCT).
Therefore in this work we have used these SWCT analytical forces, that are obtained as follows.

SWCT was originally introduced by \citet{UMRIGAR:1989tq} for an efficient calculation of the forces, but  using only finite-difference derivatives.
Within SWCT, a displaced ${\bf D}_a$ of the nucleus $a$ is followed by a displacement of the electrons.
Each electron $i$ is translated, in the direction ${\bf D}_a$, of a quantity  that depends on its distance $r_{ia}=\|{\bf r}_i-{\bf R}_a\|$ with the nucleus $a$.
If $r_{ia}\sim 0$  the displacement of electron $i$ is $\sim {\bf D}_a$; if 
$r_{ia}\to \infty$  the displacement is $\sim 0$.
In this way the electronic coordinates $\bar {\bf r}$ mimic the displacement of the charge around the nucleus ${\bf R}_a$.
More in detail, following refs. \cite{Casula:2004p12689,Sorella:2010p23644}, SWCT is described by the following transformation of the nuclear and electronic coordinates:
\begin{eqnarray} \label{eq.SWCT}
{\bf R}_b &\to& {\bf R}'_b = {\bf R}_b +  {\bf D}_a \\
{\bf r}_i &\to& {\bf r}'_i = {\bf r}_i + \omega(r_{ia}) {\bf D}_a  \nonumber 
\end{eqnarray}
for $b=1,\ldots,M$ and $i=1,\ldots,N$.
In the above equation the weight that quantifies the amount of electronic displacement is chosen to be:
\begin{equation}
\omega(r_{ia}) = {{r_{ia}}^{-4} \over \sum_{b=1}^{M} {r_{ib}}^{-4}} \,,
\end{equation}
according to refs.\cite{Filippi:2000p25406,Casula:2004p12689,Sorella:2010p23644}.

The variational energy 
${\cal E}_{\bar {\bf R}'}[\psi_{\bar \alpha,\bar {\bf R}'}]$
calculated in the nuclear coordinates $\bar {\bf R}'$,  for an infinitesimal displacement ${\bf D}_a$, considering also that the displacement of the parameters 
$\Delta \bar \alpha = \bar \alpha_{\bar {\bf R}'} - \bar \alpha_{\bar {\bf R}}$ 
is negligible at the first order as discussed previously, is given by:
\begin{equation}
{\cal E}_{\bar {\bf R}'}[\psi_{\bar \alpha,\bar {\bf R}'}] = 
  { \int H_L^{\bar {\bf R}'}[\psi_{\bar \alpha,\bar {\bf R}'}(\bar {\bf r}')] {\psi_{\bar \alpha,\bar {\bf R}'}(\bar {\bf r}')}^2 d\bar {\bf r}' 
    \over 
    \int {\psi_{\bar \alpha,\bar {\bf R'}}(\bar {\bf r}')}^2 d\bar {\bf r}'} \,,
\end{equation}
where the integrated electronic coordinates $\bar {\bf r}'$ can be substituted by the SWCT expression in Eq.~(\ref{eq.SWCT}), and  
$d\bar {\bf r}' = \det\left( {\partial \bar {\bf r}' \over \partial \bar {\bf r}} \right) d\bar {\bf r}$.
We obtain in this way an expression that we call ${\cal E}_{\bar {\bf R}'}^\textrm{SWCT}$.

The SWCT analytic force ${\bf F}_a^\textrm{SWCT}$ is then obtained by differentiating the energy ${\cal E}_{\bar {\bf R}'}^\textrm{SWCT}$ over ${\bf D}_a$, and evaluatingit  at  ${\bf D}_a=0$:
\begin{equation}\label{eq.force.SWCT}
{\bf F}_a^\textrm{SWCT} = \left. {d 
    {\cal E}_{\bar {\bf R}'}^\textrm{SWCT}
    %[\psi_{\bar \alpha,\bar {\bf R}'}]
    \over d{\bf D}_a} \right|_{{\bf D}_a=0} \,.
\end{equation}
It has to be noted that, in this case, we have also an implicit dependence 
of the electronic coordinates  on ${\bf D}_a$, yielding  additional force terms arising from the derivative of the wave function over the electrons coordinates,
${\partial \psi \over \partial {\bf r}_i}$, and to the derivative of the Jacobian of the SWCT. 
The calculation in Eq.~(\ref{eq.force.SWCT}) leads straightforwardly to an expression for the force analogous to Eq.~(\ref{eq.force.VMC})
where the Hellmann-Feynman and the Pulay terms can be easily identified:
\begin{eqnarray} \label{eq.force.VMC.SWCT}
{\bf F}_a^\textrm{SWCT}      &=& {\bf F}_a^{\textrm{SWCT}-HF} + {\bf F}_a^{\textrm{SWCT}-P} \\
{\bf F}_a^{\textrm{SWCT}-HF} &=& - { \int ({\nabla_a^{\textrm{SWCT}} H_L}) 
                               \psi^2 d\bar r \over 
                       \int \psi^2 d\bar r}  \nonumber \\ 
{\bf F}_a^{\textrm{SWCT}-P}  &=&  -2 { \int (H_L - {\cal E}[\psi]) 
                           ({\nabla_a^{\textrm{SWCT}} \log|\psi| }) 
                              \psi^2 d\bar r \over 
                         \int \psi^2 d\bar r} \nonumber             
.
\end{eqnarray}
Indeed, the above expression is almost identical to Eq.~(\ref{eq.force.VMC})
with the difference that we have introduced here a generalized gradient $\nabla_a^{\textrm{SWCT}}$, defined in the following way: 
\begin{eqnarray}\label{eq.deriv.SWCT}
{\nabla_a^{\textrm{SWCT}} H_L} 
  &\equiv& {\partial H_L \over \partial {\bf R}_a} 
                 + \sum_{i=1}^{N} \omega(r_{ia}) {\partial H_L \over \partial {\bf r}_i} \\
{\nabla_a^{\textrm{SWCT}} \log|\psi| } 
  &\equiv& {\partial \log|\psi| \over \partial {\bf R}_a}
                        + \sum_{i=1}^{N} \left( 
                            \omega(r_{ia}) {\partial\log|\psi|\over \partial {\bf r}_i} +
                            {1\over 2} {\partial \omega(r_{ia}) \over \partial {\bf r}_i}
                                        \right) \nonumber
\end{eqnarray}
for the Hellmann-Feynman and the Pulay terms, respectively.

% Adjoint Algorithmic Differentiation
As discussed exhaustively by \citet{Sorella:2010p23644}, 
the implementation of the computational technique of the
{\em adjoint algorithmic differentiation} (AAD)
allows a computationally very efficient evaluation of all the terms appearing in Eq.~(\ref{eq.deriv.SWCT}), that roughly can be evaluated in $\propto N^3$ operations.
This technique leads to a computational cost for the evaluation of the energy and all the force components amounting to about four times the time required for the calculation of the variational energy alone.
The computational gain is substantial, especially if compared with finite difference methods  on large systems\cite{Sorella:2010p23644,Coccia:2012kz}.

% infinite variance
At this point we have the exact expressions for the analytical forces, and the technical instruments to calculate all the components efficiently.
But there is still a point that has to be addressed: do these expressions lead to quantities that can be  efficiently evaluated within a stochastic approach, for a wave function $\psi$ that in general  only approximates the exact GS solution?
As discussed in Section~\ref{sec.stochastic.eval}, this  implies that we have to choose  the  appropriate weight $W$ allowing the stochastic evaluation of the expectation value of the force, 
i.e., the variance in Eq.~(\ref{eq.sigma.MC}) has to be finite.

% divergences in terms of the forces
% in HF - ee and en coalescence
Let us start considering the terms containing divergences, which could lead to an infinite variance, starting from the Hellmann-Feynman force.
We can easily recognize the two problematic terms  $\partial H_L \over \partial {\bf R}_a$ and $\partial H_L \over \partial {\bf r}_i$, respectively in the cases of electron-nucleus and of electron-electron coalescence. 
Indeed the derivatives of the potential energy $\partial V$, included in $\partial H_L$, contains  terms which would give an infinite variance, namely 
${\partial V\over \partial {\bf r}} \propto \delta_{ee}^{-2}$ for the  electron-electron distance $\delta_{ee} \ll 1$
and 
${\partial V\over \partial {\bf R}} \propto \delta_{en}^{-2}$ for the electron-nucleus distance $\delta_{en} \ll 1$.
However in our case we can handle these divergences because we are using wave functions that satisfy the cusp conditions, producing a divergence in the kinetic term of the same amount but of opposite sign with respect to the divergence of the potential, regularizing in this way the divergence of $H_L$ and of its derivatives.

% in HF - nodal surf.
Nevertheless $\partial H_L \over \partial {\bf R}_a$ and $\partial H_L \over \partial {\bf r}_i$ remain  divergent in proximity of the nodal surface.
We have already mentioned in Section~\ref{sec.energy} that in general $\psi\propto \delta$ and $H_L\propto \delta^{-1}$ at a distance $\delta \ll 1$ from the nodal surface, hence $\partial H_L\propto \delta^{-2}$. 
Using the standard sampling technique these divergences would lead to a variance that in proximity of the nodal surface is $\propto \int_0^{1} \delta^{-2} d\delta$, therefore unbounded.
However, with the {\em reweighting sampling} method 
%mentioned in Section~\ref{sec.energy} and described in detail in Section~\ref{sec.reweight}, 
described in Section~\ref{sec.reweight},
the variance becomes 
$\propto \int_0^{1} \delta^{0} d\delta$,
thus its divergence is completely under control and the variance is finite.

% in P
Also in the Pulay force there is a similar problematic behavior in proximity of the nodal surface, because both $H_L$ and $\partial \log|\psi|$ diverge as $\delta^{-1}$, giving an infinite variance if the standard sampling is used. The use of the reweighting sampling regularizes also this term, giving a finite variance.

% Reweighting Method 
%\cite{Attaccalite:2008p12639}

\subsection{The reweighting method for open systems  } \label{sec.reweight}

%\footnote{$A$ is the same as the matrix ${\bf M}^{AGP}$ in section~\ref{sec.AGP}.}
%C.~Attaccalite and S.~Sorella 
\citet{Attaccalite:2008p12639}
proposed a reweighting method to solve the infinite variance issue in the proximity of the nodal surface
by using a different probability distribution $P(\bar {\bf x})\propto W( \bar {\bf x})=\psi_G(\bar {\bf x})^2$, defined in terms of a {\em guiding function} $\psi_G(\bar {\bf x})$, rather than the standard sampling $W(\bar {\bf x})= \psi(\bar {\bf x})^2$. 

The guiding function $\psi_G(\bar {\bf x})$ is defined in terms of the wave function $\psi(\bar {\bf x})$ as follows:
\begin{equation}
  \psi_G(\bar {\bf x}) = 
    { R^\epsilon(\bar {\bf x}) \over R(\bar {\bf x})} \psi(\bar {\bf x})
\end{equation}
where $R(\bar {\bf x})$ is proportional to the distance $\delta$ from the nodal surfaand for $\delta \ll 1$, vanishes in the same way $\psi(\bar {\bf x})$ does, namely 
 $\psi(\bar {\bf x}) \propto R(\bar {\bf x})$.
The $R^{\epsilon}(\bar {\bf x})$ is the function that regularizes $\psi_G$ in the vicinity of the nodal surface, namely for 
$ \delta \propto R(\bar {\bf x}) < \epsilon$, 
and it is defined as:
\begin{equation}
  R^\epsilon(\bar {\bf x}) =
    \left\{\begin{array}{l l}
                  R(\bar {\bf x}) & \textrm{if~}R(\bar {\bf x}) \ge \epsilon \,,\\
                  \epsilon[R(\bar {\bf x})/\epsilon]^{R(\bar {\bf x})/\epsilon} &\textrm{if~}R(\bar {\bf x}) < \epsilon\,,
                  \end{array} \right.
\end{equation}
where the nontrivial  regularization for $R(\bar {\bf x}) < \epsilon$ is introduced in order to satisfy the continuity of the first derivative of $\psi_G(\bar {\bf x})$. The guiding function $\psi_G(\bar {\bf x})$ defined in this way and 
its corresponding probability density function $P(\bar {\bf x}) \propto \psi_G(\bar {\bf x})^2$ 
define a {\em reweighting factor}
\begin{equation}
\left({\psi(\bar {\bf x})\over \psi_G(\bar {\bf x})}\right)^2 = 
  \left({R(\bar {\bf x})\over R^\epsilon(\bar {\bf x})}\right)^2 =
    \min\left[ 1, \left({R(\bar {\bf x})\over \epsilon}\right)^{2 \left(1-{R(\bar {\bf x})\over \epsilon}\right)} \right]
\end{equation}
that 
in the proximity of the nodal surface, i.e. $R(\bar {\bf x}) \to 0$, 
is $\propto \delta^2$, whereas the probability density function 
$P(\bar {\bf x})\propto \epsilon^2 \left( \psi(\bar {\bf x}) \over R(\bar {\bf x}) \right)^2 \propto \epsilon^2$
remains constant but finite.
This $P(\bar {\bf x})$ slightly enhances the sampling in the vicinity of the nodal surface where $\Pi(\bar {\bf x})$ vanishes. So far, our reweighting method removes the singularities up to $\delta^{-2}$ and provides finite variance.

The regularization scheme which \citet{Attaccalite:2008p12639} proposed to evaluate $R(\bar {\bf x})$ is based on the matrix ${\bf A}$ that appears in the determinantal (antisymmetric) part of the QMC wave function, Eq.~(\ref{equ:psiQMC}).
For the AGP wave functions used in this work, we can identify the matrix ${\bf A}$ with the ${\bf M}^{AGP}$ described in Section~\ref{sec.AGP}.
As soon as the configuration of electrons approaches the nodal surface, $\det({\bf A})\rightarrow 0$ and the elements of the ${\bf A}^{-1}$ grow extremely large. 
According to this feature, the regularizing is choosen 
to be controlled by ${A^{-1}}_{ij}$ in the following way:
\begin{equation}
  R(\bar {\bf x})=\left(\sum_{i,j} \left|{A^{-1}}_{ij}\right|^2 \right)^{-1/2}\,. \label{equ:old_reg}
\end{equation}

However, within this scheme~\ref{equ:old_reg} does not take into account 
the case of open systems like isolated atoms and molecules  (type 4 in ref.~\cite{Trail:2008ft}). %\cite{PhysRevE.77.016703} -> this is the same of Trail:2008ft; remove from biblio to avoid redundancy
As an electron $i$ samples a region very far from the center of mass ${\bf r}_0$ of the nuclei, 
namely $r_{i0}=\|{\bf r}_{i}-{\bf r}_0\| \gg 1$,
the decay of the many-body wave-function 
is dominated by the determinantal part as the Jastrow correlation is identically 
one in this limit.
A simple inspection shows that $\det({\bf A})$ behaves as
$\propto \exp(-\tilde z_{min} r_{i0})$ [$\propto \exp(-\tilde z_{min} r_{i0}^2)$], 
where $\tilde z_{min}$ is the minimum exponent in the Slater [Gaussian] basis. 
The old regularization in Eq.~(\ref{equ:old_reg}) vanishes clearly in the same way.
To this purpose it is enough to apply the  Rouch\'e-Capelli theorem   
stating that the inverse matrix elements $A^{-1}_{ij}$ can be expressed 
with the ratio of a cofactor matrix determinant ($\det C_{ji} $) 
 and the determinant itself, namely:
$${A^{-1}}_{ij} = { \det C_{ji} \over \det({\bf A}) }.$$
Now we immediately arrive to the bad conclusion that the probability distribution 
$P(\bar {\bf x})$ is ill defined as it converges to a constant in the limit when $r_{i0} \gg 1$, 
because, $R(\bar {\bf x}) \to 0$ in the same way as $\psi(\bar {\bf x}) \to 0$ (as discussed above), 
and the resulting distribution $P(\bar {\bf x})$ is not normalizable.
In practice this means that the random walk for long enough simulation will be 
unstable, and all electrons are pushed to very large distance from the atoms, 
providing unpredictable and certainly biased results.

In order to overcome this clear instability 
we replace the ${\bf A}$ in~\ref{equ:old_reg} with ${\bf A}'$.
The new matrix ${\bf A}'$ is defined by changing its asymptotic behavior for 
large $r_{i0}$:
\begin{equation}
  A'_{ij}=A_{ij}\exp(z r_{i0}+z r_{j0})\,,
  %% z_{i/j}&=&\left\{\begin{array}{ll}
  %%                   z & \textrm{for $N^\uparrow+N^\downarrow$ occupied orbitals}\\
  %%                   0 & \textrm{for $N^\uparrow-N^\downarrow$ unoccupied orbitals}
  %%                 \end{array}
  %%            \right.,
\end{equation}
where $z$ can be any positive value. 
In fact the new regularization will act in the same way close to the nodes 
of $\psi$, whereas when $r_{i0} \gg 1$, $\det({\bf A}')$  decays as 
$\exp[ - \tilde z_{min} r_{i0}^2 + z r_{i0}] $
for a Gaussian basis, and
for a Slater basis, 
if $\tilde z_{min} > z$, 
it decays as 
$\exp [- (\tilde z_{min} -z) r_{i0}] $, 
and diverges otherwise.
Therefore $P(\bar {\bf x})$, by using this new definition of $R(\bar {\bf x})$,  
will decay as $\exp (-2 z r_{i0})$ in the former cases, or as $\Psi^2$ 
itself in the latter case,
 yielding {\em  in any case} a perfectly defined and normalizable distribution.

%but the optimal one is the exact asymptotic coefficient $\tilde z$.
In practice, 
if $z$ is too small, ${\bf A}'$ behaves too much like ${\bf A}$ and the instability remains.
On the other hand  if $z$ is too large, the probability distribution 
$P(x)$, as we have seen,  remains too close to the original one $\simeq \Psi^2$  for electron-ion distances $\gg 1/z$, and therefore in this region the singularities in the nodal surfaces remain, 
and the regularization is not effective also in this case.
Therefore, with this simple trick, and a reasonable value of $z\simeq 1/\xi$, where $\xi$ 
is the linear dimension of the  important region of non vanishing charge density,  
this numerical  instability, present  
in  open  systems, is readily removed, and the singularities around the nodal 
surfaces are perfectly controlled, because
the proposed regularization works exactly as the previous one\cite{Attaccalite:2008p12639} adopted for PBC.
Indeed,  if electrons are close to this  nodal surface $\det({\bf A})=0$
and $r_{i0}$ are all  finite, the following equality
\begin{equation}
  \det({\bf A}')=\prod^N_{i}\exp(z r_{i0})\det({\bf A})
\end{equation}
implies that 
the new regularization works as well as the previous one, 
being the factor $\prod^N_{i}\exp(z r_{i0})$ just an irrelevant term.

\subsection{ Charge density, Dipole and Quadrupole evaluation }\label{sec:dipole.quadrupole}

Several important properties of the molecular systems, 
as the dipole and the quadrupole,
derive from the charge density distribution:
\begin{equation}\label{eq:density}
\rho({\bf r}) \equiv 
  \sum_a^M Z_a \delta({\bf r} - {\bf R}_a) 
  - \left< \sum_i^N \delta({\bf r} - {\bf r}_i) \right>_\Pi
\end{equation}
where the first term in the right hand side is due to the nuclear charges $Z_a$ centered in their cartesian coordinate ${\bf R}_a$, in agreement with the Born-Oppenheimer approximation and classical nuclei.
The second element in the right hand side, that is due to the electronic charges, is averaged over the distribution of the electrons $\Pi \propto \psi^2$.

From the definition in Eq.~(\ref{eq:density}) of the charge density, it is straightforward to obtain the expression for the dipole ${\bf D}$:
\begin{eqnarray}\label{eq:dipole}
D^{\alpha} & \equiv & 
    \int r^\alpha \delta({\bf r}) d{\bf r} \\
  &=& \sum_a^M Z_a R_a^\alpha 
  - \left< \sum_i^N r_i^\alpha \right>_\Pi
\end{eqnarray}
and for the traceless quadrupole tensor:
\begin{eqnarray}\label{eq:quadrupole}
Q^{\alpha \beta} & \equiv & 
  {1\over 2}
    \int \left( 3 r^\alpha r^\beta - \|{\bf r}\|^2 \delta^{\alpha \beta} \right) \delta({\bf r}) d{\bf r} \\
  &=& {1\over 2} \sum_a^M Z_a \left( 3 R_a^\alpha R_a^\beta - \|{\bf R}_a\|^2 \delta^{\alpha \beta} \right)
  - {1\over 2} \left< \sum_i^N \left( 3 r_i^\alpha r_i^\beta - \|{\bf r}_i\|^2 \delta^{\alpha \beta} \right)\right>_\Pi
\end{eqnarray}
where $\alpha$ and  $\beta$ label the three cartesian axes and $\delta^{\alpha \beta}$ is the Kronecker's delta.
% pseudo
%Notice that, in presence of pseudo potentials, the value of $Z_a$ that appears in Eqs.~(41) to (44) % Eq.~(\ref{eq:density,eq:dipole,eq:quadrupole}) 
%has to be considered the effective charge of the nucleus, i.e. the atomic number minus the number of pseudo electrons for that specific atom.

The dipole depends on the choice of the reference frame, unless the total charge of the molecule is zero, 
and the quadrupole depends on the choice of the reference frame, unless the dipole is zero.
For the case of the water molecule, considered in this paper, the total charge is zero, but the dipole is not zero.
Therefore
we have to define the reference frame, 
in order to compare with the experimental and other calculated values of the quadrupole.
%In this work we have considered the reference frame of the center of the mass.
%In particular, the water molecule is in the xy-plane, with the bisector of the HOH angle along the y-axis, with the oxygen in the y axis and with negative value, and the hydrogens with positive y values.
%Thus, for symmetry reasons the only non negative coordinate of the dipole is the one along the y axis, and it is positive because the oxygen is more electronegative than the hydrogens.

The electronic part of the dipole and of the quadrupole have been calculated by averaging within a VMC scheme the quantities of interest.
We are aware that more sophisticated improved estimators for the density and related quantities are available in literature\cite{Assaraf:2007p19843,Coccia:2012fi}, however they are not necessary for this work.

\subsection{ Energy evaluation by fixed node projection Monte Carlo }\label{sec.FN}

%XXX scrivere brevemente idea del projection e mettere ref XXX

Using the projection Monte Carlo approaches, it is possible %to improve the variational ansatz 
to access  the lowest possible energy, with the constraint that $\Phi$ has the same nodal surface of an appropriately chosen guiding function $\Psi$ ({ fixed node approximation})\cite{Reynolds:1982en,Foulkes:2001p19717}.
Therefore, it is of fundamental importance to choose a  guiding function with a reliable nodal surface, and, for this purpose, it is usually considered the 
variational wave function with minimum possible energy within a given ansatz.

Among the different projection methods, we have considered in this work the 
lattice regularized diffusion Monte Carlo\cite{Casula:2005p14138,Casula:2010p14082}.
LRDMC is based on the spatial discretization of the molecular Hamiltonian on a lattice of mesh size $a$, and it resorts to the projection scheme used also in the Green function Monte Carlo algorithm\cite{Sorella:2000p17651,Buonaura:1998p25304}.
This method has two very interesting properties: 
it maintains its efficiency even  for systems with a large number of electrons\cite{Casula:2010p14082};
and it preserves the variational principle even when used in combination with nonlocal pseudo potentials\cite{Casula:2010p14082}.
The error induced by the finite mesh size $a$ is analogous to the time step error appearing in standard DMC calculations. It can be controlled by performing several calculations with different values of the mesh $a$ and finally extrapolating to the continuum limit $a\to 0$.

%%%%%%%%%%%%%%%%%%%%%%%%%%%%%%%%%%%%%%%%%%%%%%%%%%%%%%%%%%%%%%%%%%%%%
%%%%%%%%%%%%%%%%%%%%%%%%%%%%%%%%%%%%%%%%%%%%%%%%%%%%%%%%%%%%%%%%%%%%%

%  \input{4.Computational_Details}
%%%%%%%%%%%%%%%%%%%%%%%%%%%%%%%%%%%%%%%%%%%%%%%%%%%%%%%%%%%%%%%%%%%%%
\section{Computational Details}\label{sec.comput.det}
%%%%%%%%%%%%%%%%%%%%%%%%%%%%%%%%%%%%%%%%%%%%%%%%%%%%%%%%%%%%%%%%%%%%%
%%%%%%%%%%%%%%%%%%%%%%%%%%%%%%%%%%%%%%%%%%%%%%%%%%%%%%%%%%%%%%%%%%%%%

%%% QMC
{\bf QMC package. }
The QMC energy and force calculations  have been carried out using the {\em TurboRVB} package developed by S. Sorella and coworkers\cite{TurboRVB}, that includes a complete suite of  variational and  diffusion quantum Monte Carlo calculations on molecules and solids, and for wave function and geometry optimization.

%%% AE and PSEUDO POT.
%{\bf Description of core electrons: all-electron calculations or pseudo potentials } \\
{\bf Description of the core electrons. }
The results that are presented here have been obtained both by all electrons ({\bf AE}) calculations, and by calculations where the two core electrons of the oxygen atom have been described using a pseudo potential.
In order to appreciate the reliability of the calculations with  the  pseudo potential  versus the AE calculations, two different pseudo potentials have been used and compared in this work:
the scalar-relativistic energy consistent pseudo potential ({\bf ECP}) of Burkatzki et al.\cite{Burkatzki:2007p25447}, %Filippi 
and the smooth relativistic norm-conserving pseudo potential  ({\bf NCP}) of Trail and Needs\cite{Trail:2005iw}. %Needs 
%a norm-conserving pseudopotential generated through a Hartree?Fock Hamiltonian\cite{Trail:2005ix} (HF-NCP)

%%% wave functions
{\bf Wave function ansatzes. }
In this work we have considered several many-body wave functions, which have been constructed starting from the terms described in Section~\ref{sec.qmc.wf}:
%Namely, the wave functions that we have considered are:
\begin{itemize}
\item[\bf JAGP]:
a Jastrow correlated  AGP wave function, with the Jastrow factor and the determinanatal part described in Section~\ref{sec.Jastrow} and in  Section~\ref{sec.AGP}, respectively;
\item[\bf JSD]: 
a Jastrow correlated single determinant wave function, with the Jastrow factor and the determinanatal part described in Section~\ref{sec.Jastrow} and in  Section~\ref{sec.SD}, respectively; 
\item[\bf JAGPn*]:
a Jastrow correlated constrained AGPn* wave function, with the Jastrow factor and the determinanatal part described in Section~\ref{sec.Jastrow} and in  Section~\ref{sec.AGPn}, respectively;
%

%\item[\bf DFT]: 
%a single determinant wave function obtained from the Kohn-Sham orbitals of a DFT calculation within local-density approximation (LDA) as described in ref. \cite{Azadi:2010p14081} and implemented in the {\em TurboRVB} package\cite{TurboRVB}. 
%%%Note that the  parameters of  this wave function are not optimized minimizing the variational energy, and actually only the MOs coefficients are changed {\bf cambiati rispetto a che?}, while the exponents and the coefficient in the homogeneous electron-nucleus term are fixed. 
%
\item[\bf JDFT]: 
combination of the Jastrow factor described in Section~\ref{sec.Jastrow} with
%the DFT wave function 
a single determinant wave function, obtained by the Kohn-Sham orbitals of a DFT calculation within local-density approximation (LDA) as described in ref. \cite{Azadi:2010p14081} and implemented in the {\em TurboRVB} package\cite{TurboRVB}. 
This wave function, also studied in ref.\cite{TurboRVB}, is actually a JSD, but it is called differently to highlight that in JDFT, at variance of  JSD, the determinantal part has been optimized by a DFT calculation and only the parameters of the Jastrow term have been variationally optimized by QMC. 
\end{itemize}

% BASIS SET
{\bf The basis set. }
As discussed in Sections~\ref{sec.AGP},~\ref{sec.SD},~\ref{sec.AGPn},~\ref{sec.Jastrow}, both the determinantal and the Jastrow part of the wave function use atomic orbitals (see description in Section~\ref{sec.orbitals}).
The number and the type of the atomic orbitals is a nontrivial choice for QMC calculations, as for other quantum chemical methods, because if the basis set is too small the results are biased.
Anyway, in QMC a large basis set introduce a large number of parameters that are computationally expensive to optimize, leading, in the worst cases, to instabilities in the optimization. 
In this work the basis set convergence for the Jastrow and the determinantal terms is studied.

% determinant
The determinantal term is functionally similar to the wave functions used in HF, DFT or post-HF calculations, therefore  we constructed and used  several basis that are inspired by some of the standard basis used in quantum chemistry, and in particular the Dunning's basis\cite{DUNNING:1989uk,KENDALL:1992vx}.
However, the peculiarities of the QMC wave functions, namely the presence of the Jastrow term, 
and the use of particularly smooth pseudo-potentials, 
allows a large reduction of the size of the basis set and, as a consequence, the number of parameters required for the optimization of the energy.
For instance, the largest exponents (suitable to correctly describe the core) can be eliminated, because they are already  described with a reasonable accuracy by the electron-nucleus interaction term in the Jastrow, satisfying exactly the electron-nucleous cusp condition. 
Conversely  the most diffusive gaussian exponents can be safely replaced by very few but tunable STO orbitals (one for each angular momentum) introduced in the atomic basis of the determinantal part. 
The list of the basis considered in this work for the determinantal part, with the source basis, the filter criteria, and the number of parameters introduced by each basis are reported in Tab.~\ref{table:AGPbasis}.
Most of the orbitals are GTO, as the source basis are GTO, but in some cases an extra STO orbital was introduced, in order to better describe the diffusion part of the orbital and to have the theoretical long range exponential decay of the wave function.
Clearly, the filter is slightly different if the pseudo-potential is or is not used.
The basis set convergence for the determinantal part is discussed in Section~\ref{sec.BSC.det}.
% table basis det
%\input{Table_basis+np}

% Jastrow
The choice of the basis set for the Jastrow term is more challenging, because this term is a peculiar feature of the QMC calculations, and we do not have a pre-optimized or pre-characterized basis coming from other methods.
Moreover, the choice of a large enough basis set for the Jastrow is very important for the JAGP and JAGPn* ansatzes, not only for the improvement in the dynamical correlation of the wave function, but also because only in the limit of a complete Jastrow factor the unphysical charge fluctuations of the AGP are suppressed and the wave function becomes size consistent, as discussed by \citet{Sorella:2007p12646} and recently by \citet{Neuscamman:2012hm}.
In this work we only tested several GTO atomic orbitals for the Jastrow, both uncontracted, contracted and with hybrid contraction.
The performances of the different choices are discussed in Section~\ref{sec.BSC.J}.

%OPTIMIZATION
{\bf Wave function parameters. }
The different wave function ansatzes used in this work
%, namely the JSD, JAGPn* and JAGP, 
depends on several parameters, that have to be optimized variationally as explained in Section~\ref{sec.wf.opt}. 
Four main classed of such parameters can be identified:
\begin{enumerate}
% basis sets
\item the coefficients and the exponents appearing both in the determinantal basis set $\{ \phi^{a}_{\mu_a} \}$ and in the Jastrow basis set $\{ \chi^a_{\mu_a} \}$;
% J1/2
\item the elements of the Jastrow matrix $f^{a,b}_{\nu_a,\nu_b}$ of the inhomogeneous electron-electron term in Eq.~(\ref{equ:3BJas});
% matrix J
\item the Jastrow parameters $b_1$ and $b_2$ of the homogenous one-electron and two-electrons interaction terms, respectively in Eqs.~(\ref{equ:1BJas}) and (\ref{equ:2BJas});
% matrix det
\item for the JAGP ansatz: 
the elements of the $\Lambda$ matrix, see Eq.~(\ref{equ:phiG});  
or for the JSD and the JAGPn* ansatz:  
the leading eigenvectors and eigenvalues of $\Lambda$, i.e. the MOs and their weights, see Eq.~(\ref{equ:phiG_2}).
\end{enumerate}

As already mentioned, for a JDFT ansatz only the Jastrow terms have to be variationally optimized, because the determinantal part is directly obtained by a  DFT-LDA calculation.
However, the remaining ansatzes, 
namely the JSD, JAGPn* and JAGP, 
differ by the number and kind of parameters to be optimized, thus also the optimization schemes are different.
% optimization schemes
The optimization protocols are described in the Supporting Information, Section~1.

% Z det - opt
It has to be observed that the exponents appearing in the determinantal part are already pre-optimized by other computational approaches, although
their values are not  the optimal ones for a QMC calculation, as they can be further improved by 
minimizing the variational energy.
%Indeed,  they represent a good starting point, and we may wonder what is the improvement that we have in the wave function of the exponents in the determinantal part are or are not optimized.
%Although they are not so many in number, their optimization is often quite challenging. This is due to  the highly non linear way the exponents determine the wave function, so that a small change in their values may produce large changes in the wave function. 
Their optimization is often quite challenging due to the non linear way they  determine the wave function.
Consequently, they have to be optimized using a large statistics, and by moving slowly and carefully during the optimization. 
If they are not optimized, the energy minimization is more stable and easier, and this generally  leads to a computational gain.
For this reason both the cases are considered in this work, and they are marked using the following labels:
\begin{itemize}
\item[\em Opt:noZ]: the wave function optimization was carried on the determinantal matrix, the contraction coefficients in the determinantal basis set, and all the Jastrow terms, including the exponent values in the Jastrow basis; 
\item[\em Opt:all]: all the parameters are optimized, including the exponents in the determinantal part.
\end{itemize}

% structure water
{\bf Reference structure. }
The reported single point calculations  are referred to  
the experimental structure of the water molecule\cite{Benedict:1956id}, having the oxygen-hydrogen distance of $r_{OH}=0.95721(3)\AA$ and the angle between hydrogen-oxygen-hydrogen of $\phi_{HOH}=104.522(5)$ degrees.
Moreover, we have chosen the reference frame of the center of the mass (this is relevant for the quadrupole calculation).
The water molecule is in the $xy$-plane, with the bisector of the HOH angle along the $y$-axis, with the oxygen in the $y$ axis and with negative value, and the hydrogens with positive $y$ values.
Thus, for symmetry reasons the only non negative coordinate of the dipole is the one along the y axis, and it is positive because the oxygen is more electronegative than the hydrogens.

% frequencies
{\bf Evaluation of the equilibrium structure and the frequencies. }
In Section~\ref{sec.freq} we will report reported the values of the nuclear configuration at the minimum of the potential energy surface (PES), of the harmonic vibrational frequencies and of the anharmonic corrections, relative to VMC calculations for several different wave function ansatzes.
The accurate determination of this quantities, and in particular of the frequencies, is challenging for methods like QMC, that are affected by a stochastic error that is several orders of magnitudes larger than the numerical error present in non stochastic methods.
In order to control the propagation of the errors on the predicted quantities it is important to adopt a method that takes explicitly into consideration the presence of the stochastic error.
% ref. moltidimensional fit
In a recent work\cite{Zen:2012br} some of us have shown how this can be achieved, by performing several single point calculations of the energies and the forces in a grid centered around a good guess of the minimum of the PES.
The values of the energies or the forces are then used to perform a multidimensional fit of the PES, to obtain a better estimate of the minimum and of the vibrational properties.
The choice of the grid is very important in this approach,  in order to have  reasonably small  stochastic errors of the frequencies, of the order of a few cm$^{-1}$.
% grid
The results reported in Section~\ref{sec.freq} are  obtained using a grid of 59 points, % i.e. 59 configurations for the nuclei and single point calculations of energy and forces, 
and the displacements between these points are $\Delta r=0.08$a.u.  for the OH distance and $\Delta \phi=10$ degrees for the HOH angle (corresponding to ``mesh-4'' in ref.~\cite{Zen:2012br}).
% center
The experimental configuration of the molecule was taken as the initial guess of the PES structural minimum, which has a residual force of the order of $10^{-3}$a.u. 
%
% optimization w.f.
Although the same wave function is used to describe each of the 59 point in the grid, the nuclear coordinates are changed and consequently the wave function parameters have to be optimized independently.
This has been done in a computationally convenient way by taking as initial guess the already optimized wave function for the configuration at the center of the grid. 
Moreover, we have carefully checked for some points in the grid that this procedure does not introduce any bias, by comparing with a standard optimization ``from scratch''. 
%To this purpose we have verified that the wave function optimized ``from scratch'', by means of the methods described before,   provides consistent  energy and forces.  

%%% ref TURBO
% wave function \cite{Casula:2003p12694}
% LRDMC \cite{Casula:2005p14138}
% calcoli water \cite{Sterpone:2008p12640}
% orbitali molecolari \cite{Marchi:2009p12614}
% forces \cite{Casula:2004p12689}
% reweighting method \cite{Attaccalite:2008p12639}
% forces algorithmic differentiation \cite{Sorella:2010p23644}
% structural optimization \cite{Barborini:2012iy}

%%%%%%%%%%%%%%%%%%%%%%%%%%%%%%%%%%%%%%%%%%%%%%%%%%%%%%%%%%%%%%%%%%%%%
%%%%%%%%%%%%%%%%%%%%%%%%%%%%%%%%%%%%%%%%%%%%%%%%%%%%%%%%%%%%%%%%%%%%%

%  \input{5.Results_and_discutions}
%%%%%%%%%%%%%%%%%%%%%%%%%%%%%%%%%%%%%%%%%%%%%%%%%%%%%%%%%%%%%%%%%%%%%
\section{Results and Discussion}\label{sec.results}
%%%%%%%%%%%%%%%%%%%%%%%%%%%%%%%%%%%%%%%%%%%%%%%%%%%%%%%%%%%%%%%%%%%%%
%%%%%%%%%%%%%%%%%%%%%%%%%%%%%%%%%%%%%%%%%%%%%%%%%%%%%%%%%%%%%%%%%%%%%

Irrespectively of the  ansatz (JDFT, JSD, JAGPn*  or JAGP),
%as emerges from Section~\ref{sec.qmc.wf}, 
in a QMC wave function two distinct and  adequately large basis sets have to be chosen, respectively for the determinantal part and for the Jastrow factor.
Too small bases may introduce a bias on the results, but too large bases make the wave function difficult or impossible to optimize, due to the stochastic nature of the approach and because the parameters become highly correlated.
In Section~\ref{sec.BSC.J} we discuss the basis set convergence for the Jastrow factor, while 
in Section~\ref{sec.BSC.det} we discuss the convergence for the determinantal part, in the different ansatzes.
In Section~\ref{sec.IE} we discuss the ionization and the atomization energies, and finally 
in Section~\ref{sec.freq} we consider the properties of the PES obtained with different QMC approaches.

\subsection{ The basis set convergence for the Jastrow factor }\label{sec.BSC.J}

% As already mentioned, for the choice of the determinantal basis, we can start from the comparison with other methods.
Since the Jastrow factor is peculiar of the QMC wave functions,
little help for the choice of the basis set for its inhomogeneous part comes from other computational methods. %, differently  from the determinantal part.
Therefore, we have tested several basis for the Jastrow factor, in a wave function whose determinantal part was kept fixed.
The considered ansatz is a JAGP function,
with ECP pseudo potential for the two core electrons of the oxygen atom,
and with a basis for the AGP part that is 
O(4s,4p,1d)/[2s,2p,1d], H(4s,1p)/[2s,1p],
where the initial guess for the values of the exponents was inspired from the Dunning's cc-pVDZ basis.
%the PS-DZ described in Tab.~\ref{table:AGPbasis} 
Despite this basis is relatively small, it is able to provide reliable results, as shown for instance in \citet{Zen:2012br} for the  equilibrium structure of the water molecule.

\begin{figure}
\begin{center}
\includegraphics[width=.7\textwidth]{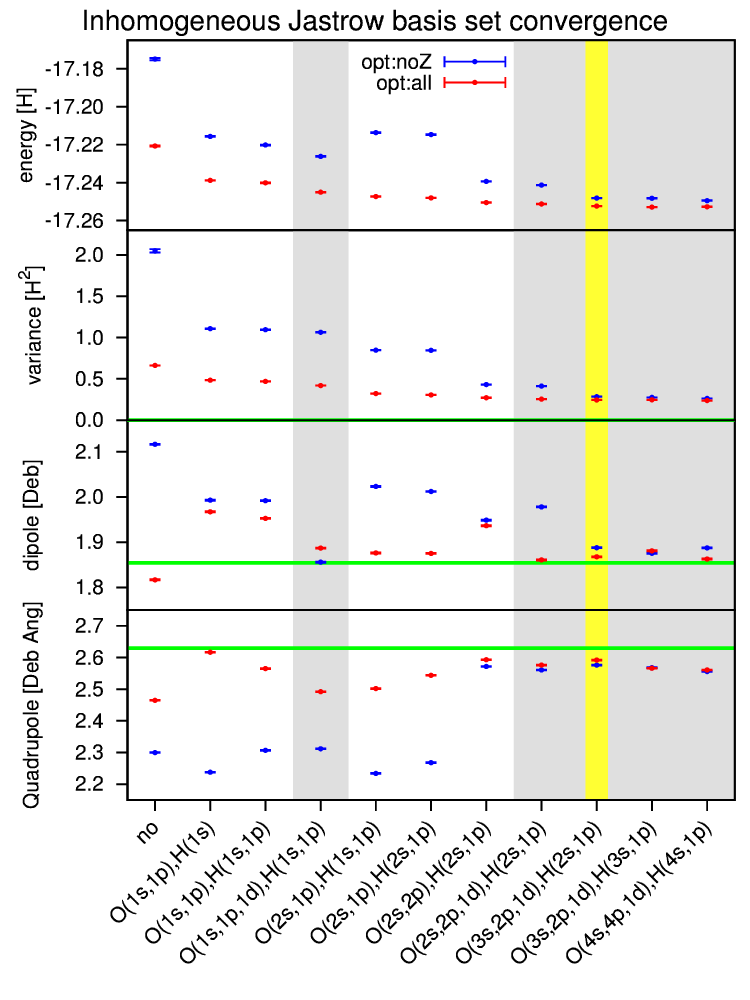}
\end{center}
\caption{
The basis set convergence 
of the water molecule VMC energy,  variance,  dipole and  Q$^{xx}$ quadrupole %(considering the molecule in the xy-plane and  in the reference frame of the center of mass, with the bisector of the HOH angle parallel to the y-axis).
for the Jastrow inhomogeneous term is shown here.
The considered wave function is a JAGP, with ECP pseudo potential for the two core electrons of the oxygen, and 
O(4s,4p,1d)/[2s,2p,1d], H(4s,1p)/[2s,1p]
determinantal basis.
The Jastrow basis set is reported in the abscissa, and the basis with $d$-orbitals have been highlighted with a grey background.
A yellow background has been used to indicate the basis considered for the following calculations.
The results corresponding to {\em Opt:noZ} and {\em Opt:all} are reported in blue and red, respectively.
In green we report the expected exact value, corresponding to zero variance, and the experimental values of the dipole and quadrupole.
} \label{fig:cbs_JBSC}
\end{figure}

We have considered uncontracted, contracted and hybrid atomic contracted basis, both with the optimization schemes {\em opt:noZ} and {\em opt:all}.
The complete list of all the obtained values for the energy, the variance, the dipole and the quadrupole are reported in Tables \ref{table:JBSC} and \ref{table:JBSC_prop}.
%Supporting Information, in tables~S2 and S3.
Looking at the values of energy and variance, it is quite evident that the uncontracted orbitals in the Jastrow provides much better results than  the contracted or the hybrid atomic contracted orbitals.
This is probably due to the fact that the 3-body term, see Eq.~(\ref{equ:3BJas}), gain a considerable variational advantage by the flexibility of an uncontracted basis. 
% fig J BSC
Thus, the choice of the optimal basis for the Jastrow factor should be an uncontracted basis.
Focusing only on the latter, in Fig.~\ref{fig:cbs_JBSC} 
we show the basis set convergence of the energy, the variance, the dipole and the $Q^{xx}$ component of the quadrupole.

%%%%%%%%%%%%%%%%%%%%%%%%%%%%%%%%%%%%%%%%%%%%%%%%%%%%%%%%%%%%%%%%%%%%%%%%%%%%%%%%%%%%%%%%%%%%%
%\input{Fig_cbs_JBSC}
%%%%%%%%%%%%%%%%%%%%%%%%%%%%%%%%%%%%%%%%%%%%%%%%%%%%%%%%%%%%%%%%%%%%%%%%%%%%%%%%%%%%%%%%%%%%%

% commenti
Several observations can be done. % from Fig.~\ref{fig:cbs_JBSC}.
% Tab En Var
% observe: opt:all
First: it is clear that the optimization of the exponents, {\em opt:all}, leads to a large improvement in the wave function, as reflected in all the properties considered.
This improvement is particularly significant if the  basis is rather small, whereas it is relatively small for a large basis.
Second, it is interesting to note that the presence of the $d$ orbitals in the oxygen basis of the Jastrow (highlighted with a grey background in the figure) highly improves the dipole and the quadrupole. %, so the charge distribution.
%
% observe: better energy - low variance
Third, we can observe the expected correlation between the energy and the variance: a lower energy is connected to a lower variance,
see also Fig.~\ref{fig:JBSC}(a). %Figure~S1(a) of the Supporting Information). 
%Therefore, by using larger basis sets,  the wave function get closer to the exact one, so the variational energy and the variance decrease.
A similar correlation is also expected with the charge distribution, and with the dipole in particular.
A general improvement of the dipole is observed with the large  basis sets, with low variances and low energies, but the convergence seems much more noisy than in the case of the energy, see Fig.~\ref{fig:JBSC}(b). %(see Figure~S1(b) of the Supporting Information). 
This  is due to the fact that the dipole is not a function of the total energy, thus the improvement in the variational energy, which is enforced during the wave function optimization, does not necessarily imply an improvement in the charge distribution.

The Jastrow basis that we have selected for the following calculations is 
O(3s,2p,1d), H(2s,1p),
corresponding to the results highlighted in yellow in Fig.~\ref{fig:cbs_JBSC}.
It represents an optimal balance between the accuracy of the results and the number of variational parameters, so the computational cost and the stability of the optimization.

\subsection{ The basis set convergence for the determinantal part and the wave function ansates }\label{sec.BSC.det}

Having defined the basis set for the Jastrow factor, we investigate now the different wave function ansatzes, namely JDFT, JSD, JAGPn* and JAGP, with different description of the core electrons of the oxygen:
using the ECP\cite{Burkatzki:2007p25447} or NCP\cite{Trail:2005iw} pseudo potentials, or all electron calculations.
As for the Jastrow factor, also here we have explored several basis sets for each wave function type, studying the basis set convergence.
The complete list of the attempted combinations is reported in the Supporting Information, where in Tab.~\ref{table:BSC} %Tables~S4 
we show the convergence of the energy and the variance, and in Tabs.~\ref{table:BSC_propSI} and \ref{table:BSC_AE_prop} %Tables~S5 and S6 
we consider also the dipole and the quadrupole.
% observations
Some interesting aspects can be observed from these results.
Some features were already observed in the previous section, %for the Jastrow convergence, 
for instance the advantage of the  {\em opt:all} scheme versus the {\em opt:noZ} one, and 
the correlation between energy, variance, and dipole (see Figs.~\ref{fig:BSC_ECP} and \ref{fig:BSC_AE}). %(see Figures~S2 and S3 of the Supporting Information).

%%%%%%%%%%%%%%%%%%%%%%%%%%%%%%%%%%%%%%%%%%%%%%%%%%%%%%%%%%%%%%%%%%%%%%%%%%%%%%%%%%%%%%%%%%%%%
%\input{Table_dip_quad}
\begin{table}
\caption{ VMC evaluation of the energy [H], the variance [H$^2$], the dipole [Deb] and the diagonal elements $Q^{xx}$, $Q^{yy}$, $Q^{zz}$ of the traceless quadrupole tensor [Deb$\cdot$\AA] for the water molecule, compared with other accurate {\em ab initio} evaluations and experimental results.$^a$}
\label{table:BSC_prop}
%{\tiny
%{\scriptsize
{\footnotesize
%{\small
\begin{tabular}{ l c  l l  l  l l l }  

\hline \hline
% function/core/basis $^b$ & parameters $^c$ & Energy & Variance & Dipole & $Q^{xx}$ & $Q^{yy}$ & $Q^{zz}$ \\
 && \bf Energy & \bf Variance & \bf Dipole & \bf Q$^{xx}$ & \bf Q$^{yy}$ & \bf Q$^{zz}$ \\
\hline \hline
 {\bf function/core/basis }$^b$ & {\bf \# parameters }$^c$ & \multicolumn{6}{c}{\em VMC evaluations  $^d$} \\
\hline
  JDFT/ECP/hybrid 		& (33)+724+210  &  -17.24548(8)  &  0.3606(3)  &  1.9059(8)  &  2.5796(9)  &  -0.1551(9)  &  -2.4245(9)  \\ % PS-VTZ-h20 
\hline
  JSD/ECP/uncontracted 	&  14+0+666     &  -17.24820(5)  &  0.2734(2)  &  1.8881(4)  &  2.5842(5)  &  -0.1711(5)  &  -2.4131(5)  \\ % PS-UNC
  JSD/ECP/contracted		&  40+10+14878  &  -17.2482(1)  &  0.2668(10)  &  1.8755(10)  &  2.596(1)  &  -0.155(1)  &  -2.441(1)  \\ %  PS-aQZ+ 
  JSD/ECP/hybrid  		&  33+1086+465  &  -17.24824(7)  &  0.2692(1)  &  1.8877(6)  &  2.5819(7)  &  -0.1445(7)  &  -2.4374(6)  \\ % PS-VTZ-h30  
\hline
  JAGPn*/ECP/contracted 	&  40+10+14878  &  -17.2513(1)  &  0.2489(7)  &  1.8629(10)  &  2.570(1)  &  -0.149(1)  &  -2.421(1)  \\ % PS-aQZ+ 
\hline
  JAGP/ECP/uncontracted 	&  14+0+666     &  -17.2536(2)  &  0.239(1)  &  1.8881(10)  &  2.579(1)  &  -0.174(1)  &  -2.406(1)  \\ % PS-UNC
  JAGP/ECP/contracted 	&  (26)+4+4186  &  -17.2529(1)  &  0.2665(7)  &  1.8609(10)  &  2.580(1)  &  -0.147(1)  &  -2.433(1)  \\ % PS-aTZ
  JAGP/ECP/contracted 	&  26+4+4186    &  -17.25397(10)  &  0.2330(10)  &  1.8710(10)  &  2.583(1)  &  -0.145(1)  &  -2.438(1)  \\ % PS-aTZ
  JAGP/ECP/hybrid 		&  33+724+210   &  -17.25383(4)  &  0.2308(1)  &  1.8648(6)  &  2.5740(7)  &  -0.1500(7)  &  -2.4240(7)  \\ % PS-VTZ-h20 

%\hline
\hline

  JSD/NCP/hybrid 		&  33+1086+465  &  -17.20239(5)  &  0.3303(2)  &  1.8949(4)  &  2.5808(5)  &  -0.1498(5)  &  -2.4310(5)  \\ %  PS-VTZ-h30 
  JAGP/NCP/hybrid 		&  33+724+210   &  -17.20803(6)  &  0.2786(1)  &  1.8704(7)  &  2.5765(8)  &  -0.1559(8)  &  -2.4206(8)  \\ % VTZ-h20

\hline
%  DFT/AE/hybrid       	&  (45)+1189+231  &  -76.0507(4)  &  17.7(1)  &  1.8645(4)  &  2.6042(5)  &  -0.1481(5)  &  -2.4561(5)  \\ % ANO4-h21
  JDFT/AE/hybrid 		&  (45)+1189+231  &  -76.39914(6)  &  1.1881(3)  &  1.9152(3)  &  2.6122(3)  &  -0.1460(3)  &  -2.4663(3)  \\ % ANO4-h21$^{b}$
  JSD/AE/hybrid   		&  36+1038+231  &  -76.40052(5)  &  1.1579(7)  &  1.8973(2)  &  2.5740(3)  &  -0.1362(3)  &  -2.4377(3)  \\ % ANO2-h21$^b$
  JAGP/AE/hybrid 		&  43+1163+231  &  -76.40741(2)  &  1.01531(9)  &  1.8894(1)  &  2.5875(1)  &  -0.1466(1)  &  -2.4409(1)  \\ % ANO5-h21$^b$
\hline \hline

\multicolumn{2}{l}{\bf method/basis} \\

%%\multicolumn{2}{l}{ Ref.\cite{Feller:1987dm}: MRSD-CI/120CGTO } & -76.3861 && 1.860 & 2.5535 \\
\multicolumn{2}{l}{ MRSD-CI/140CGTO $^e$ } & -76.3963 && 1.870 & 2.5556 \\ % Ref.\cite{Feller:1987dm}

\multicolumn{2}{l}{ HF/aug-cc-pCV6Z $^f$ }   &&& 1.9813 \\ %  0.77952
\multicolumn{2}{l}{ CCSD/aug-cc-pCV6Z $^f$ } &&& 1.8808 \\ % 0.73995
\multicolumn{2}{l}{ CCSD(T)/aug-cc-pCV6Z $^f$ } &&& 1.8578 \\ % 0.73091
\multicolumn{2}{l}{ CCSD(T)/CBS $^f$ }       &&& 1.858(12) \\ %Lodi:2008ic tab.VIII, non-relativistic all electron 0.7310(5); relativistic correction + vibrational averaging -> 0.7294(6)
%\multicolumn{2}{l}{ IC-MRCI,CAS1/aug-cc-pCV6Z $^f$ }   &&& 1.8761 \\ % 0.73811
%\multicolumn{2}{l}{ IC-MRCI+Q,CAS1/aug-cc-pCV6Z $^f$ } &&& 1.8559 \\ % 0.73018

\hline \hline
%\multicolumn{2}{l}{\bf Exact $^e$ }      & -76.438 \\ % Ref.\cite{Feller:1987dm}
\multicolumn{2}{l}{{\bf Experiment} $^g$   }                          &&& 1.8546(6)  & 2.63(2) & -0.13(3) & -2.50(2) \\
%\hline \hline

\\ %caption
\multicolumn{8}{ p{15cm} }{ 
$^a$ For the quadrupole calculation, the molecule is in the xy-plane, with the bisector of the HOH angle parallel to the y-axis, and in the reference frame of the center of mass. 
$^b$ This column reports the wave function ansatz for the VMC calculations, the description of the two core electrons of the oxygen atom, and the basis set type for the determinantal part. The Jastrow basis is O(4s,2p,1d) H(2s,1p) for the all electrons calculations, and is O(3s,2p,1d) H(2s,1p) for the ECP and NCP cases. Further details in the text. 
$^c$ Reports the number of parameters for the determinantal basis set, as the summation of the number of exponents (first number, that is written between parenthesis in case of {\em opt:noZ}), number of contraction coefficients (second number, that is zero for uncontracted basis), and the independent elements of the AGP matrix (third number).
$^d$ The VMC expectation values for the  dipole and the quadrupole have been calculated as described in Section~\ref{sec:dipole.quadrupole}. 
$^e$ From ref.\cite{Feller:1987dm}. 
$^f$ From Tables I and VIII of ref.\cite{Lodi:2008ic}. % Non-relativistic Born-Oppenheimer estimate of the dipole % From Tables I and VIII of ref.\cite{Lodi:2008ic}. IC-MRCI stands for internally contracted multireference configuration-interaction approach;  +Q denote the use of Davidson- or Pople-corrected energies; CAS1 is  8 by 8 CAS. % Non-relativistic Born-Oppenheimer estimate of the dipole 
$^g$ Dipole from ref.\cite{Clough:1973bh}, quadrupole from ref.\cite{Verhoeven:1970jd}. }\\
\end{tabular}
}
\end{table}
%%%%%%%%%%%%%%%%%%%%%%%%%%%%%%%%%%%%%%%%%%%%%%%%%%%%%%%%%%%%%%%%%%%%%%%%%%%%%%%%%%%%%%%%%%%%%

% selection of results here
A selection of the results, representing the largest basis sets (that we can consider at convergence) are reported in Tab.~\ref{table:BSC_prop}, compared with others highly accurate {\em ab initio} calculations and the experimental evaluations.
% hybrid
Considering both the JSD/ECP and the JAGP/ECP results, with uncontracted, contracted, and hybrid atomic basis, for the basis set convergence the computational advantage of the latter compared with the others can be appreciated. Indeed, calculations with large basis sets are problematic because with the increase of the number of variational parameters a large statistics and computational time are required to obtain a stable optimization. It is therefore crucial to reduce the number of variational parameters in the wave function without missing the important polarization and diffuse terms.
%To this aim, the use of atomic hybrid orbitals is particularly useful.

% ECP / NCP
A parallel comparison between similar wave function ansatzes in Tab.~\ref{table:BSC_prop} shows that  using of ECP pseudo potential leads to lower variances than using NCP pseudo potentials. Following the same trend, the dipoles obtained with ECP are slightly closer to experiments than these calculated using NCP pseudo potentials. 
% JAGP
In summary, concerning the wave function ansatz,
the general trend in accuracy is, as expected: 
$$\textrm{ JDFT < JSD < JAGPn* < JAGP. }$$
The JSD wave function has a significant difference in energy and variance with respect to  JAGPn* and JAGP, and the quality of the wave function is also reflected in the accuracy of the dipole moment.
We have also observed that, if large basis sets are used, JSD and JDFT are very stable in the optimization, whereas the JAGP wave function requires a larger statistics in the optimization, otherwise it can be unstable. 
% AE
%Using basis sets with large exponents we have obtained relatively small variances also for the JDFT functions optimized with the {\em opt:noZ} scheme, and energies up to -76.39914(6)H.
%For the JSD we obtain an energy of -76.40052(5)H with a corresponding variance of 1.1579(7)H$^2$, while our best result with the JAGP has an energy of -76.40741(2)H and variance of 1.01531(9)H$^2$.
%As can be observed in Tab.~\ref{table:BSC_AE_prop} and Figure~S3, JDFT does not give good dipoles, while JAGP gives accurate dipoles only for the larger basis.

% AE e confronto con altri metodi
Comparing the AE versus the ECP or NCP calculations, it emerges that the all electron calculations provide a value for the dipole that is slightly larger than the one obtained with pseudo potentials.
The difference could arise from the fact that the basis set convergence in the all electron calculation is more difficult to reach, or to the relativistic effects, that are not considered  in the all electron calculations, while are implicitly taken into consideration both in the ECP and in the NCP calculations, through the scalar relativistic correction in the pseudo potentials.
According to \citet{Lodi:2008ic}, the relativistic correction to the dipole can be estimated of the order of -0.0043~Deb, that is not enough to completely account for  the  difference between AE and pseudo potential results, but it is in the right direction.

The accuracy of the VMC evaluations of the dipole appears to be comparable to the CCSD calculations, or better, depending on the wave function ansatz, whereas CCSD(T) calculations with large enough basis (or CBS extrapolation) are closer to the experimental values with respect to our VMC results.
For a comparison between the computational and  the experimental results, it is important to  estimate the order or magnitude of all the theoretical/computational approximations. Beside the already mentioned relativistic effect, there are also the quantum nuclei effects.
These effects can be accounted by  averaging the dipole moment over the ground-state roto-vibrational nuclear-motion, and according to \citet{Lodi:2008ic}, the correction is of the order of {+0.0003~Deb}, thus rather small.
In conclusion, the best VMC description of the dipole moment appears to be provided by the JAGP ansatz, with ECP core and hybrid basis for the determinantal part.

\subsection{ The ionization and the atomization energies}\label{sec.IE}

The Ionization Energy ($IE$) of the water molecule can be estimated from the energy difference $\Delta E = E_{H_2O} - E_{H_2O^+}$ between the energy $E_{H_2O}$ of the neutral  molecule $H_2O$ and the energy $E_{H_2O^+}$ of the cation $H_2O^+$, both in their relaxed geometries. 
Similarly to the previous section, we have tried several wave function types and several basis, in order to study the basis set convergence.
The complete list of the results are reported in Tab.~\ref{table:IE_VMC} for the VMC calculations, and in Tab.~\ref{table:IE_LRDMC} for the LRDMC results.
%the Supporting Information (in Table~S7 for the VMC calculations, and in Table~S8 for the LRDMC calculations).
In Tab.~\ref{table:IE} we report a selection of the results for the largest (more converged) basis sets, and a comparison with other {\em ab initio} calculations and  experiments.
% effetto ZPE
In order to compare the computational results with the experimentally measured value\cite{NISTweb}  $IE_{exp}=12.621(2)eV$, % = 0.4638(1) H$, 
we have to take into account the difference $\Delta_{ZPE}$ between the vibrational zero-point energy (ZPE) of $H_2O$ and of $H_2O^+$, that can be estimated by CCSD(T)/aug-cc-pVTZ calculatons\cite{cccbdb} to be of the order of $0.067eV$. %\sim0.0025 H$.
%Thus, we can compare the exact estimation of energy difference 
%$\Delta E_{exp} = IE_{exp}+\Delta_{ZPE} \sim 0.4663 H$
%with our computational results.

%%%%%%%%%%%%%%%%%%%%%%%%%%%%%%%%%%%%%%%%%%%%%%%%%%%%%%%%%%%%%%%%%%%%%%%%%%%%%%%%%%%%%%%%%%%%%
%\input{Table_IE_VMC}
% Ionization Energy Water Molecule
\begin{table}
\caption{ VMC and LRDMC evaluation of the Ionization Energy (IE) for the water Molecule, in comparison with other accurate {\em ab initio} evaluations and experimental results.$^a$ }  
\label{table:IE}
%{\tiny
%{\scriptsize
%{\footnotesize
%{\small
\begin{tabular}{ l   l l  l  }  
%\hline \hline
%method/function/core/opt & $E_{H_2O}$[H] & $E_{H_2O^+}$[H] & IE[eV] \\
 & $E_{H_2O}$[H] & $E_{H_2O^+}$[H] & IE[eV] \\
\hline \hline
{\bf method/function/core/opt}  \\
\hline
%{\em Pseudo Filippi}\\
VMC/JSD/ECP/all & -17.2481(1) & -16.7795(1)  & 12.684(5)\\ 
\hline
VMC/JAGPn*/ECP/all & -17.2513(1) & -16.7824(1) & 12.692(5) \\ 
\hline
VMC/JAGP/ECP/noZ & -17.2520(1) & -16.7823(1) & 12.714(3) \\
VMC/JAGP/ECP/all & -17.2538(1) & -16.7842(1) & 12.711(3) \\
\multicolumn{1}{l}{LRDMC(a$\to$0)/JAGP/ECP/all} & -17.2647(3) &  -16.7954(3) & 12.703(8) \\
\hline
%{\em Pseudo Needs}\\
VMC/JAGP/NCP/noZ & -17.2050(1) & -16.7253(2) & 12.986(8) \\
VMC/JAGP/NCP/all & -17.2068(1) & -16.7383(1) & 12.681(3) \\
\hline 
%{\em All Electron }\\
VMC/JAGP/AE/noZ & -76.3909(4) & -75.9191(4)   & 12.771(16) \\
VMC/JAGP/AE/all & -76.4041(3) & -75.9336(3)   & 12.736(11) \\ % AE{,_CATION}_DZ_OPTXXX/OPTvmc_x8/
%\multicolumn{1}{l}{LRDMC(a$\to$0)/JAGP/AE/all}  & -76.4265(2) &  -75.9594(2) & 12.643(8) \\ % + a=0.07
\multicolumn{1}{l}{LRDMC(a$\to$0)/JAGP/AE/all}  & -76.4266(1) &  -75.9586(2)  & 12.668(6) \\ % + a=0.07,0.05
%\hline
\hline \hline
{\bf method/basis}  \\
\hline
\multicolumn{1}{l}{ HF/aug-cc-pVQZ $^b$}      && & 10.868 \\
\multicolumn{1}{l}{ B3LYP/aug-cc-pVQZ $^b$}   && & 12.610 \\
\multicolumn{1}{l}{ MP2FC/aug-cc-pVTZ $^b$}   && & 12.709 \\
\multicolumn{1}{l}{ CCSD(T)/aug-cc-pVTZ $^b$} && & 12.505 \\
\hline \hline
\multicolumn{1}{l}{\bf Experiment $^b$}     && & 12.621(2) \\
\hline \hline
% note
\\
\multicolumn{4}{ p{13cm} }{
$^a$ The ionization energy has been calculated as the sum of the energy difference $\Delta E = E_{H_2O}-E_{H_2O^+}$ and the zero point energy difference $\Delta_{ZPE}$ between the cation and the neutral form of water.
For the QMC results we have considered the $\Delta_{ZPE}$ evaluated by a CCSD(T)/aug-cc-pVTZ calculation\cite{cccbdb}, see further details in Section~\ref{sec.IE}.
$^b$ From ref.\cite{cccbdb}.
}
\end{tabular}
%}
\end{table}
%%%%%%%%%%%%%%%%%%%%%%%%%%%%%%%%%%%%%%%%%%%%%%%%%%%%%%%%%%%%%%%%%%%%%%%%%%%%%%%%%%%%%%%%%%%%%

We observe in Tab.~\ref{table:IE} that the basis set convergence plays an important role in determining an accurate value for the IE, and also here the {\em opt:all} scheme gives a remarkable improvement compared with the {\em opt:noZ} results.
The $IE$ obtained from the VMC approach with all electron calculations is slightly larger than the result for the pseudo potentials. This is probably in part due to the difficulty in reaching the basis set convergence. 

\begin{figure}
\begin{center}
\includegraphics[width=.8\textwidth]{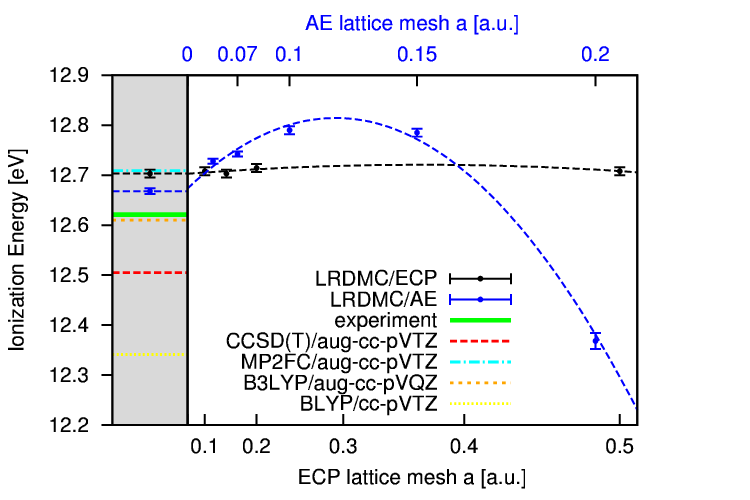}
\end{center}
\caption{
LRDMC evaluation of the ionization energy of the water molecule.
Two JAGP wave functions are considered: one corresponding to an all electrons calculation (in blue), and one with ECP pseudo potential for the two core electrons of the oxygen atom (in black).
In the right panel the extrapolation for the lattice mesh $a\to 0$ is shown, 
with a  functional form $y=c_0 + c_1 a^2 + c_2 a^4$ for the fitted lines, see ref.\cite{Casula:2010p14082}.
In the left panel with gray background the LRDMC$(a\to0)$ results are compared with other accurate {\em ab initio} computational methods and the experimental value. 
} \label{fig:IE_LRDMC}
\end{figure}

% LRDMC / ECP
The LRDMC results for JAGP function with ECP pseudo potential,  yields a minimal improvement compared to the corresponding VMC result. 
% $\Delta E_{VMC} = 0.4696(1)H$ versus $\Delta E_{LRDMC(a\to 0)} = 0.4693(4)H$. 
This is a good indication of the high quality of our variational ansatz in the description the 
electronic properties of the molecules.
In Fig.~\ref{fig:IE_LRDMC} it  also appears that the IE for the {LRDMC(a)} is almost independent of the lattice size $a$, % for the values considered of 0.1, 0.15, 0.2 and 0.5 Bohr. 
although the total energy calculated for the different $a$ have a sizable dependence on $a$, see Tab.~\ref{table:IE_LRDMC} . %(see Supporting Information, Table~S8).
This consideration can be useful for energy differences estimation, because a LRDMC calculation with $a=0.5$ is about $25$ times  computationally cheaper than the one with $a=0.1$.

% LRDMC AE
The situation for the AE calculation is rather different.
First, we have to consider very small values of the lattice size $a$, otherwise the results are meaningless.
Moreover we observe a large dependence of the $IE$ on  the mesh size $a$.
In Fig.~\ref{fig:IE_LRDMC}  we also observe that the extrapolated $a\to 0$ value of the IE is quite close to the experimental value.

%%%\input{Table_IE_LRDMC}

%%%%%%%%%%%%%%%%%%%%%%%%%%%%%%%%%%%%%%%%%%%%%%%%%%%%%%%%%%%%%%%%%%%%%%%%%%%%%%%%%%%%%%%%%%%%%
%\input{fig_IE_LRDMC}
%%%%%%%%%%%%%%%%%%%%%%%%%%%%%%%%%%%%%%%%%%%%%%%%%%%%%%%%%%%%%%%%%%%%%%%%%%%%%%%%%%%%%%%%%%%%%

%%% Tab.~\ref{table:atomization}.
In Tab.~\ref{table:atomization} we have reported some VMC estimations of the  atomization energy of the water molecule, 
which has been calculated as $E_{H_2O} - (E_{O}+2 E_{H})$.
More precisely we considered the JSD and the JAGP ansatzes for  two different basis sets: 
the PS-VTZ-h20 basis set with Jastrow basis O(3s,2p,1d) H(2s,1p), and 
the completely uncontracted basis O(4s,5p,1d) H(3s,1p) with Jastrow O(3s,2p,1d) H(2s,2p). 
In the calculation, the oxygen atom has been considered in its triplet ground state, whereas the hydrogen energy $E_{H}$ has been set to the exact 0.5H value.
% commento
All the VMC estimations are in  good agreement with the exact value of the atomization energy. 
It is interesting to note that the JAGP and the JSD estimations are almost identical, whereas there is a small difference, of the order of 2mH, between the estimations of the two different basis.
% migliora desctizione O
The improvement, in terms of variational energy,  from JSD to JAGP, both for the water molecule and for the oxygen atom, is $\sim$6mH, but the fact that JSD and JAGP give the same atomization energy indicates that this improvement is due only to a better  description of the oxygen atom by the JAGP.
% ma migliora anche legame
However this does not imply that going from JSD to JAGP produces just a vertical shift of the energy and that they provide an equivalent description of the molecular bonds.
A better description of the oxygen atom could turn out to an improvement in the description of the OH bond in  water, and consequently of the potential energy surface.  
In the following sections we will see that this is actually the case, as JAGP yields an equilibrium structure and vibrational frequencies that appears more accurate than the JSD ones.

%%%%%%%%%%%%%%%%%%%%%%%%%%%%%%%%%%%%%%%%%%%%%%%%%%%%%%%%%%%%%%%%%%%%%%%%%%%%%%%%%%%%%%%%%%%%%
%\input{Table_atomization}
% structural minimum and frequencies
%\begin{sidewaystable}

\begin{table}
%\begin{turn}{90}
\caption{ Atomization Energy of the water molecule.$^a$ }  
\label{table:atomization}

%{\tiny
%{\scriptsize
%{\footnotesize
%{\small

\begin{tabular}{ c c c c }  
%\hline \hline
   & $E_{H_2O}$ [H] & $E_O$ [H] & AE [H] \\
\hline %\hline 

%\multicolumn{4}{c}{\em basis: O(4s,5p,1d) H(3s,1p); Jastrow: O(3s,2p,1d) H(2s,2p) } \\
\multicolumn{4}{l}{ {\em VMC calculation / ECP core / uncontracted determinantal basis} $^b$  }\\

JSD  	& -17.24819(5) & -15.87586(9) & 0.3723(1) \\
JAGP 	& -17.2536(2)  & -15.8811(1)  & 0.3725(2) \\
$E_{JAGP}-E_{JSD}$ & 0.0054(2)  & 0.0052(1) \\

\hline

%\multicolumn{4}{c}{\em basis: PS-VTZ-h20; Jastrow: O(3s,2p,1d) H(2s,1p)} \\
\multicolumn{4}{l}{ {\em VMC calculation / ECP core / hybrid determinantal basis} $^c$  }\\

JSD  	& -17.2471(1)  & -15.8769(1) &  0.3702(2) \\
JAGP 	& -17.25383(4) & -15.8838(2) &  0.3700(2) \\
$E_{JAGP}-E_{JSD}$ & 0.0067(1)  & 0.0069(2) \\

\hline %\hline
{\bf Exact} $^d$ & -76.438 & -75.0673 & 0.3707 \\
%{\em exact} & -76.4396$^a$ & -75.0673$^b$ & 0.3723$^c$ \\
%\hline \hline
\\
\multicolumn{4}{ p{10cm} }{ 
$^a$ The atomization energy (AE) is calculated as $E_{H_2O} - (E_{O}+2 E_{H})$ for different wave functions and basis sets. The hydrogen atom energy $E_H$ is 0.5H, with a negligible stochastic error.
$^b$ Determinantal basis: O(4s,5p,1d) H(3s,1p); Jastrow basis: O(3s,2p,1d) H(2s,2p).
$^c$ Determinantal basis: O(9s,9p,2d,1f)/\{12\} H(6s,5p,1d)/\{4\}; Jastrow basis: O(3s,2p,1d) H(2s,1p).
$^d$ All electron evaluation of $E_{H_2O}$ from ref.\cite{Feller:1987dm} and of  $E_O$ from ref.\cite{Chakravorty:1993gg}.}\\
\end{tabular}
%}
%\end{turn}
\end{table}

%\end{sidewaystable}

%%%%%%%%%%%%%%%%%%%%%%%%%%%%%%%%%%%%%%%%%%%%%%%%%%%%%%%%%%%%%%%%%%%%%%%%%%%%%%%%%%%%%%%%%%%%%

\subsection{ The PES properties: equilibrium structure,  harmonic and fundamental frequencies }\label{sec.freq}

\begin{figure}
\begin{center}
\includegraphics[width=.8\textwidth]{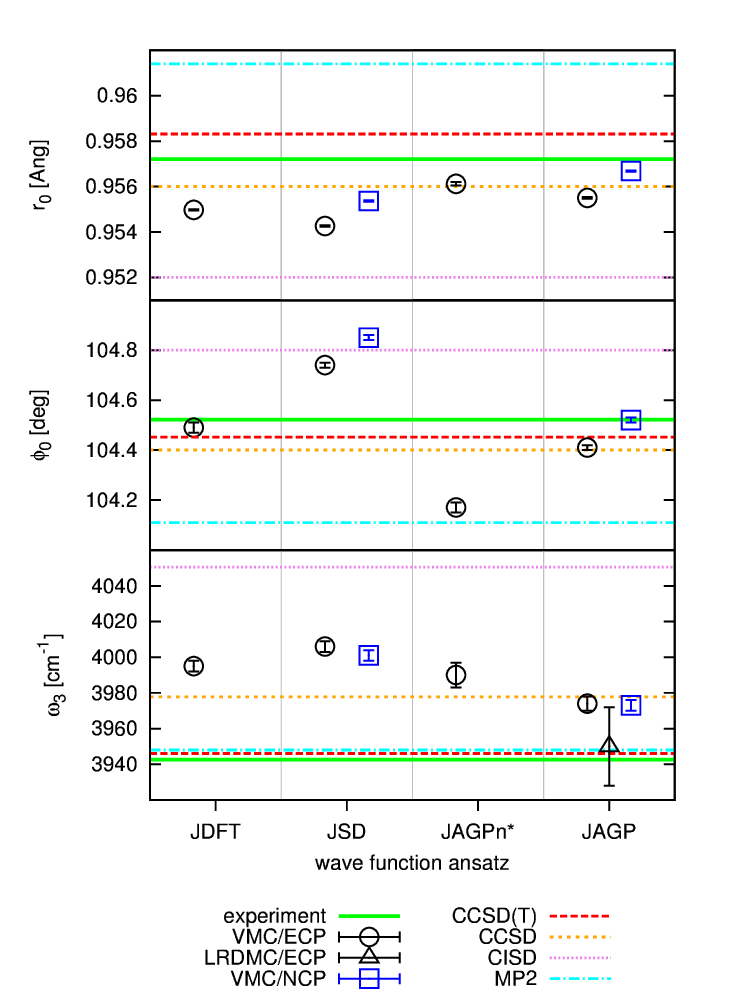}
\end{center}
\caption{
The following properties of the ground state PES of the water molecule around the equilibrium structure are reported: 
the bond length $r_0$, the bond angle $\phi_0$, and the harmonic frequency of the asymmetric stretching $\omega_3$.
VMC results for 
JDFT, JSD, JAGPn*, and JAGP wave functions are reported, 
using ECP (in black) and NCP (in blue) pseudo potential for the two core electrons of the oxygen.
The LRDMC$(a\to0)$ value of $\omega_3$ for the JAGP/ECP function extrapolated in Fig.~\ref{fig:omega3} is also shown.
For a comparison, the results of MP2, CISD, CCSD and CCSD(T) calculations are reported (see Tab.~\ref{table:eqfr} and references therein for details).
} \label{fig:PESprop}
\end{figure}

The equilibrium structure  and the vibrational frequencies, both harmonic and fundamental, have been calculated for several wave function types and with increasing basis sets. 
The results that comes from the fitting of the energies or of the forces (see Section~\ref{sec.comput.det}), for all the tested wave functions, are reported in Tab.~\ref{table:eqfrSI}. %the Table~S9 of the Supporting Information.
In Tab.~\ref{table:eqfr} and in Fig.~\ref{fig:PESprop} we report a selection of the results obtained for the largest basis sets.

% osservazioni su fit energy/force
In agreement with \citet{Zen:2012br}, we observe that the stochastic error for the minimum energy configuration and the frequencies obtained by the fit of the forces are much smaller than that coming from the fit of the energies.
We have also tested the correlated sampling (CS) technique for the fitting of the energies, and, in this case, the error is not much larger than the corresponding one obtained with the force fit.
Moreover, we can observe in Tab.~\ref{table:eqfrSI} %Table~S9 
that the  results for the JDFT with ECP pseudo potential, %and basis PS-VTZ-h20, 
obtained by the fitting of the CS energies are not in perfect agreement with the results coming from the fit of the forces.
The reason for this discrepancy is easily understood if we consider that in the expression used for the forces, we are neglecting the explicit derivatives of the parameters, because they vanish at the minimum, as explained in Section~\ref{sec.force}.
For a JDFT wave function this assumption is not correct, because only the parameters in the Jastrow are optimized, whereas the parameters in the determinant remain those of a DFT calculation, and are  in general not at the minimum of the VMC energy.
As a confirmation of this interpretation, we observe that the results for the JSD function with  ECP pseudo potential %and  PS-VTZ-h30 basis 
obtained by fitting of the CS energies and of the forces are compatible within the estimated stochastic errors.

It is evident that the equilibrium structure and the frequencies are clearly and smoothly converging with an increasing basis set, and both the {\em opt:all} and the hybrid atomic orbitals are very useful for this convergence.
The converged equilibrium structure is in good agreement both with other highly accurate  {\em ab initio} calculations and with the experimental values, also reported in Tab.~\ref{table:eqfr}.
The calculated frequencies are slightly overestimated compared with the experimental results of the CCSD(T) values, but they are in agreement with the CCSD results.

By comparing the JSD and the JAGP results (with the larger basis sets) we notice  that the latter ones are closer to the experimental values, see Fig.~\ref{fig:PESprop}.
This is an indication that the JAGP ansatz provides a better description of the PES and of the chemical bonds, as compared with the JSD ansatz. 

\begin{figure}
\begin{center}
\includegraphics[width=.8\textwidth]{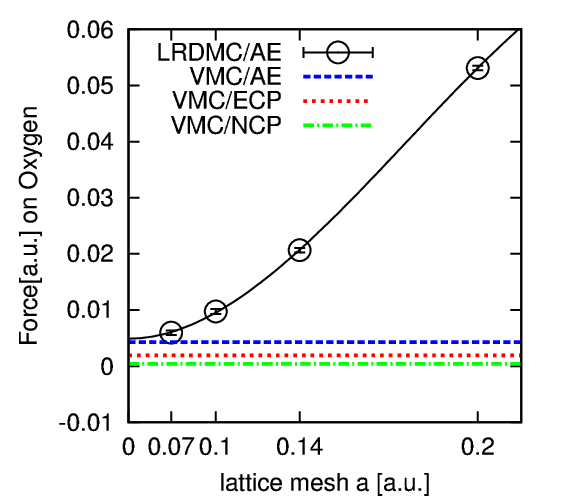}
\end{center}
\caption{
Residual force (in H/Bohr) on the oxygen atom, calculated on the experimental equilibrium structure. The reported values correspond to VMC calculations with  AE (blue), ECP (red) or NCP (green) core,  and LRDMC calculations (black) with AE core and lattice mesh size $a$ equal to 0.07, 0.1, 0.14, and 0.2 Bohr.
The fitting line $F=f_0 + f_1 a^2 + f_2 a^4$ for the LRDMC calculations is reported in black. 
} \label{fig:Fresidua}
\end{figure}

Converged results obtained using ECP or NCP results are in good agreement for the frequencies, while it appears that the OH bond for the equilibrium structure obtained from ECP is slightly smaller than the bond for NCP. 
This leads us to ask which pseudo potential is more compatible with the all-electron calculations, either  ECP or NCP.
Since all-electron calculations are computationally expensive, especially if we consider a basis that is large enough to be considered converged, we have decided to address this question by simply evaluating the residual force in the experimental equilibrium configuration with both pseudo potentials.
As can be observed from Fig.~\ref{fig:Fresidua}, the ECP pseudo potential is more compatible with the AE calculations.

%%%%%%%%%%%%%%%%%%%%%%%%%%%%%%%%%%%%%%%%%%%%%%%%%%%%%%%%%%%%%%%%%%%%%%%%%%%%%%%%%%%%%%%%%%%%%
%\input{Table_eq+freq}
% structural minimum and frequencies
%\begin{sidewaystable}
\begin{table}
%\begin{turn}{90}
\caption{ VMC evaluation of the equilibrium configuration, the harmonic and the fundamental frequencies of the water molecule, compared with other accurate {\em ab initio} evaluations and experimental results.$^a$ }  
\label{table:eqfr}

%{\tiny
{\scriptsize
%{\footnotesize
%{\small

\begin{tabular}{ l    l l   l l l   l l l }  
%\hline \hline
 & \multicolumn{2}{c}{\bf equilibrium structure} & \multicolumn{3}{c}{{\bf harmonic freq.} [cm$^{-1}$]} & \multicolumn{3}{c}{{\bf fundamental freq.} [cm$^{-1}$]} \\
 & $r_0$[\AA] & $\phi_0$[deg] & $\omega_2$ & $\omega_1$ & $\omega_3$ & $\nu_2$[010] & $\nu_1$[100] & $\nu_3$[001] \\
                                       
\hline \hline 

{\bf function/core} & \multicolumn{8}{c}{\em VMC results} \\ 

\hline
%E & DFT    & ECP & PS-VTZ-h20 &  & 0.97482(4) & 104.30(1)   & 1533(1)  & 3673(3)  & 3782(4)  & 1489(2)   & 3517(11) & 3614(13)  \\ 
%E  & JDFT   & ECP & PS-VTZ-h20 &  & 0.95560(1) & 104.548(3)  & 1667.7(6)& 3880(1)  & 3995(2)  & 1614.2(4) & 3702(2)  & 3799(4)   \\ 
JDFT/ECP &  0.95497(3) & 104.49(2)   & 1664(2)  & 3882(2)  & 3995(3)  & 1610(1)   & 3693(2) & 3787(3)    \\ 
%\hline
%E  & JSD    & ECP & PS-VTZ-h30 &  & 0.95432(3) & 104.81(1)   & 1674(2)  & 3894(3)  & 3997(6)  & 1615(2)   & 3713(6)  & 3815(11)  \\ 
JSD/ECP &  0.95426(3) & 104.74(1)   & 1670(2)  & 3892(3)  & 4006(3)  & 1617(1)   & 3702(3) & 3794(2)    \\ 
%\hline
JAGPn*/ECP &   0.95612(8) & 104.17(2)   & 1710(3)  & 3896(6)  & 3990(7)  & 1654(1)   & 3710(3)  & 3800(7)   \\ 
%\hline
%JAGP   & ECP & PS-aDZ       &  & 0.95545(7) & 104.42(2)   & 1666(3)  & 3873(5)  & 3989(9)  & 1613(2)   & 3685(5)  & 3775(12)  \\ 
JAGP/ECP &  0.95550(4) & 104.41(1)   & 1669(1)  & 3872(3)  & 3974(4)  & 1613.3(6) & 3677(2) & 3772(2)    \\ 
\hline
JSD/NCP &   0.95536(3) & 104.85(1)   & 1668(2)  & 3889(2)  & 4001(3)  & 1613.1(9) & 3700(2) & 3796(3)    \\ 
%\hline
%JAGP   & NCP & PS-aDZ       &  & 0.95665(7) & 104.54(2)   & 1664(3)  & 3884(5)  & 3990(7)  & 1609(1)   & 3678(5)  & 3782(8)   \\ 
JAGP/NCP &  0.95668(3) & 104.52(1)   & 1663(2)  & 3869(2)  & 3973(3)  & 1610.4(7) & 3679(2) & 3767(3)    \\ 

\hline \hline

{\bf method/basis} & \multicolumn{8}{c}{\em } \\ 

%\multicolumn{1}{l}{   SCF/DZP  $^b$ }       & 0.9457 & 106.16 \\ 
\multicolumn{1}{l}{ BLYP/aug-cc-pVTZ $^b$}       & 0.9719 & 104.47 & 1596 & 3655 & 3757 & 1543 & 3480 & 3567 \\ % & 1.8033 dipole 
%\multicolumn{1}{l}{ B97-2/aug-cc-pVTZ  $^b$ }       & 0.9567 & 104.77 & 1645 & 3860 & 3968 & 1594 & 3701 & 3794 \\ % & 1.8473 
\multicolumn{1}{l}{ B3LYP/aug-cc-pVTZ  $^b$ }       & 0.9619 & 105.08 & 1627 & 3796 & 3899 & 1575 & 3631 & 3720 \\ % & 1.8411 
\multicolumn{1}{l}{FC MP2/aug-cc-pVTZ  $^b$ }       & 0.9614 & 104.11 & 1628 & 3822 & 3948 & 1578 & 3653 & 3767 \\ % & 1.8701 

\multicolumn{1}{l}{    CISD/(13,8,4,2/8,4,2) $^c$ }      & 0.952 & 104.8 & 1676.1 & 3947.3 & 4050.5 \\ 
%\multicolumn{1}{l}{   QCISD/(13,8,4,2/8,4,2) $^c$ }      & 0.956 & 104.4 & 1659.4 & 3868.0 & 3976.2 \\ 
%\multicolumn{1}{l}{QCISD(T)/(13,8,4,2/8,4,2) $^c$ }      & 0.959 & 104.2 & 1646.5 & 3827.0 & 3937.3 \\ 
\multicolumn{1}{l}{    CCSD/(13,8,4,2/8,4,2) $^c$ }      & 0.956 & 104.4 & 1662.5 & 3870.9 & 3977.8 \\ 
%\multicolumn{5}{l}{ CCSD(T)/(13,8,4,2/8,4,2) $^c$ }      & 0.959 & 104.1 & 1648.4 & 3831.7 & 3941.5 \\ 
\multicolumn{1}{l}{FC CCSD(T)/aug-cc-pV7Z $^d$   }      & 0.95831 & 104.452 & 1649.83 & 3835.55 & 3946.05 & 1595.58  & 3659.31 & 3757.45 \\

\hline \hline
% experiment
\multicolumn{1}{l}{\bf Experiment $^e$ }          & 0.95721(30) & 104.522(50) & 1648.47  & 3832.17  & 3942.53  & 1594.59   & 3656.65 & 3755.79    \\                                                       
\hline \hline

\\ % caption
\multicolumn{9}{ p{15cm} }{ 
$^a$ For the VMC results, the equilibrium configuration, the harmonic frequencies $\omega_i$ and the fundamental frequencies $\nu_i$ have been evaluated from the PES fitted using the VMC forces; see details in the text and in \citet{Zen:2012br}. 
$^b$ From ref.\cite{Barone:2005p24347}.
$^c$ From ref.\cite{KIM:1995p23441}.
$^d$ From ref.\cite{Feller:2009p23440}.
$^e$ From ref.\cite{Benedict:1956id}.
}\\
\end{tabular}
}
%\end{turn}
\end{table}

%\end{sidewaystable}
%%%%%%%%%%%%%%%%%%%%%%%%%%%%%%%%%%%%%%%%%%%%%%%%%%%%%%%%%%%%%%%%%%%%%%%%%%%%%%%%%%%%%%%%%%%%%

%%%%%%%%%%%%%%%%%%%%%%%%%%%%%%%%%%%%%%%%%%%%%%%%%%%%%%%%%%%%%%%%%%%%%%%%%%%%%%%%%%%%%%%%%%%%%
%\input{Fig_PESprop}
%%%%%%%%%%%%%%%%%%%%%%%%%%%%%%%%%%%%%%%%%%%%%%%%%%%%%%%%%%%%%%%%%%%%%%%%%%%%%%%%%%%%%%%%%%%%%

%%%%%%%%%%%%%%%%%%%%%%%%%%%%%%%%%%%%%%%%%%%%%%%%%%%%%%%%%%%%%%%%%%%%%%%%%%%%%%%%%%%%%%%%%%%%%
%\input{Fig_Fresidua}
%%%%%%%%%%%%%%%%%%%%%%%%%%%%%%%%%%%%%%%%%%%%%%%%%%%%%%%%%%%%%%%%%%%%%%%%%%%%%%%%%%%%%%%%%%%%%

\begin{figure}
\begin{center}
%(a)\includegraphics[width=.5\textwidth]{FIGnew/omega3_LRDMC}
%(b)\includegraphics[width=.4\textwidth]{FIGnew/omega3lrdmc}
(a)\includegraphics[width=.5\textwidth]{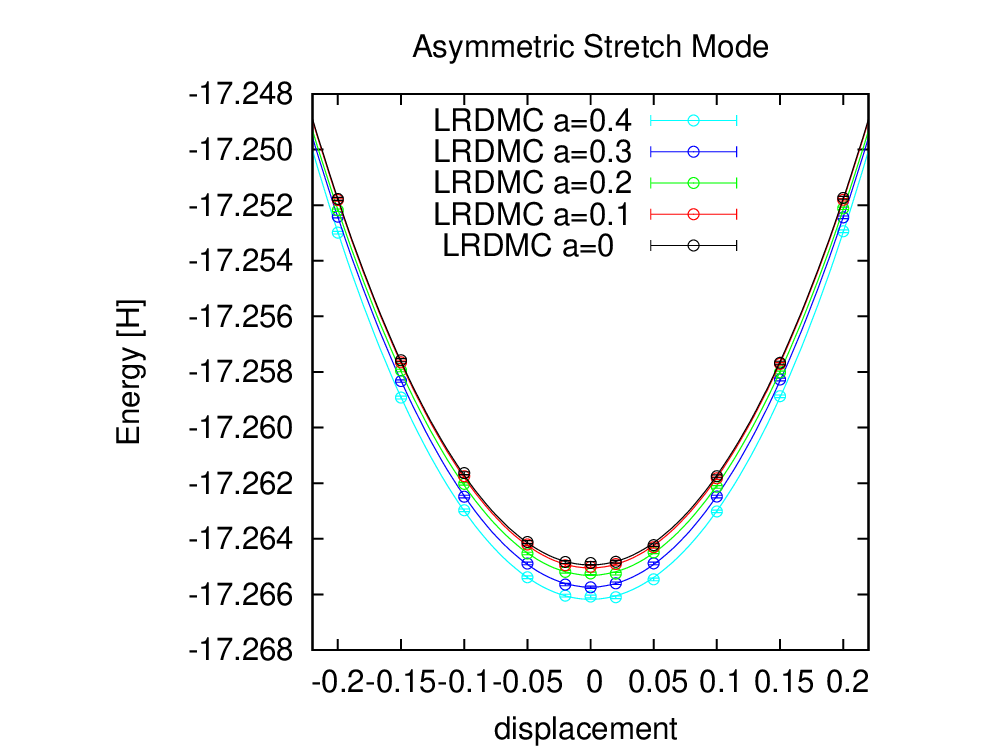}
(b)\includegraphics[width=.4\textwidth]{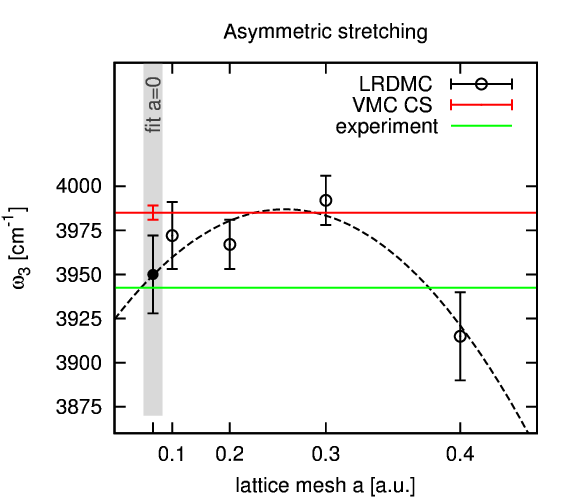}
\end{center}
\caption{
%functions: JAGP ECP PS-aDZ 
%center: fitted minimum structure, see Tab.~\ref{table:eqfr}: 0.95545(7) 104.42(2).
%asymmetric stretching
%fitting function: $y(x) = a + b x^2 + c x^4$
%corresponding frequencies $\omega_3$ reported in Tab.~\ref{table:omega3SI}.
We consider a JAGP/ECP wave function.
In panel (a) we report the 
LRDMC values of energy, calculated for lattice mesh sizes $a$ of 0.4, 0.3, 0.2, 0.1, and extrapolation to zero.
%, PS-aDZ basis set for the determinantal part and O(3s,2p,1d) H(2s,1p) for the $J_3$.
The displacement is along the asymmetric stretching mode, and the center corresponds to the VMC minimum structure for this wave function. 
%: $r=0.95545$\AA~ and $\phi=104.42^o$,  
The fitting functions $y(x) = c_0 + c_2 x^2 + c_4 x^4$ are represented in the plot as color coded continuous lines. 
%of the same color of the corresponding lattice size $a$.
In panel (b) the values for the harmonic frequency are reported versus the corresponding values of the lattice mesh $a$.
For a comparison, the experimental evaluation and the value obtained from the correlated sampling of the VMC energies is also shown, in green and red respectively.
} \label{fig:omega3}
\end{figure}

It is clear that, by  comparing the VMC frequencies with the experimental or the CCSD(T) ones, there is  still room to improve the accuracy of the QMC variational wave function as far as the vibrational properties  are concerned. We have explored the possibility to go beyond the variational scheme  using LRDMC calculations. Since these calculations are much more computationally demanding than VMC, we have only evaluated at the LRDMC level the frequency of vibration of the asymmetric stretching of the molecule by interpolating the one-dimensional energy profile computed along the mode eigenvector (computed by the VMC/JAGP/ECP). 
The analysis of different mesh sizes $a$, together with the corresponding harmonic frequencies $\omega_3$ are reported in Fig.~\ref{fig:omega3}(a). 
% interpolation e stima delle frequenze
The interpolating lines have been used to extrapolate an estimation of the frequency, and are reported in Fig.~\ref{fig:omega3}(b) and in Tab.~\ref{table:omega3SI}. %Table~S10 of the Supporting Information.

% considerazione sui risultati
\ref{fig:omega3}(b) shows for the LRDMC estimates a general improvement in the value of the frequency, from the VMC  
$\omega_3^{VMC}=3989(9) cm^{-1}$ towards the experimental value 
$\omega_3^{exp}=3942.53 cm^{-1}$, because the LRDMC $a\to 0$ extrapolation 
$\omega_3^{a\to 0}=3950(22) cm^{-1}$
differs from $\omega_3^{exp}$ only by $\sim 7 cm^{-1}$, i.e., comfortably within one sigma.
%However, also $\omega_3^{VMC}$ is within two sigma from $\omega_3^{a\to 0}$.
Therefore, the size of the stochastic error does not allow to definitively conclude that  LRDMC, within the fixed node approximation, provides a very accurate frequency, but it is likely that it improves the VMC calculation.
To definitely solve this issue it is necessary to further decrease the stochastic error, which is  at least one order of magnitude computationally more expensive than the corresponding VMC calculations. Moreover we have seen that we need a careful  $a\to 0$ extrapolation, with almost prohibitive computations  with small $a$ values.
% DMC
It could be that, always within the fixed node approximation, the DMC approach gives more accurate results.
However the most convenient way to enhance the precision of the frequency estimation for a fixed node calculation is probably to use the forces, as for our VMC calculations.
To this aim several other issues have to be tackled, such as having a consistent estimation of the force with finite variance, and eliminating any possible bias due to the mesh size $a$ for LRDMC or to the time step $\tau$ for DMC.

%%%%%%%%%%%%%%%%%%%%%%%%%%%%%%%%%%%%%%%%%%%%%%%%%%%%%%%%%%%%%%%%%%%%%%%%%%%%%%%%%%%%%%%%%%%%%
%\input{Fig_omega3}
%%%%%%%%%%%%%%%%%%%%%%%%%%%%%%%%%%%%%%%%%%%%%%%%%%%%%%%%%%%%%%%%%%%%%%%%%%%%%%%%%%%%%%%%%%%%%

%%%%%%%%%%%%%%%%%%%%%%%%%%%%%%%%%%%%%%%%%%%%%%%%%%%%%%%%%%%%%%%%%%%%%
%%%%%%%%%%%%%%%%%%%%%%%%%%%%%%%%%%%%%%%%%%%%%%%%%%%%%%%%%%%%%%%%%%%%%

%  \input{6.Conclutions}
%%%%%%%%%%%%%%%%%%%%%%%%%%%%%%%%%%%%%%%%%%%%%%%%%%%%%%%%%%%%%%%%%%%%%
\section{Conclusions}\label{sec.conclusion}
%%%%%%%%%%%%%%%%%%%%%%%%%%%%%%%%%%%%%%%%%%%%%%%%%%%%%%%%%%%%%%%%%%%%%
%%%%%%%%%%%%%%%%%%%%%%%%%%%%%%%%%%%%%%%%%%%%%%%%%%%%%%%%%%%%%%%%%%%%%

In this paper we have taken the water molecule as a test system to challenge the abilities of QMC approaches to evaluate several molecular properties: 
the energy, the dipole and the quadrupole momenta, the ionization and the atomization energies, the structural minimum, and the harmonic and the fundamental frequencies of vibrations.
For each of these quantities we have performed and compared several calculations corresponding to different setups for the QMC algorithm, namely: different ansatzes, different basis sets and contraction schemes, different ways to tackle the core electrons. Most of the investigation reported are based on VMC calculations, but we have also carried out   several LRDMC calculations in order to go beyond the variational ansatz.

% ansatz and Jastrow
It is known\cite{Brown:2007gh,Sorella:2007p12646,Marchi:2009p12614,LopezRios:2012cg,Clark:2011ie,Morales:2012kk} that the accuracy of QMC evaluations, both in the variational and diffusion approaches, strongly depends  on the wave function ansatz. However, systematic studies on the differences between the various ansatzes are typically limited to the total energy evaluation, whereas molecular properties can be more sensitive to these choices.
Here we have compared the JDFT, JSD, JAGPn* and JAGP ansatzes versus a larger set of properties, always bearing in mind that a good ansatz has to provide reliable results using a compact wave function with a limited number of variational parameters.
This is important, because  the optimization of the wave function could otherwise be difficult, especially if large molecules are taken into consideration. 

% basis convergence
In the first part of our work, we focused on the study of the basis set convergence, underlining the importance to optimize also the exponents of the orbitals in order to have a better chemical description with a lower number of parameters. Inspired by the strong interplay between the building of the AGP wave function and the atomic orbitals, we have introduced a new kind of orbital contraction, that we have termed atomic hybrid orbitals, which are specifically constructed for QMC calculations, and 
are somehow similar to the natural hybrid orbitals expansion\cite{NYO}. % well known in standard quantum chemistry methods
In particular, the atomic hybrid orbitals allow us to introduce diffusive and polarization orbitals in the wave function, with an impact in terms of   number of parameters much lower than the one introduced by ordinary contracted or uncontracted orbitals. Even if orbital exponents are optimized, a converged basis set for the molecular properties, such as dipole and vibrational frequencies, requires the presence of diffusive and polarization orbitals. Atomic hybrid orbitals result therefore in a remarkable computational advantage, especially for large systems. According to our observations, it emerges that the JAGP  ansatz with hybrid orbitals represents the best balance between accuracy of the results and compactness of the wave function. The reduced number of variational parameters allow us an easy management of the wave function optimization procedure,
and open perspectives for the application of VMC to large molecules using large basis sets.

% pseudo
We have also considered the impact of the description of the core electrons of the oxygen by energy consistent ECP and norm conserving NCP pseudo potentials, versus an all electron AE calculation.
Although some differences between these three approaches were observed, we have  noticed that for converged basis the differences are quite small, and they reach almost the same level of accuracy.
Thus, the most convenient choice seems the ECP, because it is computationally  cheaper than AE, and  it gives {\em ceteris paribus} a smaller variance with respect to NCP.

% diffusion
The LRDMC calculations reported in this work demonstrate that the projections schemes with fixed node approximation can partly improve the VMC results, although the computational cost is often high. 
This confirms the quality of the JAGP wave function, not only for the description of the electronic properties of relaxed molecules, but also of forces and potential energy surfaces.
On the other hand, projection methods are computationally more demanding and, as  we have seen,  often they are limited by a very difficult and expensive  extrapolation to the continuous limit $a\to 0$ (that is the analogous of the extrapolation of the time step $\tau \to 0$ for the DMC). %otherwise no improvement in the results can be appreciated.
In conclusion, the use of the JAGP wave function in combination with the hybrid orbital contraction scheme represents a promising way for an accurate many body calculation of properties for large molecules.
%%%%%%%%%%%%%%%%%%%%%%%%%%%%%%%%%%%%%%%%%%%%%%%%%%%%%%%%%%%%%%%%%%%%%
%%%%%%%%%%%%%%%%%%%%%%%%%%%%%%%%%%%%%%%%%%%%%%%%%%%%%%%%%%%%%%%%%%%%%

%%%%%%%%%%%%%%%%%%%%%%%%%%%%%%%%%%%%%%%%%%%%%%%%%%%%%%%%%%%%%%%%%%%%%
%% Acknoledgements
%%%%%%%%%%%%%%%%%%%%%%%%%%%%%%%%%%%%%%%%%%%%%%%%%%%%%%%%%%%%%%%%%%%%%

\section*{Acknowledgement}
The authors thank Michele Casula for useful comments and Matteo Barborini for illuminating discussions and critical reading of the manuscript.
The authors acknowledge funding provided by the European Research Council project n. 240624 within the VII Framework Program of the European Union and by MIUR within PRIN 2011 project. Computational resources were supplied by CINECA (ISCRA award N. HP10AOW1FU), PRACE infrastructure, and the Caliban-HPC centre at the Universit\'a de L'Aquila.

\appendix

\section{ VMC Optimization Schemes}

\subsection{ The optimization scheme for JAGP}
In this work, the optimization of the JAGP function has been obtained along the following intermediate steps:
\begin{itemize}
\item[{\em (i)}] optimization of the AGP, namely the matrix and the contractions in the basis set, with $b_1=b_2=1$ and without the other Jastrow terms; 
\item[{\em (ii)}] optimization of the Jastrow terms, keeping the AGP parameters fixed; % to the values obtained in step {\em i};
\item[{\em (iii)}] optimization of the overall JAGP, keeping fixed only the exponents in the basis set of the AGP;
\item[{\em (iv)}] optimization/relaxation of the AGP exponents, keeping fixed all the other terms; % to the values obtained in step {\em iii};
\item[{\em (v)}] optimization of all the AGP part, with Jastrow fixed;
\item[{\em (vi)}] optimization of all the parameters. \\
\end{itemize}
The end of  {\em Opt:noZ} is the step {\em (iii)}
and the end of {\em Opt:all} is the step {\em (vi)}.
In some cases, namely when we have used the hybrid orbitals in the AGP part, we followed a slightly different scheme, as we optimized also the exponents already in steps {\em (i)} and {\em (iii)}, obtaining the {\em Opt:all} at step {\em (iii)}.

\subsection{ The optimization scheme for JSD and JDFT}
The optimization of JSD starts from the DFT function, obtained from a DFT-LDA calculation, and then proceeds going through the following intermediate steps:
\begin{itemize}
\item[{\em (i)}] optimization of the Jastrow, keeping fixed the determinantal part;
\item[{\em (ii)}] optimization of the MOs, those number $n$ remains fixed to $N/2$;
\item[{\em (iii)}] optimization of all the parameters, always keeping fixed to $N/2$ the number of MOs.
\end{itemize}
Notice  that after the step {\em (i)} we obtain the JDFT function.
The constrained optimization of the MOs with $n$ fixed is described in ref.\cite{Marchi:2009p12614} for the case where all the orbitals in the determinantal part are uncontracted, while for the case of contracted orbitals, as in this work, a straightforward generalization of the method is used.

\subsection{ The optimization scheme for JAGPn*}
JAGPn* differs from JSD only for the number of MOs considered and the presence of the weights. Therefore the optimization is very similar, and actually proceeds going through the three steps of JSD, with the difference that here the number of MOs is $n^*$. 
The presence of these extra MOs makes the optimization less  stable. Therefore the statistics has to be increased and optimization speed has to be decreased with an increased computational cost.

%\newpage

%\input{Table_basis+np}
% AGP basis list
\begin{table}[p!]
\caption{ Atomic basis sets }
\label{table:AGPbasis}
%{\tiny
{\scriptsize
%{\footnotesize
%{\small
\begin{tabular}{ c  c c  c c  c c  c c }  

\hline
\hline
name & source & filter $^a$ & Oxygen & Hydrogen & $n_z$ & $n_c$ & $n_{orb}$ & $n_m$  \\
\hline
\hline
PS-UNC       & none            &       & (4s,5p,1d)                        & (3s,1p)                     & 14 &  0 &  36 &  666 \\
PS-DZ        & cc-pVDZ         & R$_c$ + Z$^{20}$   & (4s,4p,1d)/[2s,2p,1d]             & (4s,1p)/[2s,1p]             & 14 &  6 &  23 &  276 \\ % (9+5)  (4+2) 20
PS-aDZ       & aug-cc-pVDZ     & R$_c$ + Z$^{20}$   & (5s,5p,2d)/[3s,3p,2d]             & (5s,2p)/[3s,2p]             & 19 &  6 &  40 &  820 \\ % (12+7) (4+2) 25
PS-aDZ+      & aug-cc-pVDZ     &   Z$^{50}$   & (6s,5p,2d)/[4s,3p,2d]             & (5s,2p)/[3s,2p]             & 20 & 10 &  41 &  861 \\ % (13+7) (8+2) 30
PS-TZ-c      & cc-pVTZ         & R$_c$ + R$_l$ + Z$^{20}$ & (5s,4p,2d)/[3s,3p,2d]             & (4s,2p)/[3s,2p]             & 17 &  4 &  40 &  820 \\ % (11+6) (3+1) 21
PS-TZ        & cc-pVTZ         & R$_c$ + Z$^{20}$   & (5s,4p,2d,1f)/[3s,3p,2d,1f]       & (4s,2p,1d)/[3s,2p,1d]       & 19 &  4 &  57 & 1653 \\ % (12+7) (3+1) 23
PS-aTZ-a     & aug-cc-pVTZ     & R$_c$ + R$_o$ + Z$^{20}$ & (6s,5p,3d,2f)/[1s,1p,1d,1f]       & (5s,3p,2d)/[1s,1p,1d]       & 26 & 19 &  34 &  595 \\ % (16+10) (12+7) 45 % 34=16+2*9  
PS-aTZ-b     & aug-cc-pVTZ     & R$_c$ + R$_o$ + Z$^{20}$ & (6s,5p,3d,2f)/[3s,2p,1d,1f]       & (5s,3p,2d)/[2s,1p,1d]       & 26 & 37 &  41 &  861 \\ % (16+10) (26+11) 63 & 41=21+2*10 
PS-aTZ-c     & aug-cc-pVTZ     & R$_c$ + R$_l$ + Z$^{20}$ & (6s,5p,3d)/[4s,4p,3d]             & (5s,3p)/[4s,3p]             & 22 &  4 &  57 & 1653 \\ % (14+8) (3+1) 26
PS-aTZ       & aug-cc-pVTZ     & R$_c$ + Z$^{20}$   & (6s,5p,3d,2f)/[4s,4p,3d,2f]       & (5s,3p,2d)/[4s,3p,2d]       & 26 &  4 &  91 & 4186 \\ % (16+10) (3+1) 30
PS-aTZ+      & aug-cc-pVTZ     &   Z$^{50}$   & (7s,6p,3d,2f)/[5s,4p,3d,2f]       & (6s,3p,2d)/[4s,3p,2d]       & 29 & 48 &  92 & 4278 \\ % (18+11) (34+14) 77
PS-aQZ+      & aug-cc-pVTZ     &   Z$^{70}$   & (8s,7p,4d,3f,2g)/[6s,5p,4d,3f,2g] & (7s,4p,3d,2f)/[5s,4p,3d,2f] & 40 & 10 & 172 &14878 \\ % (24+16) (8+2) 50 % (80+2*46)
PS-VTZ-h20   & VTZ $^b$        &   Z$^{0.1-15/5}$   & (9s,9p,2d,1f)/\{12\}              & (6s,5p,1d)/\{4\}            & 33 &724 &  20 &  210 \\ % (21+12) (624+100) 757
PS-VTZ-h30   & VTZ $^b$        &   Z$^{0.1-15/5}$   & (9s,9p,2d,1f)/\{18\}              & (6s,5p,1d)/\{6\}            & 33 &1086&  30 &  465 \\ % (21+12) (936+150) 1119
PS-VTZ+h20   & VTZ $^b$        &   Z$^{0.01-50/15}$   & (10s,10p,2d,1f)/\{12\}            & (10s,8p,1d)/\{4\}           & 42 &824 &  20 &  210 \\ % (23+19) (672+152) 866

\hline

DZ           & cc-pVDZ         &  Z$^{50}$    & (5s,4p,1d)/[3s,2p,1d]             & (4s,1p)/[2s,1p]             & 15 &  10 &  24 &  300 \\ % (10+5) (8+2) 25
aDZ          & aug-cc-pVDZ     &  Z$^{50}$    & (6s,5p,2d)/[4s,3p,2d]             & (5s,2p)/[3s,2p]             & 20 &  10 &  41 &  861 \\ % (13+7) (8+2) 30
%aDZ-h5      & aug-cc-pVDZ     &  Z$^{50}$    & (6s,5p,2d)/\{3\}                  & (5s,2p)/\{1\}               & 20 & 100 &   5  &  15 \\ % (13+7) (90+10)
%aDZ-h6      & aug-cc-pVDZ     &  Z$^{50}$    & (6s,5p,2d)/\{4\}                  & (5s,2p)/\{1\}               & 20 & 130 &   6 &   21 \\ % (13+7) (120+10)
aDZ-h8       & aug-cc-pVDZ     &  Z$^{50}$    & (6s,5p,2d)/\{4\}                  & (5s,2p)/\{2\}               & 20 & 140 &   8 &   36 \\ % (13+7) (120+20) 160
VTZ-h21      & VTZ $^b$        &  Z$^{0.01-50/15}$ + A$_{sto}$  & (11s,10p,2d,1f)/\{13\}            & (10s,8p,1d)/\{4\}           & 43 & 893 &  21 &  231 \\ % (24+19) (741+152) 936
ANO1-h21     & Roos aug-TZ-ANO &  Z$^{50}$    & (8s,7p,4d,3f)/\{13\}             & (6s,4p,3d)/\{4\}            & 35 &1025 &  21 &  231 \\ % (22+13) (897+128) 1060
ANO2-h21     & Roos aug-TZ-ANO &  Z$^{120}$    & (9s,7p,4d,3f)/\{13\}             & (6s,4p,3d)/\{4\}            & 36 &1038 &  21 &  231 \\ % (23+13) (910+128) 1074
ANO3-h21     & Roos aug-TZ-ANO &  Z$^{1000}$    & (11s,9p,4d,3f)/\{13\}            & (8s,4p,3d)/\{4\}            & 42 &1150 &  21 &  231 \\ % (27+15) (1014+136) 1192
ANO4-h21     & Roos aug-TZ-ANO &       & (14s,9p,4d,3f)/\{13\}            & (8s,4p,3d)/\{4\}            & 45 &1189 &  21 &  231 \\ % (30+15) (1053+136) 1234 
ANO5-h21     & Roos aug-TZ-ANO & Z$^{1000}$ + A$_{sto}$ & (12s,9p,4d,3f)/\{13\}            & (8s,4p,3d)/\{4\}            & 43 &1163 &  21 &  231 \\ % (28+15) (1027+136) 1206

\hline
\hline
% cut off types
\multicolumn{8}{p{15cm}}{$^a$  the symbols reported in the column correspond to the following filtering criteria of the orbitals in the source basis:
R$_c$: removed the core orbital;   % 0
R$_l$: removed the largest L orbitals for both the oxygen and the hydrogen; % 2
R$_o$: reduced number of contracted orbitals; % 11
Z$^{20}$: considered only the orbitals with exponent $\zeta<20$;   % 1
Z$^{50}$: $\zeta<50$; % 3
Z$^{70}$: $\zeta<70$; % 4
Z$^{0.1-15/5}$: $0.1<\zeta<15$ for the oxygen orbitals and $0.1<\zeta<5$ for the hydrogen orbitals; % 5
Z$^{0.01-50/15}$: $0.01<\zeta<50$ for the $s$ orbitals and $\zeta<15$ for the $p$ orbitals; % 6
Z$^{120}$: $\zeta<120$; % 7
Z$^{1000}$: $\zeta<1000$; % 8
A$_{sto}$: added one STO orbital $s$ in the oxygen atom and one in the hydrogen atom. } \\ % 9+10
\multicolumn{8}{l}{ $^b$ from http://people.sissa.it/$\sim$sorella/web/epaps\_material1.html} \\
\\
%caption
\multicolumn{8}{p{15cm}} { The atomic basis sets used in this work are here summarized. 
Square parenthesis are used for the contracted orbitals, and 
brace parenthesis are used to indicate the number of atomic hybrid orbitals.
Basis those name starts with ``PS-'' have been used with a pseudo potential for the 2 core electrons in the oxygen atom.
For the creation of our basis sets we took as source some standard basis sets, reported in the second column (in most of the cases Dunning's basis), to which we applied the filtering criteria reported in the third column (those meaning is listed in note $^a$ below the table).
The filtering is used because we do not need large atomic exponents since in you QMC calculations there is a Jastrow 1-body term, and in many cases also a pseudo potential.

The number of parameters for each basis are the sum of the atomic parameters $n_a$ and the matrix parameters $n_m$. 
% atomic par
The $n_a$ atomic parameters are the sum of the $n_z$ exponents and  the $n_c$ contraction coefficients.
Each (standard) contracted orbital $x$ contribute to $n_c$ with the number of contracted exponents for orbital $x$ minus one (the overall normalization is given by the matrix).
Each atomic hybrid orbitals $h$ contribute to $n_c$ with $n_c(h) = \sum_l (2l+1)*n_z^l(h) -1$, being  $n_z^l(h)$ the number of exponents with angular momentum $l$ in the atom of orbital $h$.
Moreover, both for hybrid and contracted orbitals, $n_a$  depends on the choice of using or not the same exponents and coefficients for the 2 hydrogens. 
The first case is assumed for the  calculated  value of $n_a$ reported in the 6th column.
% matrix par
The number $n_m$ of matrix parameters is $n_{orb}*(n_{orb}+1)/2$, being $n_{orb}$ the total number of orbitals in the basis (i.e., for a standard contracted orbital, each orbital of angular momentum $l$ is counted a number of times equal to its multiplicity $2l-1$). 
If a symmetric configuration of the water molecule is studied, several constraints can be imposed in order to decrease the number on matrix parameters to $\sim 1/3$.
}  \\

\end{tabular}
}
\end{table}

%\input{Table_JBSC}
% JBSC
% Jastrow Basis Set Convergence
\begin{table}[p!]
\caption{ Basis set convergence of the Jastrow term.}  
\label{table:JBSC}
\begin{tabular}{ c c | c c |  c c  }  

\hline
\hline
\multicolumn{2}{c}{Jatrow basis} & \multicolumn{2}{|c|}{\em Opt:noZ } & \multicolumn{2}{c}{\em Opt:all} \\ 
Oxygen & Hydrogen & Energy & Variance & Energy & Variance \\
\hline
\hline
  no$^a$     &   no$^a$  & -17.1749(6) &  2.05(2) & -17.2207(3) & 0.662(1)  \\
\hline
  (1s,1p)    &   (1s)    & -17.2157(2) & 1.107(3) &	 -17.2388(1) & 0.483(1) \\
  (1s,1p)    &   (1s,1p) & -17.2202(2) & 1.095(2) &	 -17.2401(1) & 0.468(1) \\
  (1s,1p,1d) &   (1s,1p) & -17.2262(2) & 1.063(3) &	 -17.2451(1) & 0.419(1) \\
  (2s,1p)    &   (1s,1p) & -17.2260(2) & 0.878(1) &	 -17.2473(1) & 0.321(1) \\ 
  (2s,1p)    &   (2s,1p) & -17.2274(2) &	0.789(2) &	 -17.2481(1) & 0.305(1) \\ 
  (2s,2p)    &   (2s,1p) & -17.2394(1) & 0.430(1) &	 -17.2505(1) & 0.272(3) \\
  (2s,2p,1d) &   (2s,1p) & -17.2413(1) & 0.412(1) &	 -17.2513(1) & 0.255(2) \\
  (3s,2p,1d) &   (2s,1p) & -17.2482(1) & 0.283(2) &	 -17.2524(1) & 0.244(1) \\
  (3s,2p,1d) &   (3s,1p) & -17.2483(1) & 0.275(1) &	 -17.2529(1) & 0.245(1) \\
  (4s,4p,1d) &   (4s,1p) & -17.2495(1) & 0.262(1) &	 -17.2527(1) & 0.2372(7)\\
\hline
(3s,2p,1d)/[1s,1p,1d] & (2s,1p)/[1s,1p] 
                         & -17.2351(2) & 0.849(2) & -17.2491(1) & 0.296(1)\\
\hline
(3s,2p,1d)/\{3\} & (2s,1p)/\{1\} 
                         & -17.2254(1) & 0.463(8) & -17.2441(1) & 0.332(1) \\
%\hline
(6s,5p,2d)/\{2\}&(5s,2p)/\{1\}& -17.2427(1) & 0.344(1)  & -17.2480(1) & 0.294(1) \\ % VMC8 VMCallz8
(6s,5p,2d)/\{4\}&(5s,2p)/\{1\}& -17.2441(1) & 0.381(2)  & -17.2482(1) & 0.291(1) \\ % VMC8 VMCallz8
(6s,5p,2d)/\{4\}&(5s,2p)/\{2\}& -17.2468(1) & 0.333(1)  & -17.2509(1) & 0.269(1) \\ % VMC8 VMCallz8
\hline
\hline
\multicolumn{6}{l}{$^a$ includes in the Jastrow term only the homogeneous e-n and e-e interactions.} \\
\\
% caption
\multicolumn{6}{ p{15cm} }{
VMC energy and variance, in a.u., for a JAGP wave function of the water molecule, with ECP pseudo-potential used to describe the two core electrons of the oxygen.
The basis set for the AGP part is the PS-DZ, see \ref{table:AGPbasis}. 
The basis set for the inhomogeneous terms in the Jastrow is indicated in the first two columns of the table.
Among the considered cases, the first is without these terms, then  several basis are constituted by uncontracted orbitals, one with the standard contracted orbitals and the last four with contracted atomic hybrid orbitals.
For each Jastrow basis set, two different cases have been considered: 
in the two rightmost columns -{\em Opt:all}-
optimization of all the parameters of the wave function; 
and in the two central columns -{\em Opt:noZ}- optimization of all the parameters except  the exponents of the AGP basis, that are kept fixed to the original values of the Dunning cc-pVDZ basis.
}
\end{tabular}
\end{table}

%\input{Table_JBSC_prop}
% JBSC_prop 
% Jastrow Basis Set Convergence
\begin{table}[p!]
\caption{ Jastrow basis and VMC Energy, Variance, Dipole and Quadrupole.}  
\label{table:JBSC_prop}

%{\tiny
%{\scriptsize
{\footnotesize
%{\small

\begin{tabular}{ c c c c  c  c c c }  

\hline
\hline
\multicolumn{2}{c}{Jastrow basis} & Energy & Variance & Dipole & \multicolumn{3}{c}{Quadrupole} \\
Oxygen & Hydrogen & [H] & [H$^2$] & [D] & xx [D\AA] & yy [D\AA] & zz [D\AA] \\
\hline
\hline
\\
\multicolumn{8}{c}{\em Opt:noZ}\\
\\
  no$^a$     & no$^a$ &  -17.1749(6)  &  2.05(2)  &  2.1165(9)  &  2.300(1)  &  -0.079(1)  &  -2.2211(10)  \\
\\
    (1s,1p)  &  (1s)  &  -17.2157(2)  &  1.107(3)  &  1.9929(9)  &  2.238(1)  &  -0.089(1)  &  -2.1485(10)  \\
    (1s,1p)  &  (1s,1p)  &  -17.2202(2)  &  1.095(3)  &  1.9919(9)  &  2.307(1)  &  -0.079(1)  &  -2.2280(10)  \\
    (1s,1p,1d)  &  (1s,1p)  &  -17.2262(2)  &  1.064(3)  &  1.8558(9)  &  2.312(1)  &  -0.117(1)  &  -2.1951(10)  \\
    (2s,1p)  &  (1s,1p)  &  -17.2137(2)  &  0.848(1)  &  2.0232(9)  &  2.234(1)  &  -0.071(1)  &  -2.1636(10)  \\
    (2s,1p)  &  (2s,1p)  &  -17.2147(2)  &  0.845(1)  &  2.0122(9)  &  2.268(1)  &  -0.067(1)  &  -2.2016(10)  \\
    (2s,2p)  &  (2s,1p)  &  -17.2394(1)  &  0.430(2)  &  1.9487(9)  &  2.572(1)  &  -0.147(1)  &  -2.425(1)  \\
    (2s,2p,1d)  &  (2s,1p)  &  -17.2413(1)  &  0.412(1)  &  1.9782(9)  &  2.561(1)  &  -0.132(1)  &  -2.429(1)  \\
    (3s,2p,1d)  &  (2s,1p)  &  -17.2482(1)  &  0.283(2)  &  1.8880(9)  &  2.576(1)  &  -0.144(1)  &  -2.432(1)  \\
    (3s,2p,1d)  &  (3s,1p)  &  -17.2483(1)  &  0.275(1)  &  1.8757(10)  &  2.568(1)  &  -0.159(1)  &  -2.409(1)  \\
    (4s,4p,1d)  &  (4s,1p)  &  -17.2495(1)  &  0.262(1)  &  1.8876(9)  &  2.556(1)  &  -0.157(1)  &  -2.399(1)  \\
    \\
    (3s,2p,1d)/[1s,1p,1d]  &  (2s,1p)/[1s,1p]  &  -17.2351(2)  &  0.849(2)  &  1.8882(10)  &  2.526(1)  &  -0.108(1)  &  -2.418(1)  \\
    \\
    (3s,2p,1d)/\{3\}  &  (2s,1p)/\{1\}  &  -17.2254(1)  &  0.463(8)  &  2.0280(9)  &  2.296(1)  &  -0.111(1)  &  -2.1847(10)  \\
    (6s,5p,2d)/\{2\}  &  (5s,2p)/\{1\}  &  -17.2427(1)  &  0.344(1)  &  1.8783(10)  &  2.542(1)  &  -0.156(1)  &  -2.386(1)  \\
    (6s,5p,2d)/\{3\}  &  (5s,2p)/\{1\}  &  -17.2437(1)  &  0.328(1)  &  1.8876(10)  &  2.526(1)  &  -0.142(1)  &  -2.384(1)  \\
    (6s,5p,2d)/\{4\}  &  (5s,2p)/\{1\}  &  -17.2441(1)  &  0.381(2)  &  1.8664(10)  &  2.528(1)  &  -0.154(1)  &  -2.374(1)  \\
    (6s,5p,2d)/\{4\}  &  (5s,2p)/\{2\}  &  -17.2468(1)  &  0.3327(8)  &  1.8721(10)  &  2.540(1)  &  -0.137(1)  &  -2.403(1)  \\
\\
\hline
\\
\multicolumn{8}{c}{\em Opt:all} \\
\\
  no$^a$     & no$^a$ &  -17.2207(3)  &  0.662(1)  &  1.8168(10)  &  2.465(1)  &  -0.121(1)  &  -2.344(1)  \\
\\
    (1s,1p)  &  (1s)  &  -17.2388(1)  &  0.484(2)  &  1.9673(10)  &  2.617(1)  &  -0.144(1)  &  -2.472(1)  \\
    (1s,1p)  &  (1s,1p)  &  -17.2401(2)  &  0.469(1)  &  1.9528(10)  &  2.565(1)  &  -0.138(1)  &  -2.427(1)  \\
    (1s,1p,1d)  &  (1s,1p)  &  -17.2451(1)  &  0.4187(10)  &  1.8871(9)  &  2.492(1)  &  -0.208(1)  &  -2.285(1)  \\
    (2s,1p)  &  (1s,1p)  &  -17.2473(1)  &  0.321(1)  &  1.8762(10)  &  2.502(1)  &  -0.156(1)  &  -2.346(1)  \\
    (2s,1p)  &  (2s,1p)  &  -17.2481(1)  &  0.3046(9)  &  1.8752(10)  &  2.544(1)  &  -0.142(1)  &  -2.403(1)  \\
    (2s,2p)  &  (2s,1p)  &  -17.2505(1)  &  0.272(3)  &  1.9366(10)  &  2.593(1)  &  -0.122(1)  &  -2.471(1)  \\
    (2s,2p,1d)  &  (2s,1p)  &  -17.2513(1)  &  0.255(2)  &  1.8608(10)  &  2.576(1)  &  -0.150(1)  &  -2.427(1)  \\
    (3s,2p,1d)  &  (2s,1p)  &  -17.2524(1)  &  0.244(1)  &  1.8678(10)  &  2.592(1)  &  -0.133(1)  &  -2.459(1)  \\
    (3s,2p,1d)  &  (3s,1p)  &  -17.2529(1)  &  0.245(1)  &  1.8811(10)  &  2.566(1)  &  -0.139(1)  &  -2.428(1)  \\
    (4s,4p,1d)  &  (4s,1p)  &  -17.25271(10)  &  0.2372(7)  &  1.8629(10)  &  2.561(1)  &  -0.160(1)  &  -2.402(1)  \\
    \\
    (3s,2p,1d)/[1s,1p,1d]  &  (2s,1p)/[1s,1p]  &  -17.2491(1)  &  0.296(1)  &  1.8747(10)  &  2.554(1)  &  -0.110(1)  &  -2.444(1)  \\
    \\
    (3s,2p,1d)/\{3\}  &  (2s,1p)/\{1\}  &  -17.2441(1)  &  0.3320(8)  &  1.9206(10)  &  2.544(1)  &  -0.117(1)  &  -2.426(1)  \\
    (6s,5p,2d)/\{2\}  &  (5s,2p)/\{1\}  &  -17.2480(1)  &  0.294(1)  &  1.8832(10)  &  2.570(1)  &  -0.140(1)  &  -2.430(1)  \\
    (6s,5p,2d)/\{4\}  &  (5s,2p)/\{1\}  &  -17.2482(1)  &  0.2912(8)  &  1.8725(10)  &  2.566(1)  &  -0.154(1)  &  -2.413(1)  \\
    (6s,5p,2d)/\{4\}  &  (5s,2p)/\{2\}  &  -17.2509(1)  &  0.2691(9)  &  1.8728(10)  &  2.587(1)  &  -0.135(1)  &  -2.453(1)  \\
\hline
\hline
\multicolumn{8}{l}{ $^a$ includes in the Jastrow term only the homogeneous e-n and e-e interactions.}\\
\\
\multicolumn{8}{p{15cm}}{
Jastrow basis set convergence on the VMC energy, variance, dipole, quadrupole.
The wave functions are the same considered in \ref{table:JBSC}.
}

\end{tabular}

}

\end{table}

%\input{Table_BSC}
% AGPBSC
% AGP Basis Set Convergence
\begin{table}[p!]
\caption{Convergence of VMC energy and variance.} 
\label{table:BSC}

%{\tiny
{\scriptsize
%{\footnotesize
%{\small
\begin{tabular}{ c c | c c | c c }  

\hline
\hline

\multirow{2}{*}{\em function} & \multirow{2}{*}{\em basis set } & \multicolumn{2}{|c|}{\em Opt:noZ } & \multicolumn{2}{c}{\em Opt:all} \\
&& Energy & Variance & Energy & Variance\\
%\cline{2-3} \cline{4-5}

\hline
\hline
\multicolumn{6}{l}{\em core: ECP} \\ 
\hline
JDFT  & PS-VTZ\{20\}& -17.24548(8) & 0.3606(3) &-&- \\

JSD   & PS-aDZ+     &-&-&         -17.2478(1) & 0.280(1) \\
JSD   & PS-aTZ+     &-&-&         -17.2478(1) & 0.270(2) \\
JSD   & PS-aQZ+     &-&-&         -17.2481(1) & 0.269(1) \\

JSD   &  PS-VTZ-h20 &-&-&         -17.2471(1)  &  0.3049(6)  \\
JSD   & PS-VTZ-h30  &-&-&         -17.24824(7)& 0.2692(1)\\ 

JSD   &  PS-UNC  	&-&-&  	-17.24820(5)  &  0.2734(2) \\

JAGPn* & PS-aDZ+     &-&-&         -17.2512(1) & 0.260(2) \\ % JAGPn*(n=6)
JAGPn* & PS-aTZ+     &-&-&         -17.2513(1) & 0.252(1) \\
JAGPn* & PS-aQZ+     &-&-&         -17.2513(1) & 0.249(1) \\

JAGP  & PS-DZ       & -17.2482(1) & 0.283(2) & -17.2524(1) & 0.244(1) \\
JAGP  & PS-aDZ      & -17.2520(1) & 0.263(1) & -17.2538(1) & 0.231(1) \\
JAGP  & PS-TZ-c     & -17.2502(1) & 0.313(1) & -17.2532(1) & 0.242(1) \\
JAGP  & PS-aTZ-c    & -17.2527(1) & 0.256(1) & -17.2536(1) & 0.239(1) \\
JAGP  & PS-TZ       & -17.2503(1) & 0.308(1) & -17.2533(1) & 0.239(1) \\
JAGP  & PS-aTZ      & -17.2529(1) & 0.266(1) & -17.2540(1) & 0.233(1) \\ 

JAGP & PS-VTZ-h20            &-&-& -17.25383(4) & 0.2308(1) \\

JAGP & PS-UNC  				&-&-& -17.2536(2)  &  0.239(1) \\

\hline
\hline
\multicolumn{6}{l}{\em core: NCP} \\
\hline
JSD  & PS-VTZ-h30        &-&-& -17.20239(5) & 0.3303(2) \\ % VMC_2_0
JAGP & PS-DZ       & -17.2007(2) & 0.584(3) & -17.2052(1) & 0.489(4) \\
JAGP & PS-aDZ   & -17.2050(1) & 0.587(1) & -17.2068(1) & 0.325(1) \\
JAGP & PS-VTZ-h20        &-&-& -17.20803(6) & 0.2786(1) \\

\hline
\hline
\multicolumn{6}{l}{\em all electrons} \\ 
\hline
JDFT & VTZ-h21  & -76.3444(6)  & 5.98(2)   &-&-\\ % AE/JDFT/pip0.d
JDFT & ANO1-h21 & -76.3970(1)  & 3.317(1)  &-&-\\ % AE/AO/WRONG/JDFT
JDFT & ANO2-h21 & -76.39806(7) & 1.6171(5) &-&-\\ % AE/AO/JDFT
JDFT&ANO2-h21$^b$&-76.39881(6) & 1.5746(8) &-&-\\ % AE/AO/JDFT_j4s2p1d
JDFT&ANO3-h21$^b$&-76.39902(5) & 1.1774(2) &-&-\\ % AE/AO/JDFT_zless1000
JDFT&ANO4-h21    &-76.39841(7) & 1.5013(4) &-&-\\ % AE/AO/JDFT_fullbasis/j3s2p1d 
JDFT&ANO4-h21$^b$&-76.39914(6) & 1.4482(3) &-&-\\ % AE/AO/JDFT_fullbasis 
JDFT&ANO4-h21$^{b,e}$&-76.39914(6) & 1.1881(3) &-&-\\ % AE/AO/JDFT_fullbasis/moreconvergedDFT

JSD & ANO2-h21$^b$ & - & - &  -76.40052(5)  &  1.1579(7) \\

%JAGP & DZ           & -76.3909(4) & 3.02(1)  & -76.4038(3) & 1.56(1) \\
JAGP & DZ           & -76.3909(4) & 3.02(1)  & -76.4041(3) & 1.534(1) \\
%JAGP & aDZ $^a$     & -76.4026(3) & 2.59(2)  & -76.4055(3) & 1.74(1) \\ % opt_allz8; further opt is possible
JAGP & aDZ $^a$     & -76.4026(3) & 2.59(2)  & -76.4066(3) & 1.470(4) \\ 
JAGP & VTZ-h21  & -76.4011(4) & 2.578(3) & -76.40668(9) & 1.1137(8) \\ % noz/pip0.d - optz/pip0.d
JAGP&VTZ-h21$^c$ &-&-& -76.40724(3) & 1.0640(2) \\ % AE/JAGP_j4s2p1d/vmc
JAGP & ANO2-h21  &-&-& -76.40663(5) & 1.1944(5) \\ % AE/AO/JAGP
JAGP&ANO2-h21$^b$&-&-& -76.40720(5) & 1.1184(4) \\ % AE/AO/JAGP_j4s2p1d
JAGP&ANO5-h21$^c$&-&-& -76.40741(2) &1.01245(9) \\ % AE/AO/JAGP_zless1000_j4s2p1d/vmc

\hline
\hline
\multicolumn{6}{p{12cm}}{ 
Basis set for the inhomogeneous Jastrow terms: 
$^a$~O(4s,3p,1d) H(2s,1p); 
$^b$~O(4s,2p,1d) H(2s,1p); 
$^c$~O(4s,2p,1d) H(3s,1p); 
$^d$~none.} \\
\multicolumn{6}{l}{ $^e$ more converged DFT calculation around the nuclei. } \\
\\
%caption
\multicolumn{6}{p{12cm}} {
Convergence of energy and variance for VMC calculations, studied for several different wave functions and for different description of the core electrons of the oxygen atom (namely ECP and NCP pseudo potentials or all electrons calculations).
The functional form of wave function, namely JDFT, JSD, JAGPn* or JAGP, are reported in the first column. 
In the second column we report the basis set, see \ref{table:AGPbasis}, used for the determinantal part of the wave function.
The basis set for the Jastrow term is O(3s,2p,1d) H(2s,1p), unless explicitly indicated.
%{\em Opt:noZ} and {\em Opt:all} have the same meaning explained in \ref{table:JBSC}.
} 
\end{tabular}
}
\end{table}

%\input{Table_BSC_dip_quad}
% AGPBSC_prop 
% AGP Basis Set Convergence
\begin{table}[p!]
\caption{ VMC Energy, Dipole and Quadrupole of water molecule.}
\label{table:BSC_propSI}
%{\tiny
{\scriptsize
%{\footnotesize
%{\small
\begin{tabular}{ c c c c  c c  c  c c c }  

\hline
\hline
 function & core & basis set & Opt & Energy[H] & Variance[H$^2$] & dipole[Deb] & \multicolumn{3}{c}{quadrupole[ Deb \AA ]} \\
\hline
\hline

  JDFT  	&  ECP  &  PS-aDZ  	&  noZ  &  -17.2296(2)  &  1.351(4)  &  1.8935(10)  &  2.595(1)  &  -0.147(1)  &  -2.448(1)  \\
  JDFT  	&  ECP  &  PS-aTZ  	&  noZ  &  -17.2431(2)  &  0.529(5)  &  1.8742(10)  &  2.633(1)  &  -0.156(1)  &  -2.477(1)  \\
  JDFT  	&  ECP  &  PS-aQZ  	&  noZ  &  -17.2452(1)  &  0.444(2)  &  1.8767(10)  &  2.635(1)  &  -0.165(1)  &  -2.469(1)  \\
  JDFT  	&  ECP  &  PS-VTZ-h20  	&  noZ  &  -17.24548(8)  &  0.3606(3)  &  1.9059(8)  &  2.5796(9)  &  -0.1551(9)  &  -2.4245(9)  \\

  JSD 		&  ECP  &  PS-aDZ  	&  all  &  -17.2478(1)  &  0.2796(8)  &  1.8813(10)  &  2.599(1)  &  -0.155(1)  &  -2.444(1)  \\
  JSD  		&  ECP  &  PS-aTZ  	&  all  &  -17.2478(1)  &  0.270(2)  &  1.8729(10)  &  2.597(1)  &  -0.155(1)  &  -2.442(1)  \\
  JSD  		&  ECP  &  PS-aQZ  	&  all  &  -17.2482(1)  &  0.2668(10)  &  1.8755(10)  &  2.596(1)  &  -0.155(1)  &  -2.441(1)  \\
  JSD  		&  ECP  &  PS-VTZ-h20  	&  all  &  -17.2471(1)  &  0.3049(6)  &  1.8907(7)  &  2.5676(8)  &  -0.1419(8)  &  -2.4256(8)  \\

  JSD  		&  ECP  &  PS-VTZ-h30  	&  all  &  -17.24824(7)  &  0.2692(1)  &  1.8877(6)  &  2.5819(7)  &  -0.1445(7)  &  -2.4374(6)  \\
  JSD  	        &  ECP  &  PS-UNC  	&  all  &  -17.24820(5)  &  0.2734(2)  &  1.8881(4)  &  2.5842(5)  &  -0.1711(5)  &  -2.4131(5)  \\

  JAGPn*  	&  ECP  &  PS-aDZ  	&  all  &  -17.2512(1)  &  0.261(2)  &  1.8606(10)  &  2.584(1)  &  -0.148(1)  &  -2.435(1)  \\
  JAGPn*  	&  ECP  &  PS-aTZ  	&  all  &  -17.2513(1)  &  0.2517(9)  &  1.8499(10)  &  2.573(1)  &  -0.150(1)  &  -2.423(1)  \\
  JAGPn*  	&  ECP  &  PS-aQZ  	&  all  &  -17.2513(1)  &  0.2489(7)  &  1.8629(10)  &  2.570(1)  &  -0.149(1)  &  -2.421(1)  \\

  JAGP  	&  ECP  &  PS-DZ  	&  noZ  &  -17.2482(1)  &  0.283(2)  &  1.8880(9)  &  2.576(1)  &  -0.144(1)  &  -2.432(1)  \\
  JAGP  	&  ECP  &  PS-TZ-C  	&  noZ  &  -17.2502(1)  &  0.3130(9)  &  1.8631(10)  &  2.541(1)  &  -0.142(1)  &  -2.399(1)  \\
  JAGP  	&  ECP  &  PS-TZ  	&  noZ  &  -17.2504(1)  &  0.3082(8)  &  1.8635(10)  &  2.537(1)  &  -0.144(1)  &  -2.393(1)  \\

  JAGP  	&  ECP  &  aPS-DZ  	&  noZ  &  -17.2520(1)  &  0.2631(10)  &  1.8702(10)  &  2.568(1)  &  -0.156(1)  &  -2.412(1)  \\
  JAGP  	&  ECP  &  aPS-TZ-C  	&  noZ  &  -17.2527(1)  &  0.2559(7)  &  1.8656(10)  &  2.580(1)  &  -0.153(1)  &  -2.427(1)  \\
  JAGP  	&  ECP  &  aPS-TZ  	&  noZ  &  -17.2529(1)  &  0.2665(7)  &  1.8609(10)  &  2.580(1)  &  -0.147(1)  &  -2.433(1)  \\

  JAGP  	&  ECP  &  PS-DZ  	&  all  &  -17.2524(1)  &  0.244(1)  &  1.8678(10)  &  2.592(1)  &  -0.133(1)  &  -2.459(1)  \\
  JAGP  	&  ECP  &  PS-TZ-C  	&  all  &  -17.2532(1)  &  0.242(1)  &  1.8718(10)  &  2.557(1)  &  -0.144(1)  &  -2.412(1)  \\
  JAGP  	&  ECP  &  PS-TZ  	&  all  &  -17.2533(1)  &  0.239(2)  &  1.8680(10)  &  2.556(1)  &  -0.147(1)  &  -2.409(1)  \\

  JAGP  	&  ECP  &  aPS-DZ  	&  all  &  -17.25378(9)  &  0.2314(8)  &  1.8729(10)  &  2.574(1)  &  -0.154(1)  &  -2.420(1)  \\
  JAGP  	&  ECP  &  aPS-TZ-C  	&  all  &  -17.2536(1)  &  0.239(1)  &  1.8597(10)  &  2.580(1)  &  -0.152(1)  &  -2.428(1)  \\
  JAGP  	&  ECP  &  aPS-TZ  	&  all  &  -17.25397(10)  &  0.2330(10)  &  1.8710(10)  &  2.583(1)  &  -0.145(1)  &  -2.438(1)  \\
  JAGP  	&  ECP  &  PS-VTZ-h20  	&  all  &  -17.25383(4)  &  0.2308(1)  &  1.8648(6)  &  2.5740(7)  &  -0.1500(7)  &  -2.4240(7)  \\

  JAGP  	&  ECP  &  PS-UNC  	&  all  &  -17.2536(2)  &  0.239(1)  &  1.8881(10)  &  2.579(1)  &  -0.174(1)  &  -2.406(1)  \\

\hline

  JSD  		&  NCP  &  PS-VTZ-h30  	&  all  &  -17.20239(5)  &  0.3303(2)  &  1.8949(4)  &  2.5808(5)  &  -0.1498(5)  &  -2.4310(5)  \\

  JAGP  	&  NCP  &  PS-DZ  	&  noZ  &  -17.2007(2)  &  0.585(3)  &  1.8944(10)  &  2.567(1)  &  -0.151(1)  &  -2.416(1)  \\
  JAGP  	&  NCP  &  PS-aDZ  	&  noZ  &  -17.2050(2)  &  0.587(10)  &  1.8909(10)  &  2.548(1)  &  -0.164(1)  &  -2.384(1)  \\
  JAGP  	&  NCP  &  PS-DZ  	&  all  &  -17.2052(1)  &  0.489(4)  &  1.8847(10)  &  2.588(1)  &  -0.141(1)  &  -2.447(1)  \\
  JAGP  	&  NCP  &  PS-aDZ  	&  all  &  -17.2068(1)  &  0.325(2)  &  1.8826(10)  &  2.563(1)  &  -0.158(1)  &  -2.405(1)  \\
  JAGP  	&  NCP  &  VTZ-h20  	&  all  &  -17.20803(6)  &  0.2786(1)  &  1.8704(7)  &  2.5765(8)  &  -0.1559(8)  &  -2.4206(8)  \\

\hline
\hline
% other methods and experimental results

\multicolumn{4}{l}{ Ref.\cite{Sterpone:2008p12640}: VMC   JAGP {\em core:}ECP  }   & -17.2435(1) && 1.890(8) \\
\multicolumn{4}{l}{ Ref.\cite{Sterpone:2008p12640}: LRDMC JAGP {\em core:}ECP  }   & -17.2620(1) && 1.87(1)  \\

%\multicolumn{4}{l}{ MRSD-CI/120CGTO (from \citet{Feller:1987dm}, Tab.~IV ) } & -76.3861 && 1.860 & 2.5535 \\
%\multicolumn{4}{l}{ MRSD-CI/140CGTO (from \citet{Feller:1987dm}, Tab.~IV ) } & -76.3963 && 1.870 & 2.5556 \\
%\multicolumn{4}{l}{ Exact (from \citet{Feller:1987dm}, Tab.~IV ) }      & -76.4396 \\

% XXX ADD from Niu, Shuqiang, Ming-Liang Tan, and Toshiko Ichiye. 2011. ?The Large Quadrupole of Water Molecules.? The Journal of Chemical Physics 134 (13): ?. doi:10.1063/1.3569563. XXX

\hline
%\multicolumn{4}{l}{ Experiment (from NIST database $^e$) }                             &&& 1.850  & 2.63(2) & -0.13(3) & -2.50(2) \\
\multicolumn{4}{l}{ Experiment\cite{Franks:1972wj}   }                          &&& 1.855 & \\
\multicolumn{4}{l}{ Experiment\cite{Clough:1973bh}   }                          &&& 1.8546(6) \\
\multicolumn{4}{l}{ Experiment\cite{Kukolich:1969es} }                          &&&  & 1.6(6) & 0.4(4) & -2.0(6) \\ % in a.u. 1.2(4) & 0.3(3) & -1.5(4) \\
\multicolumn{4}{l}{ Experiment\cite{Verhoeven:1970jd}}                          &&&  & 2.63(2) & -0.13(3) & -2.50(2) \\  % in a.u. 1.955(14)& -0.097(22)& -1.859(14)\\
\hline
\hline
\\
\\ %caption
\multicolumn{10}{ p{15cm} }{
Similarly to \ref{table:JBSC_prop}, several properties of the water molecule obtained from VMC calculations are reported here: the energy (in Hartree), the variance (in Hartree$^2$), the dipole (in Debye), the traceless quadrupole components ( in Debye~\AA).
% see  \ref{table:AGPBSC}  \ref{table:JBSC_prop}
The wave functions type is reported in the first column.
%, and the description of the core electrons of the oxygen atom in the third.  
The second column  indicates in which way the core electrons of the oxygen atom have been described, {\em i.e.} by using ECP or NCP pseudo potentials (results for AE calculations are reported in \ref{table:BSC_AE_prop}).
The basis sets for the determinantal part of the wave function are reported in the third column (description in \ref{table:AGPbasis}), while
the basis set of the inhomogeneous  Jastrow terms is O(3s,2p,1d) H(2s,1p), unless explicitly indicated.
In the fourth column we indicate the optimization of the wave function: all or noZ, respectively to indicate that all the parameters or all except the exponents of the determinantal part have been optimized. 
At the end of the table we have reported several results obtained by other computational methods and experimental results.
} 
\end{tabular}
}
\end{table}

%\input{Table_BSC_AE_dip_quad}
% AGPBSC_prop 
% AGP Basis Set Convergence
\begin{table}[p!]
\caption{ VMC Energy, Dipole and Quadrupole of water molecule.}
\label{table:BSC_AE_prop}
%{\tiny
{\scriptsize
%{\footnotesize
%{\small
\begin{tabular}{ c c c c  c c  c  c c c }  

\hline
\hline
 function & core & basis set & Opt & Energy[H] & Variance[H$^2$] & dipole[Deb] & \multicolumn{3}{c}{quadrupole[ Deb \AA ]} \\
\hline
\hline

  DFT  	&  AE  &  ANO2-h21$^d$  	&  noZ  &  -76.0496(5)  &  19.27(9)  &  1.8446(6)  &  2.5738(7)  &  -0.1744(7)  &  -2.3994(6)  \\
  DFT  	&  AE  &  ANO4-h21$^d$  	&  noZ  &  -76.0507(4)  &  17.7(1)  &  1.8645(4)  &  2.6042(5)  &  -0.1481(5)  &  -2.4561(5)  \\

  JDFT  	&  AE  &  VTZ-h21  	&  noZ  &  -76.3445(1)  &  5.986(3)  &  1.9235(2)  &  2.5757(3)  &  -0.1895(3)  &  -2.3862(3)  \\
  JDFT  	&  AE  &  ANO1-h21  	&  noZ  &  -76.3970(1)  &  3.317(1)  &  1.9083(3)  &  2.5874(3)  &  -0.1720(3)  &  -2.4154(3)  \\
  JDFT  	&  AE  &  ANO2-h21  	&  noZ  &  -76.39806(7)  &  1.6171(5)  &  1.9070(3)  &  2.5788(3)  &  -0.1757(3)  &  -2.4031(3)  \\
  JDFT  	&  AE  &  ANO2-h21$^b$  	&  noZ  &  -76.39881(6)  &  1.5746(8)  &  1.9029(2)  &  2.5839(3)  &  -0.1701(3)  &  -2.4138(3)  \\
  JDFT  	&  AE  &  ANO3-h21$^b$  	&  noZ  &  -76.39902(5)  &  1.1774(2)  &  1.9164(3)  &  2.6124(3)  &  -0.1441(3)  &  -2.4682(3)  \\
  JDFT  	&  AE  &  ANO4-h21$^b$  		&  noZ  &  -76.39914(6)  &  1.4482(4)  &  1.9155(3)  &  2.6127(3)  &  -0.1459(3)  &  -2.4669(3)  \\
  JDFT  	&  AE  &  ANO4-h21$^{b,e}$  	&  noZ  &  -76.39914(6)  &  1.1881(3)  &  1.9152(3)  &  2.6122(3)  &  -0.1460(3)  &  -2.4663(3)  \\

  JSD  		&  AE  &  ANO2-h21$^b$  	&  all  &  -76.40052(5)  &  1.1579(7)  &  1.8973(2)  &  2.5740(3)  &  -0.1362(3)  &  -2.4377(3)  \\

  JAGP  	&  AE  &  DZ  	&  noZ  &  -76.3909(4)  &  3.03(1)  &  1.8176(9)  &  2.361(1)  &  -0.229(1)  &  -2.132(1)  \\
  JAGP  	&  AE  &  aDZ$^a$  	&  noZ  &  -76.4026(3)  &  2.59(2)  &  1.9027(10)  &  2.547(1)  &  -0.158(1)  &  -2.389(1)  \\
  JAGP  	&  AE  &  VTZ-h21  	&  noZ  &  -76.4011(1)  &  2.578(3)  &  1.9193(4)  &  2.5769(5)  &  -0.1561(5)  &  -2.4208(5)  \\
  JAGP  	&  AE  &  DZ  	&  all  &  -76.4041(3)  &  1.534(3)  &  1.904(1)  &  2.517(1)  &  -0.149(1)  &  -2.368(1)  \\
  JAGP  	&  AE  &  aDZ$^a$  	&  all  &  -76.4066(2)  &  1.470(4)  &  1.8855(10)  &  2.593(1)  &  -0.160(1)  &  -2.433(1)  \\
  JAGP  	&  AE  &  VTZ-h21  	&  all  &  -76.40668(9)  &  1.1137(8)  &  1.8752(4)  &  2.5676(6)  &  -0.1287(5)  &  -2.4388(5)  \\
  JAGP  	&  AE  &  VTZ-h21$^c$  	&  all  &  -76.40724(3)  &  1.0640(2)  &  1.8778(1)  &  2.5758(2)  &  -0.1357(2)  &  -2.4401(2)  \\
  JAGP  	&  AE  &  ANO2-h21  	&  all  &  -76.40663(5)  &  1.1944(4)  &  1.8724(2)  &  2.5749(3)  &  -0.1481(3)  &  -2.4268(3)  \\
  JAGP  	&  AE  &  ANO2-h21$^b$  	&  all  &  -76.40720(5)  &  1.1184(4)  &  1.8752(2)  &  2.5778(3)  &  -0.1503(3)  &  -2.4275(3)  \\
  JAGP  	&  AE  &  ANO5-h21$^b$  	&  all  &  -76.40741(2)  &  1.01531(9)  &  1.8894(1)  &  2.5875(1)  &  -0.1466(1)  &  -2.4409(1)  \\

\hline
\hline
% other methods and experimental results

%\multicolumn{4}{l}{ VMC, core ECP (from \citet{Sterpone:2008p12640})}   & -17.2435(1) && 1.890(8) \\
%\multicolumn{4}{l}{ LRDMC, core ECP (from \citet{Sterpone:2008p12640})} & -17.2620(1) && 1.87(1)  \\

\multicolumn{4}{l}{ Ref.\cite{Feller:1987dm}: MRSD-CI/120CGTO } & -76.3861 && 1.860 & 2.5535 \\
\multicolumn{4}{l}{ Ref.\cite{Feller:1987dm}: MRSD-CI/140CGTO } & -76.3963 && 1.870 & 2.5556 \\
\multicolumn{4}{l}{ Ref.\cite{Feller:1987dm}: Exact }      & -76.438 \\

\hline
\multicolumn{4}{l}{ Experiment\cite{Franks:1972wj}    }                      &&& 1.855 & \\
\multicolumn{4}{l}{ Experiment\cite{Clough:1973bh}    }                      &&& 1.8546(6) \\
\multicolumn{4}{l}{ Experiment\cite{Kukolich:1969es}  }                      &&&  & 1.6(6) & 0.4(4) & -2.0(6) \\ % in a.u. 1.2(4) & 0.3(3) & -1.5(4) \\
\multicolumn{4}{l}{ Experiment\cite{Verhoeven:1970jd} }                      &&&  & 2.63(2) & -0.13(3) & -2.50(2) \\  % in a.u. 1.955(14)& -0.097(22)& -1.859(14)\\
\hline
\hline

\multicolumn{10}{l}{ Basis set for the inhomogeneous Jastrow terms: 
$^a$ O(4s,3p,1d) H(2s,1p); 
$^b$ O(4s,2p,1d) H(2s,1p); 
$^c$ O(4s,2p,1d) H(3s,1p); 
$^d$ none. } \\
\multicolumn{10}{l}{ $^e$ more converged DFT calculation around the nuclei. }
\\
%R. D. Nelson Jr., D. R. Lide, A. A. Maryott "Selected Values of electric dipole moments for molecules in the gas phase" NSRDS-NBS10, 1967. 
%Hellwege, KH and AM Hellwege (eds.). Landolt-Bornstein: Group II: Volume 6 Molecular Constants from Microwave, Molecular Beam, and Electron Spin Resonance Spectroscopy Springer-Verlag. Berlin. 1974. 
%D. Feller, C. M. Boyle, and E. R. Davidson, J. Chem. Phys. 86, 3424 (1986). % table IV 
%S. A. Clough, Y. Beers, G. P. Klein, and L. S. Rothman, J. Chem. Phys. 59, 2254 (1973).
%Franks, F. Water, a comprehensive treatise; Plenum: New York, 1972. 
% quadrupoles exp. Verhoeven(1070) in in units of 10*-26 esu·cm^2 for H2O: 2.63(2) -0.13(3) -2.50(2)
% quadrupoles exp. H2O S. G. Kukolich, J. Chcm. Phys. 50, 3751 (1969).:  1.6±0.6 0.4±0.4 -2.0±0.6 (Ref. 9 Verhoeven)
% conversione Debye \AA -> a.u. : (0.393430/0.5291772083)
\\ %caption
\multicolumn{10}{ p{15cm} }{
Same of \ref{table:BSC_propSI}, but for the all-electron VMC calculations.
}

\end{tabular}
}
\end{table}

%\input{Table_IE_VMCall}
% IE
% Ionization Energy Water Molecule
\begin{table}[p!]
\caption{ Ionization Energy of Water Molecule by VMC. }  
\label{table:IE_VMC}
%{\tiny
%{\scriptsize
%{\footnotesize
%{\small
\begin{tabular}{ c c c c  c c  c c }  

\hline
\hline
function & core & basis set & Opt & $E_{H_2O}$[H] & $E_{H_2O^+}$[H] & $\Delta E$[H] & IE[eV] \\
\hline
\hline
%{\em Pseudo Filippi}\\

 JSD    & ECP & PS-aDZ+     & all & -17.2478(1) & -16.7791(1) &  0.4687(2) & 12.687(5) \\
 JSD    & ECP & PS-aTZ+     & all & -17.2478(1) & -16.7794(1) &  0.4684(2) & 12.678(5) \\ 
 JSD    & ECP & PS-aQZ+     & all & -17.2481(1) & -16.7795(1) &  0.4686(2) & 12.684(5)\\ 
                                  
 JAGPn* & ECP & PS-aDZ+     & all & -17.2512(1) & -16.7821(1) &  0.4691(2) & 12.697(5) \\
 JAGPn* & ECP & PS-aTZ+     & all & -17.2513(1) & -16.7834(1) &  0.4679(1) & 12.665(5) \\ 
 JAGPn* & ECP & PS-aQZ+     & all & -17.2513(1) & -16.7824(1) &  0.4689(2) & 12.692(5) \\ 
                                  
 JAGP   & ECP & PS-DZ$^{a}$ & noZ & -17.2157(2) & -16.7610(2) & 0.4548(3) & 12.308(8) \\
 JAGP   & ECP & PS-DZ       & noZ & -17.2482(1) & -16.7775(1) & 0.4707(2) & 12.741(5) \\
 JAGP   & ECP & PS-TZ       & noZ & -17.2503(1) & -16.7789(1) & 0.4714(2) & 12.760(5) \\
 JAGP   & ECP & PS-aDZ      & noZ & -17.2520(1) & -16.7823(1) & 0.4697(1) & 12.714(3) \\
                                  
 JAGP   & ECP & PS-DZ$^{a}$ & all & -17.2388(1) & -16.7774(1) & 0.4614(2) & 12.488(5) \\
 JAGP   & ECP & PS-DZ       & all & -17.2524(1) & -16.7831(1) & 0.4693(1) & 12.703(3) \\
 JAGP   & ECP & PS-TZ       & all & -17.2529(1) & -16.7830(1) & 0.4699(1) & 12.719(3) \\
 JAGP   & ECP & PS-aDZ      & all & -17.2538(1) & -16.7842(1) & 0.4696(1) & 12.711(3) \\

\hline

%{\em Pseudo Needs}\\
 JAGP & NCP & PS-DZ  & noZ & -17.2007(2) & -16.7178(2) & 0.4829(3) & 13.073(8) \\
 JAGP & NCP & PS-aDZ & noZ & -17.2050(1) & -16.7253(2) & 0.4797(3) & 12.986(8) \\
                           
 JAGP & NCP & PS-DZ  & all & -17.2052(1) & -16.7377(1) & 0.4675(2) & 12.654(5) \\ 
 JAGP & NCP & PS-aDZ & all & -17.2068(1) & -16.7383(1) & 0.4685(1) & 12.681(3) \\

\hline 
%{\em All Electron }\\
 JAGP & AE & DZ & noZ & -76.3909(4) & -75.9191(4)   & 0.4718(6) & 12.771(16) \\
 JAGP & AE & DZ & all & -76.4041(3) & -75.9336(3)   & 0.4705(4) & 12.736(11) \\ 

\hline 
%\hline
%\multicolumn{4}{l}{ HF/aug-cc-pVQZ}      &&&  0.402 & 10.868 \\
%\multicolumn{4}{l}{ B3LYP/aug-cc-pVQZ}   &&&  0.466 & 12.610 \\
%\multicolumn{4}{l}{ MP2FC/aug-cc-pVTZ}   &&&  0.469 & 12.709 \\
%\multicolumn{4}{l}{ CCSD(T)/aug-cc-pVTZ} &&&  0.462 & 12.505 \\
\multicolumn{4}{l}{ Exact}     &&&  0.4663$^b$ & 12.621(2) \\
\hline
\hline
% note
\multicolumn{7}{l}{$^a$ here the basis set for the inhomogeneous Jastrow terms is O(1s,1p) H(1s)}\\
\multicolumn{7}{p{10cm}}{$^b$ calculated as $IE_{exp}+\Delta_{ZPE}$
} \\
\\
\multicolumn{7}{ p{13cm} }{
Ionization Energy ($IE$) of Water Molecule.
For all these calculations the method used is the VMC, while LRDMC results for a pair of basis sets are reported in \ref{table:IE_LRDMC}.
Similarly to \ref{table:BSC_propSI} we indicate the wave function type, the description of the core, the basis set and the optimization method, while 
the basis set and the optimization method, while the basis for the inhomogeneous Jastrow terms is O(3s,2p,1d) H(2s,1p), unless specifically indicated.
The energy difference $\Delta E = E_{H_2O}-E_{H_2O^+}$ is a biased estimate of $IE$ because it does not consider the ZPE. 
For a comparison between the computed and the experimental $IE$, we have considered the estimate of the difference $\Delta_{ZPE}$ in the zero point energy between the cation and the neutral form of water calculated from CCSD(T)/aug-cc-pVTZ calculatons\cite{cccbdb}.
}
\end{tabular}
%}
\end{table}

%\input{Table_IE_LRDMC}
% IE
% Ionization Energy Water Molecule
\begin{table}[p!]
\caption{ Ionization Energy of Water Molecule by LRDMC. }  
\label{table:IE_LRDMC}
%{\tiny
%{\scriptsize
%{\footnotesize
%{\small
\begin{tabular}{ c  c c  c c }  

\hline
\hline
method & $E_{H_2O}$[H] & $E_{H_2O^+}$[H] & $\Delta E$[H] & IE[eV] \\
\hline
\hline

%\\
\multicolumn{5}{l}{\bf {\em function:}JAGP {\em core:}ECP {\em basis:}PS-aDZ {\em opt:all} } \\
%\\

VMC        & -17.2538(1) & -16.7842(1) & 0.4696(1) & 12.711(3) \\

LRDMC(0.5) & -17.2672(2) & -16.7977(2) &  0.4695(3) & 12.708(8) \\ 
LRDMC(0.2) & -17.2654(2) & -16.7957(2) &  0.4697(3) & 12.714(8) \\ 
LRDMC(0.15)& -17.2651(2) & -16.7958(2) &  0.4693(3) & 12.703(8) \\ 
LRDMC(0.1) & -17.2649(2) & -16.7954(2) &  0.4695(3) & 12.708(8) \\ 

LRDMC-0    & -17.2647(3) & -16.7954(3) &  0.4693(4) & 12.703(8) \\ 

\hline

%\\
\multicolumn{5}{l}{\bf {\em function:}JAGP {\em core:}AE {\em basis:}DZ {\em opt:all} } \\
%\\

VMC           & -76.4041(3) & -75.9336(3)   & 0.4705(4)  & 12.736(11) \\ 

LRDMC(0.23)   & -76.4316(3) &  -75.9976(7)  &  0.4340(8) & 11.742(22) \\
LRDMC(0.2)    & -76.4298(2) &  -75.9727(5)  &  0.4570(6) & 12.368(16) \\
LRDMC(0.15)   & -76.4274(2) &  -75.9551(3)  &  0.4723(3) & 12.785(8) \\
LRDMC(0.1)    & -76.4265(2) &  -75.9540(2)  &  0.4725(3) & 12.790(8) \\
LRDMC(0.07)   & -76.4266(1) &  -75.9559(1)  &  0.4708(2) & 12.742(5) \\
LRDMC(0.05)   & -76.4267(1) &  -75.9564(1)  &  0.4703(2) & 12.728(5) \\
LRDMC-0       & -76.4266(1) &  -75.9586(2)  & 0.4681(2) & 12.668(6) \\ 
\hline
\hline
% attenzione: i valori seguenti hanno la correzione della zpe calcolata con il proprio metodo
\multicolumn{3}{l}{ HF/aug-cc-pVQZ $^a$}      &  0.402 & 10.868 \\
\multicolumn{3}{l}{ BLYP/cc-pVTZ $^a$}        &  0.456 & 12.341 \\
\multicolumn{3}{l}{ B3LYP/aug-cc-pVQZ $^a$}   &  0.466 & 12.610 \\
\multicolumn{3}{l}{ MP2FC/aug-cc-pVTZ $^a$}   &  0.469 & 12.709 \\
\multicolumn{3}{l}{ CCSD(T)/aug-cc-pVTZ $^a$} &  0.462 & 12.505 \\

\multicolumn{3}{l}{Exact $^a$ } &  0.4663$^b$ & 12.621(2) \\
\hline
\hline
% note
\multicolumn{5}{p{10cm}}{$^a$ from ref.\cite{cccbdb}}\\
\multicolumn{5}{p{10cm}}{$^b$ calculated as $IE_{exp}+\Delta_{ZPE}$ }\\
\\
\multicolumn{5}{ p{13cm} }{
Ionization Energy of Water Molecule by LRDMC, calculated for a two wave functions: the first with pseudo potential ECP core and the second with all electrons.
The first column indicates the method used for the calculation: VMC or LRDMC(a), the latter being the lattice regolarized diffusion Monte Carlo with lattice parameter $a$.
LRDMC-0 represents the extrapolation $a\to 0$  of the LRDMC energy. 
Similarly to \ref{table:BSC_propSI} we indicate the wave function type, the description of the core, the basis set and the optimization method, while the basis for the inhomogeneous Jastrow terms is O(3s,2p,1d) H(2s,1p).
Experimental and results from other computational methods are also reported for a comparison.
}
\end{tabular}
%}
\end{table}

%\input{Table_eq+freqALL}
% structural minimum and frequencies
%\begin{sidewaystable}

\begin{table}[p!]
%\begin{turn}{90}
\caption{ Water molecule: equilibrium configuration, harmonic and fundamental frequencies. }  
\label{table:eqfrSI}

{\tiny
%{\scriptsize
%{\footnotesize
%{\small

\begin{tabular}{ c c c c c   l l   l l l   l l l }  
\hline
\hline
\multicolumn{5}{l}{} & \multicolumn{2}{c}{equilibrium structure} & \multicolumn{3}{c}{harmonic freq.[cm$^{-1}$]} & \multicolumn{3}{c}{fundamental freq.[cm$^{-1}$]} \\
method & function & core & basis & Opt & $r_0$[\AA] & $\phi_0$[deg] & $\omega_2$ & $\omega_1$ & $\omega_3$ & $\nu_2$[010] & $\nu_1$[100] & $\nu_3$[001] \\
\hline \hline \multicolumn{5}{c}{\em fit energy} \\

CS & DFT    & ECP & PS-VTZ-h20 & noZ & 0.97482(4) & 104.30(1)   & 1533(1)  & 3673(3)  & 3782(4)  & 1489(2)   & 3517(11) & 3614(13)  \\ %PS-VTZ \{20a\} & O(3s,2p,1d)/H(2s,1p) & F & DFT LDA                

VMC & JDFT   & ECP & PS-VTZ-h20 & noZ & 0.9563(3)  & 104.63(7)   & 1691(15) & 3913(38) & 4030(41) & 1606(6)   & 3689(68) & 3800(81)  \\ %PS-VTZ \{20a\} & O(3s,2p,1d)/H(2s,1p) & F & VMC JDFT                 
CS  & JDFT   & ECP & PS-VTZ-h20 & noZ & 0.95560(1) & 104.548(3)  & 1667.7(6)& 3880(1)  & 3995(2)  & 1614.2(4) & 3702(2)  & 3799(4)   \\ %PS-VTZ \{20a\} & O(3s,2p,1d)/H(2s,1p) & F & CS JDFT                  

%NOT CONVERGED%%VMC & JSD-   & ECP & PS-VTZ-h30 & all & 0.9541(3)  & 104.78(9)   & 1669(18) & 3898(38) & 3995(43) & 1610(11)  & 3675(59) & 3785(74)  \\ %PS-VTZ \{30a\} & O(3s,2p,1d)/H(2s,1p) & F & VMC JSD $n_{opt}=1000$ 
VMC & JSD    & ECP & PS-VTZ-h30 & all & 0.9543(2)  & 104.84(6)   & 1679(10) & 3931(21) & 4021(44) & 1619(5)   & 3664(41) & 3772(84)  \\ %PS-VTZ \{30a\} & O(3s,2p,1d)/H(2s,1p) & F & VMC JSD $n_{opt}=1800$ 
CS  & JSD    & ECP & PS-VTZ-h30 & all & 0.95432(3) & 104.81(1)   & 1674(2)  & 3894(3)  & 3997(6)  & 1615(2)   & 3713(6)  & 3815(11)  \\ %PS-VTZ \{20a\} & O(3s,2p,1d)/H(2s,1p) & F & CS JSD                 

VMC & JAGPn* & ECP & PS-aTZ+      & all & 0.9556(3)  & 104.20(9)   & 1681(17) & 3897(42) & 4012(41) & 1621(10) & 3675(76)  & 3751(69)  \\ %PS-aug-pVTZ & O(3s,2p,1d)/H(2s,1p) & F & VMC mol.AGP*              

VMC & JAGP   & ECP & PS-DZ $^a$   & noZ & 0.954(3)   & 104.4(6)    & 1637(99) & 3678(302)& 4197(322)& 1724(66) & 3283(433) & 3113(465) \\ %PS-pVDZ     & O(1s,1p)/H(1s)       & F & VMC opt.NoZ               
VMC & JAGP   & ECP & PS-DZ $^a$   & all & 0.951(2)   & 104.7(5)    & 1737(87) & 4021(189)& 4028(190)& 1569(42) & 3668(1240)& 3988(1144)\\ %PS-pVDZ     & O(1s,1p)/H(1s)       & F & VMC opt.all               
VMC & JAGP   & ECP & PS-aDZ       & all & 0.9554(3)  & 104.59(9)   & 1668(23) & 3879(45) & 3988(65) & 1609(11) & 3610(78)  & 3687(117) \\ %PS-aug-pVDZ & O(3s,2p,1d)/H(2s,1p) & F & VMC opt.all               
VMC & JAGP   & ECP & PS-aTZ       & all & 0.9556(3)  & 104.46(7)   & 1655(15) & 3854(37) & 3983(73) & 1616(9)  & 3684(71)  & 3764(161) \\ %PS-aug-pVTZ & O(3s,2p,1d)/H(2s,1p) & F & VMC opt.all               
VMC & JAGP   & ECP & PS-VTZ-h20 & all & 0.9556(1)  & 104.40(4)   & 1665(7)  & 3861(16) & 4029(36) & 1612(4)   & 3694(33) & 3698(71)  \\ %PS-VTZ \{20a\} & O(3s,2p,1d)/H(2s,1p) & F & VMC JAGP               

\hline 
VMC & JSD-   & NCP & PS-VTZ-h30 & noZ & 0.9551(2)  & 104.87(7)   & 1682(11) & 3903(26) & 4005(32) & 1612(7)   & 3688(44) & 3811(58)  \\ %PS-VTZ \{30a\} & O(3s,2p,1d)/H(2s,1p) & N & VMC JSD opt.SRH        
VMC & JSD    & NCP & PS-VTZ-h30 & noZ & 0.9558(2)  & 104.87(5)   & 1675(11) & 3880(25) & 3953(40) & 1609(7)   & 3724(48) & 3888(91)  \\ %PS-VTZ \{30a\} & O(3s,2p,1d)/H(2s,1p) & N & VMC JSD opt.SR         

VMC & JAGP   & NCP & PS-aDZ       & all & 0.9568(3)  & 104.46(9)   & 1656(18) & 3874(41) & 3958(57) & 1616(14) & 3708(78)  & 3831(125) \\ %PS-aug-pVDZ & O(3s,2p,1d)/H(2s,1p) & N & VMC opt.all               
VMC & JAGP   & NCP & PS-VTZ-h20 & all & 0.9569(2)  & 104.53(5)   & 1650(11) & 3862(27) & 3965(25) & 1612(7)   & 3699(46) & 3809(39)  \\ %PS-VTZ \{20a\} & O(3s,2p,1d)/H(2s,1p) & N & VMC JAGP               
                                       
\hline \hline \multicolumn{5}{c}{\em fit forces} \\

VMC & JDFT   & ECP & PS-VTZ-h20 & noZ & 0.95497(3) & 104.49(2)   & 1664(2)  & 3882(2)  & 3995(3)  & 1610(1)   & 3693(2) & 3787(3)    \\ %PS-VTZ \{20a\} & O(3s,2p,1d)/H(2s,1p) & F & VMC JDFT  

%NOT CONVERGED%%VMC & JSD-   & ECP & PS-VTZ-h30 & all & 0.95422(4) & 104.73(1)   & 1673(1)  & 3892(2)  & 4004(3)  & 1618(1)   & 3702(1) & 3798(4)    \\ %PS-VTZ \{30a\} & O(3s,2p,1d)/H(2s,1p) & F & VMC JSD $n_{opt}=1000$ 
VMC & JSD    & ECP & PS-VTZ-h30 & all & 0.95426(3) & 104.74(1)   & 1670(2)  & 3892(3)  & 4006(3)  & 1617(1)   & 3702(3) & 3794(2)    \\ %PS-VTZ \{30a\} & O(3s,2p,1d)/H(2s,1p) & F & VMC JSD $n_{opt}=1800$ 

VMC & JAGPn* & ECP & PS-aTZ+      & all & 0.95612(8) & 104.17(2)   & 1710(3)  & 3896(6)  & 3990(7)  & 1654(1)   & 3710(3)  & 3800(7)   \\ %PS-aug-pVTZ & O(3s,2p,1d)/H(2s,1p) & F & VMC mol.AGP*              

VMC & JAGP   & ECP & PS-DZ $^a$   & noZ & 0.9552(2)  & 104.11(7)   & 1723(10) & 3920(14) & 4039(17) & 1663(6)   & 3732(10) & 3846(13)  \\ %PS-pVDZ     & O(1s,1p)/H(1s)       & F & VMC opt.NoZ               
VMC & JAGP   & ECP & PS-DZ $^a$   & all & 0.9524(1)  & 104.61(5)   & 1690(6)  & 3954(11) & 4056(11) & 1626(3)   & 3738(7)  & 3852(8)   \\ %PS-pVDZ     & O(1s,1p)/H(1s)       & F & VMC opt.all               
VMC & JAGP   & ECP & PS-aDZ       & all & 0.95545(7) & 104.42(2)   & 1666(3)  & 3873(5)  & 3989(9)  & 1613(2)   & 3685(5)  & 3775(12)  \\ %PS-aug-pVDZ & O(3s,2p,1d)/H(2s,1p) & F & VMC opt.all               
VMC & JAGP   & ECP & PS-aTZ       & all & 0.95570(17)& 104.40(5)   & 1676(7)  & 3871(16) & 4043(23) & 1613(4)   & 3681(10) & 3777(24)  \\ %PS-aug-pVTZ & O(3s,2p,1d)/H(2s,1p) & F & VMC opt.all               
VMC & JAGP   & ECP & PS-VTZ-h20 & all & 0.95550(4) & 104.41(1)   & 1669(1)  & 3872(3)  & 3974(4)  & 1613.3(6) & 3677(2) & 3772(2)    \\ %PS-VTZ \{20a\} & O(3s,2p,1d)/H(2s,1p) & F & VMC JAGP               

\hline
VMC & JSD    & NCP & PS-VTZ-h30 & all & 0.95536(3) & 104.85(1)   & 1668(2)  & 3889(2)  & 4001(3)  & 1613.1(9) & 3700(2) & 3796(3)    \\ %PS-VTZ \{30a\} & O(3s,2p,1d)/H(2s,1p) & N & VMC JSD opt.SR         

VMC & JAGP   & NCP & PS-aDZ       & all & 0.95665(7) & 104.54(2)   & 1664(3)  & 3884(5)  & 3990(7)  & 1609(1)   & 3678(5)  & 3782(8)   \\ %PS-aug-pVDZ & O(3s,2p,1d)/H(2s,1p) & N & VMC opt.all               
VMC & JAGP   & NCP & PS-VTZ-h20 & all & 0.95668(3) & 104.52(1)   & 1663(2)  & 3869(2)  & 3973(3)  & 1610.4(7) & 3679(2) & 3767(3)    \\ %PS-VTZ \{20a\} & O(3s,2p,1d)/H(2s,1p) & N & VMC JAGP               

\hline \hline

\multicolumn{5}{l}{   SCF/DZP         from \citet{Barone:2005p24347} }       & 0.9457 & 106.16 \\ 
\multicolumn{5}{l}{  BLYP/aug-cc-pVTZ from \citet{Barone:2005p24347} }       & 0.9719 & 104.47 & 1596 & 3655 & 3757 & 1543 & 3480 & 3567 \\ % & 1.8033 dipole 
\multicolumn{5}{l}{ B97-2/aug-cc-pVTZ from \citet{Barone:2005p24347} }       & 0.9567 & 104.77 & 1645 & 3860 & 3968 & 1594 & 3701 & 3794 \\ % & 1.8473 
\multicolumn{5}{l}{ B3LYP/aug-cc-pVTZ from \citet{Barone:2005p24347} }       & 0.9619 & 105.08 & 1627 & 3796 & 3899 & 1575 & 3631 & 3720 \\ % & 1.8411 
\multicolumn{5}{l}{FC MP2/aug-cc-pVTZ from \citet{Barone:2005p24347} }       & 0.9614 & 104.11 & 1628 & 3822 & 3948 & 1578 & 3653 & 3767 \\ % & 1.8701 

\multicolumn{5}{l}{    CISD/(13,8,4,2/8,4,2) from \citet{KIM:1995p23441} }      & 0.952 & 104.8 & 1676.1 & 3947.3 & 4050.5 \\ 
\multicolumn{5}{l}{   QCISD/(13,8,4,2/8,4,2) from \citet{KIM:1995p23441} }      & 0.956 & 104.4 & 1659.4 & 3868.0 & 3976.2 \\ 
\multicolumn{5}{l}{QCISD(T)/(13,8,4,2/8,4,2) from \citet{KIM:1995p23441} }      & 0.959 & 104.2 & 1646.5 & 3827.0 & 3937.3 \\ 
\multicolumn{5}{l}{    CCSD/(13,8,4,2/8,4,2) from \citet{KIM:1995p23441} }      & 0.956 & 104.4 & 1662.5 & 3870.9 & 3977.8 \\ 
\multicolumn{5}{l}{ CCSD(T)/(13,8,4,2/8,4,2) from \citet{KIM:1995p23441} }      & 0.959 & 104.1 & 1648.4 & 3831.7 & 3941.5 \\ 
\multicolumn{5}{l}{FC CCSD(T)/aug-cc-pV7Z from \citet{Feller:2009p23440}   }      & 0.95831 & 104.452 & 1649.83 & 3835.55 & 3946.05 & 1595.58  & 3659.31 & 3757.45 \\
\hline % experiment
\multicolumn{5}{l}{ Experiment (from \citet{Benedict:1956id})}          & 0.95721(30) & 104.522(50) & 1648.47  & 3832.17  & 3942.53  & 1594.59   & 3656.65 & 3755.79    \\                                                       
\hline \hline
\multicolumn{13}{l}{ $^a$ basis set for the Jastrow 3/4-body is O(1s,1p) H(1s). }\\

\\ % caption
\multicolumn{13}{ p{15cm} }{ 
Equilibrium configuration, harmonic frequencies $\omega_i$ and fundamental frequencies $\nu_i$ of water molecule,  calculated from the fitting of the PES using VMC calculation of energies or forces (see details in the text and in \citet{Zen:2012br}) for several different wave functions. 
The wave function description is the same used also in previous tables.
With CS in the method column it is meat that the correlated sampling has been used, within a VMC scheme, to calculate the energy difference between the wave functions of the water at the different configurations and at the experimental configuration.
One DFT calculation has also been executed (the first entry in the table) with SLP functional.
At the end of the table we have reported the results obtained through several other computational methods, and the experimental results.
}\\
\end{tabular}
}
%\end{turn}
\end{table}

%\end{sidewaystable}

%\input{Table_omega3_LRDMC}
% omega3 LRDMC
\begin{table}[p!]
\caption{ $\omega_3$  from energy fitting of the LRDMC results }  
\label{table:omega3SI}
%{\tiny
%{\scriptsize
%{\footnotesize
%{\small
\begin{tabular}{ c c c }  

\hline \hline
                 &    fit    \\
\hline \hline
VMC CS           &  3985( 4)   \\
\hline
LRDMC(0.4)       &  3915(25)  \\
LRDMC(0.3)       &  3992(14)  \\
LRDMC(0.2)       &  3967(14)  \\ %3963(24)  \\
LRDMC(0.1)       &  3972(19)  \\
\hline
LRDMC-0          &  3950(22)  \\ %3955(27)  \\
\hline \hline
\\
\multicolumn{3}{ p{8cm} } {
% functions: JAGP ECP PS-aDZ 
% center: fitted minimum structure, see \ref{table:eqfr}: 0.95545(7) 104.42(2).
% asymmetric stretching (computed with VMC, fit forces: 3989(9); experimental: 3942.53). 
% see \ref{fig:omega3} for a plot of the LRDMC energies plotted versus the displacement along the asymmetric stretching mode.
% fitting function: $y(x) = c_0 + c_2 x^2 + c_4 x^4$ 
% fit1: all ;
% fit2: no 0 ;
% fit3: no 0 and 0.2 
Harmonic asymmetric stretching vibrational frequency $\omega_3$, calculated for several LRDMC calculations with lattice size $a$ of 0.4, 0.3, 0.2, 0.1 Bohr, and of the $a\to 0$ extrapolation (LRDMC-0).
In the first line, VMC CS, we report the frequency obtained using the correlated sampling within a VMC approach.
The guiding function is a JAGP with ECP pseudo potential, 
PS-aDZ basis for the determinantal and O(3s,2p,1d) H(2s,1p) for the Jastrow.
$\omega_3$ is calculated from the fitting of the single point energy calculations along the asymmetric stretching mode.
The fitting function is  $y(x) = c_0 + c_2 x^2 + c_4 x^4$, and 
$\omega_3$ is obtained from $c_2$.
For a comparison, the experimentally determined  $\omega_3$ is 3942.53$cm^{-1}$ and the value calculated for the same basis with a VMC approach is 3989(9)$cm^{-1}$, see \ref{table:eqfr}, in excellent agreement with the VMC CS result.
}

\end{tabular}
%}
\end{table}

\begin{figure}[p!]
\begin{center}
(a)\includegraphics[scale = 0.5]{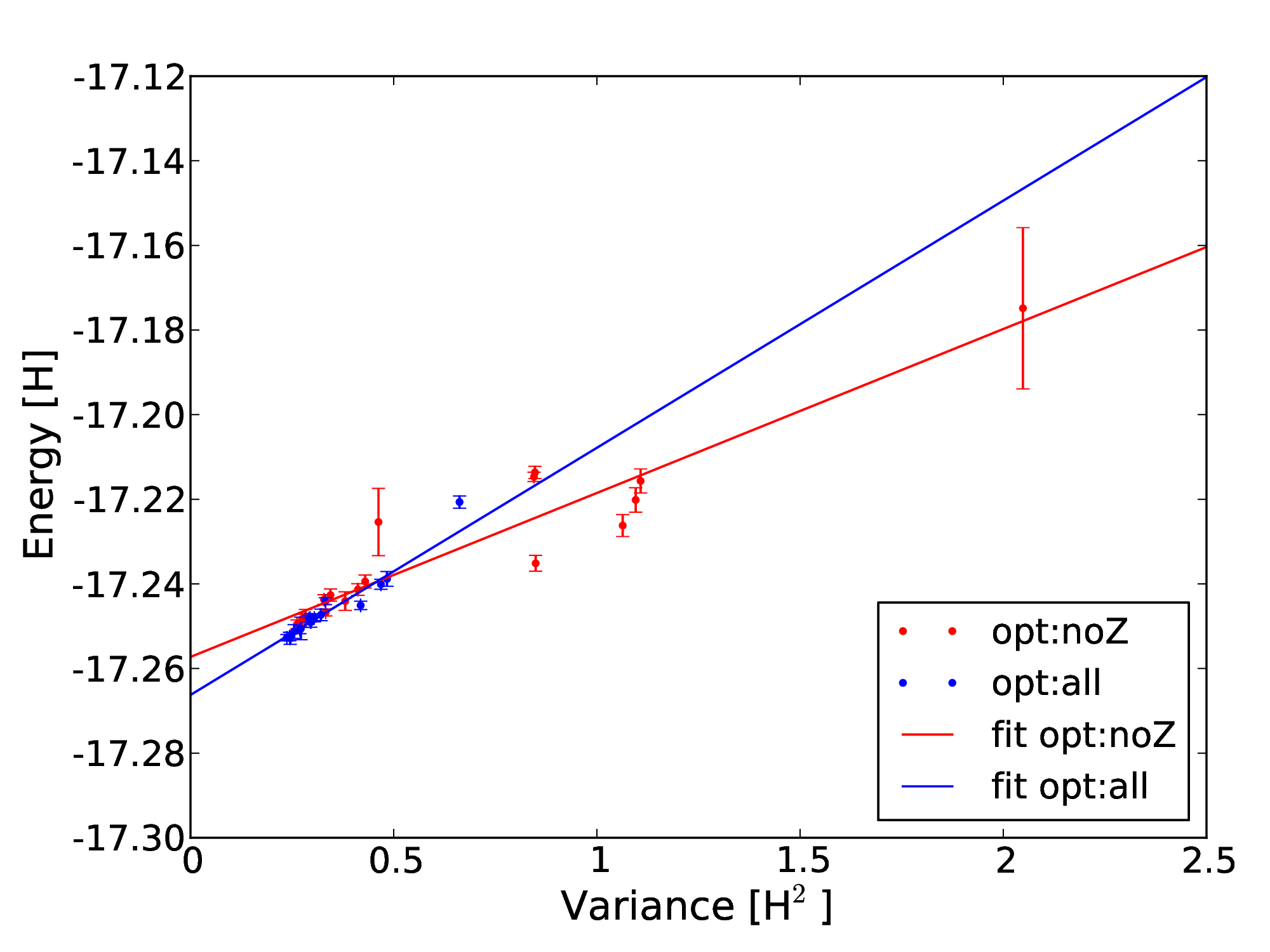}
(b)\includegraphics[scale = 0.5]{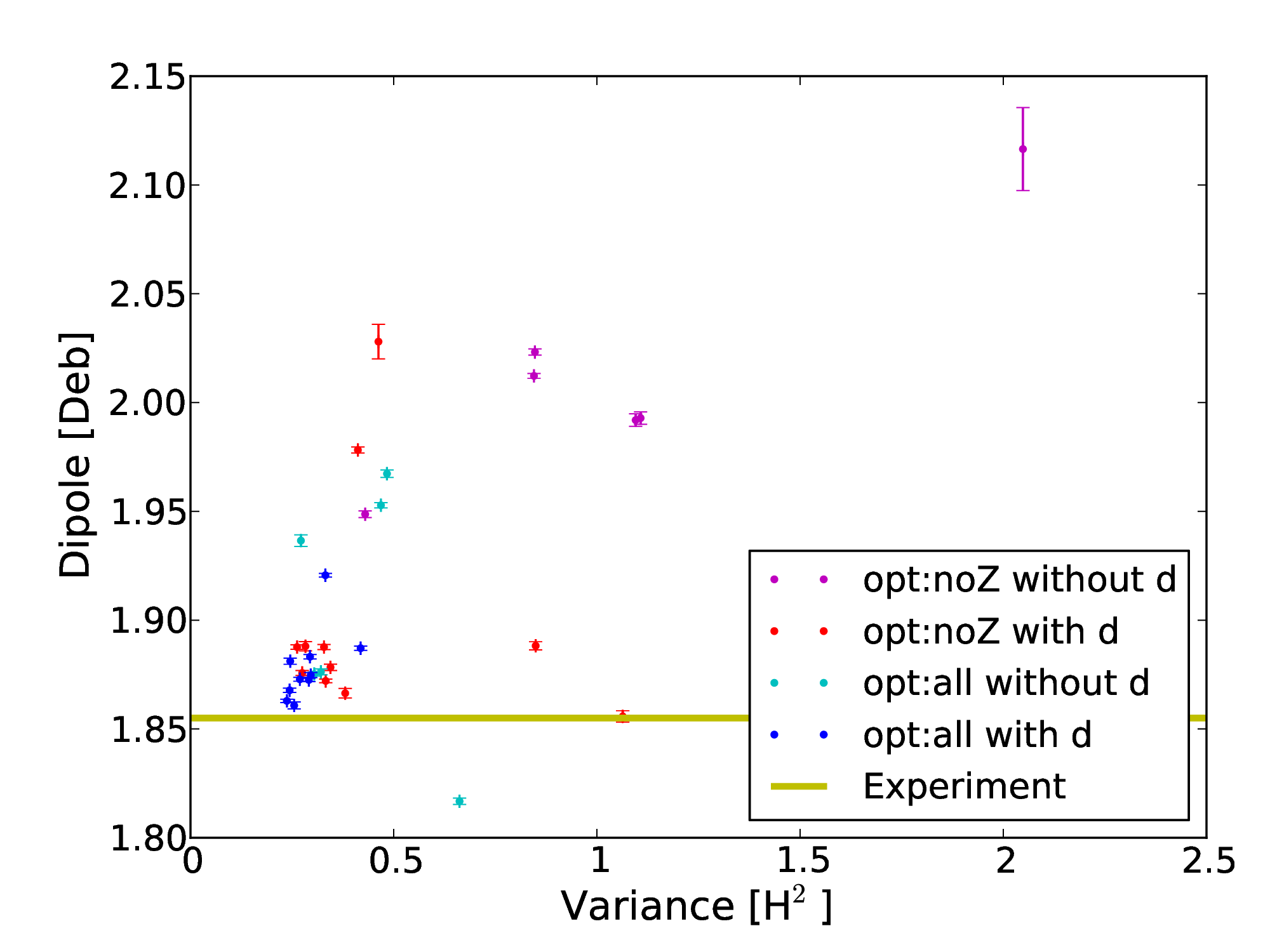}
\end{center}
\caption{
Basis set convergence of the inhomogeneous Jastrow term. 
The JAGP wave functions considered are those reported in \ref{table:JBSC}.
(a) Scatter plot of Energy versus Variance.
The {\em opt:noZ} and {\em opt:all} cases are colored respectively in red and blue. The linear fitting lines are plotted, and the extrapolated zero variance values of the energy are   
-17.257(3)H for {\em opt:noZ}, and
-17.266(1)H for {\em opt:all}.
(b) Scatter plot of the 
Dipole [Debye] versus the Variance [H$^2$].
} \label{fig:JBSC}
\end{figure}

\begin{figure}[p!]
\begin{center}
(a)\includegraphics[scale = 0.5]{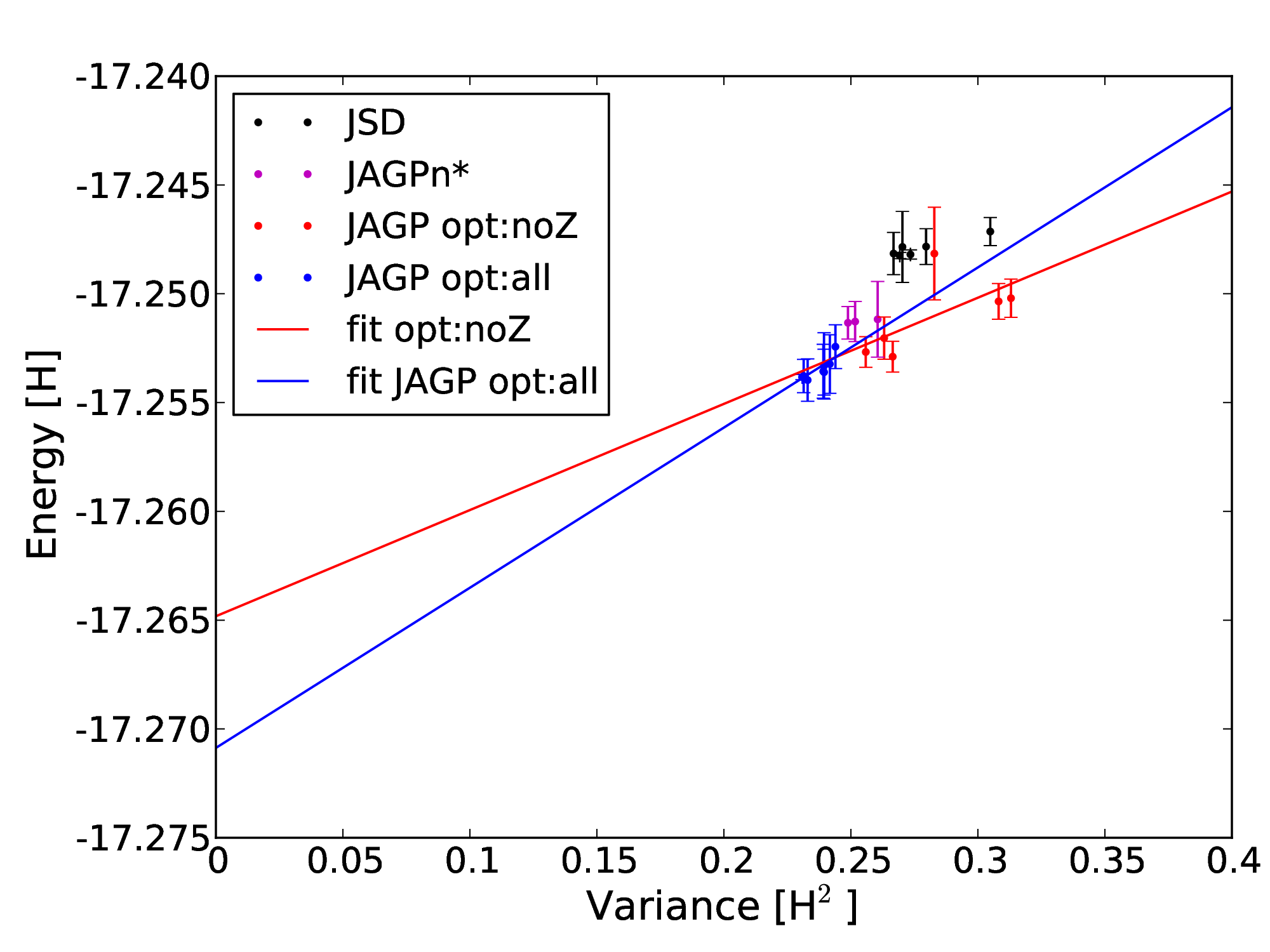}
(b)\includegraphics[scale = 0.5]{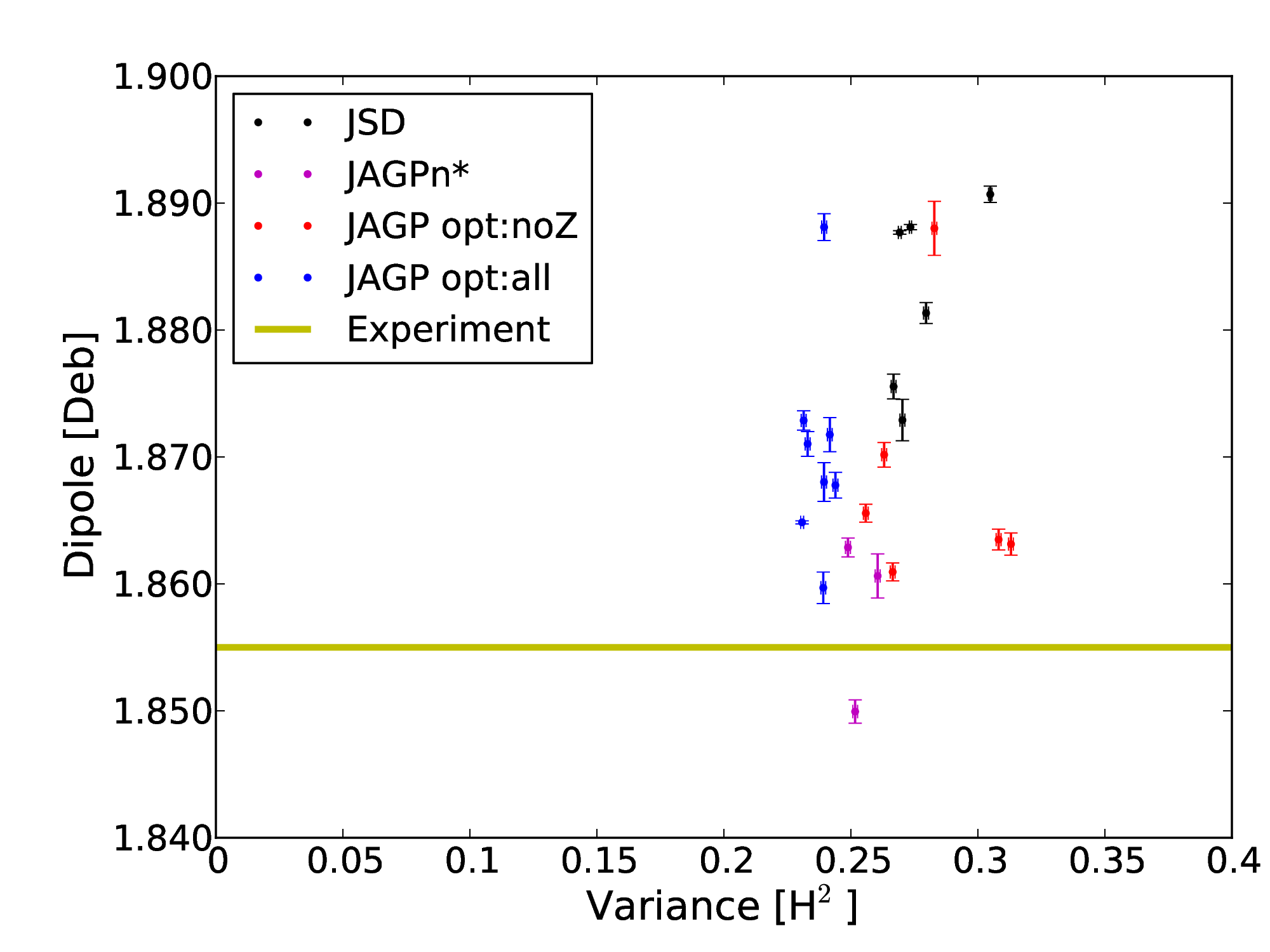}
\end{center}
\caption{
Basis set convergence of the determinantal term, for the JSD, JAGPn* and JAGP wave functions considered in \ref{table:BSC}, 
which uses ECP for the core electrons of the oxygen.
The JSD data are in black, the JAGPn* in violet and 
the JAGP {\em opt:noZ} and {\em opt:all} cases are colored respectively in red and blue.
(a)
Energy versus Variance scatter plot.
The linear fitting lines are plotted, and the extrapolated zero variance values of the energy are   
-17.265(8)H for {\em opt:noZ}, and
-17.271(4)H for {\em opt:all}.
(b)
Dipole [Debye] versus Variance [H$^2$] scatter plot.
} \label{fig:BSC_ECP}
\end{figure}

\begin{figure}[p!]
\begin{center}
(a)\includegraphics[scale = 0.5]{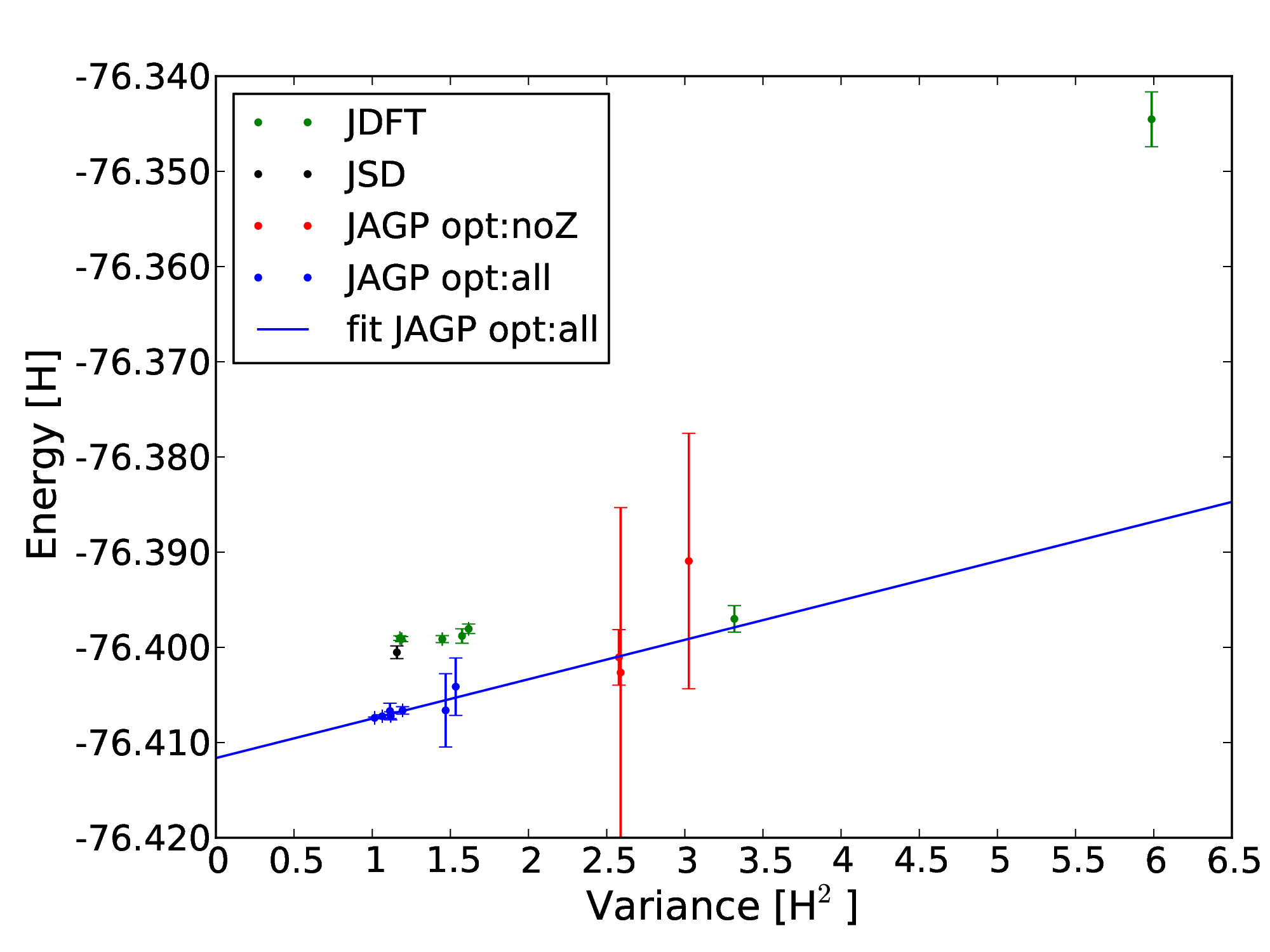}
(b)\includegraphics[scale = 0.5]{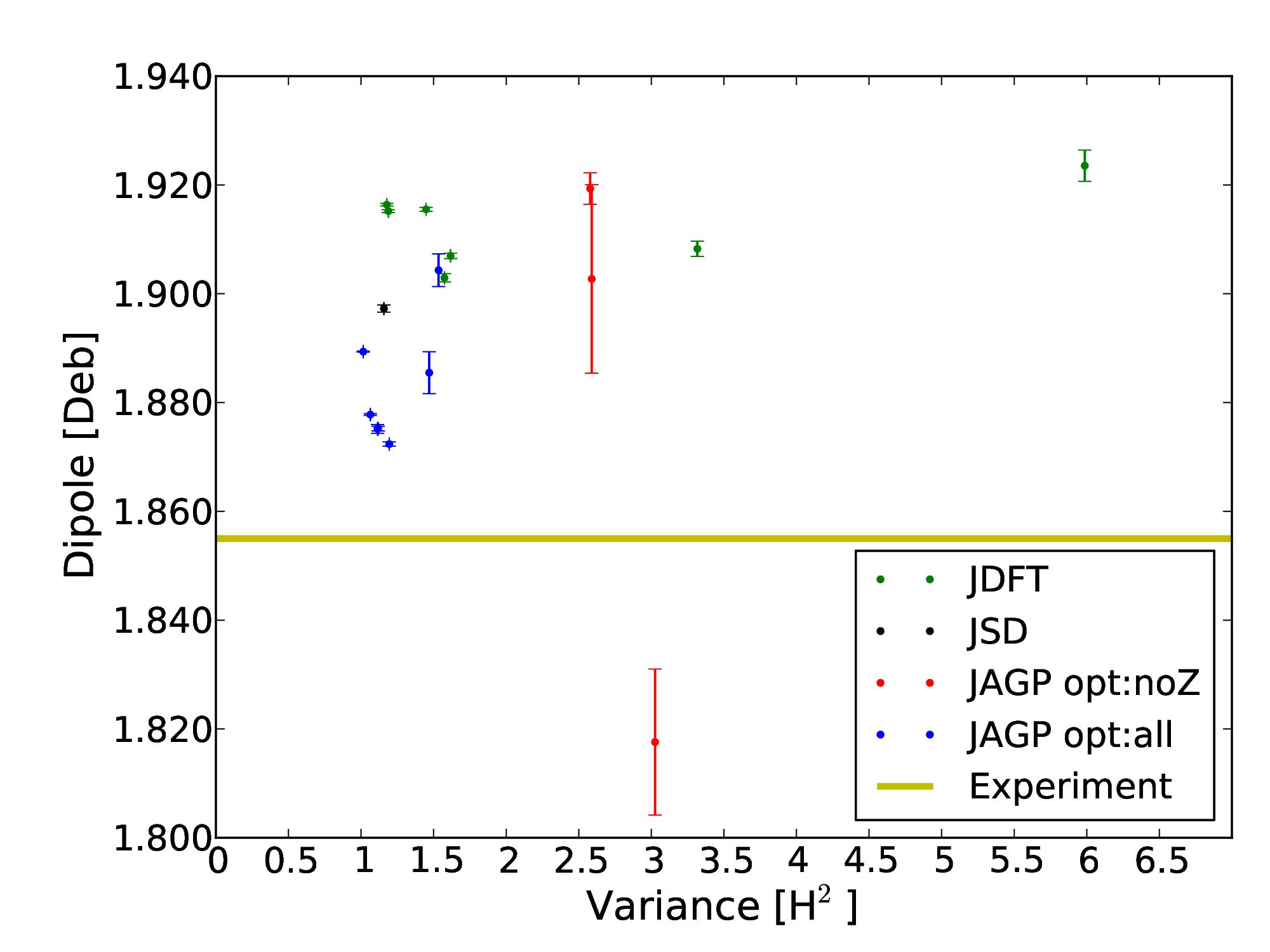}
\end{center}
\caption{
Basis set convergence of the determinantal term, for all electrons calculations and the JDFT (green), JSD (black) and JAGP ({\em opt:noZ} in red, {\em opt:all} in blue) wave functions considered in \ref{table:BSC_AE_prop}.
(a)
Scatter plot of the Energy versus the Variance. 
The linear fitting lines are plotted, and the extrapolated zero variance values of the energy for the JAGP {\em opt:all} functions  is   
-76.4116(10)H.
(b)
Scatter plot of the Dipole [Debye] versus the Variance [H$^2$],
} \label{fig:BSC_AE}
\end{figure}

\newpage

%%%%%%%%%%%%%%%%%%%%%%%%%%%%%%%%%%%%%%%%%%%%%%%%%%%%%%%%%%%%%%%%%%%%%
%% Bibliography
%%%%%%%%%%%%%%%%%%%%%%%%%%%%%%%%%%%%%%%%%%%%%%%%%%%%%%%%%%%%%%%%%%%%%

%\bibliography{Bibliography}

\begin{thebibliography}{103}
\expandafter\ifx\csname natexlab\endcsname\relax\def\natexlab#1{#1}\fi
\expandafter\ifx\csname bibnamefont\endcsname\relax
  \def\bibnamefont#1{#1}\fi
\expandafter\ifx\csname bibfnamefont\endcsname\relax
  \def\bibfnamefont#1{#1}\fi
\expandafter\ifx\csname citenamefont\endcsname\relax
  \def\citenamefont#1{#1}\fi
\expandafter\ifx\csname url\endcsname\relax
  \def\url#1{\texttt{#1}}\fi
\expandafter\ifx\csname urlprefix\endcsname\relax\def\urlprefix{URL }\fi
\providecommand{\bibinfo}[2]{#2}
\providecommand{\eprint}[2][]{\url{#2}}

\bibitem[{\citenamefont{Austin et~al.}(2012)\citenamefont{Austin, Zubarev, and
  Lester}}]{Austin:2012kt}
\bibinfo{author}{\bibfnamefont{B.~M.} \bibnamefont{Austin}},
  \bibinfo{author}{\bibfnamefont{D.~Y.} \bibnamefont{Zubarev}},
  \bibnamefont{and} \bibinfo{author}{\bibfnamefont{W.~A.~J.}
  \bibnamefont{Lester}}, \bibinfo{journal}{Chem. Rev.}
  \textbf{\bibinfo{volume}{112}}, \bibinfo{pages}{263} (\bibinfo{year}{2012}).

\bibitem[{\citenamefont{Needs et~al.}(2010)\citenamefont{Needs, Towler,
  Drummond, and Rios}}]{Needs:2010p28123}
\bibinfo{author}{\bibfnamefont{R.~J.} \bibnamefont{Needs}},
  \bibinfo{author}{\bibfnamefont{M.~D.} \bibnamefont{Towler}},
  \bibinfo{author}{\bibfnamefont{N.~D.} \bibnamefont{Drummond}},
  \bibnamefont{and} \bibinfo{author}{\bibfnamefont{P.~L.} \bibnamefont{Rios}},
  \bibinfo{journal}{J. Phys.: Condens. Matter} \textbf{\bibinfo{volume}{22}},
  \bibinfo{pages}{023201} (\bibinfo{year}{2010}).

\bibitem[{\citenamefont{Assaraf
  et~al.}(2007{\natexlab{a}})\citenamefont{Assaraf, Caffarel, and
  Khelif}}]{Assaraf:2007gn}
\bibinfo{author}{\bibfnamefont{R.}~\bibnamefont{Assaraf}},
  \bibinfo{author}{\bibfnamefont{M.}~\bibnamefont{Caffarel}}, \bibnamefont{and}
  \bibinfo{author}{\bibfnamefont{A.}~\bibnamefont{Khelif}},
  \bibinfo{journal}{J. Phys. A: Math. Theor.} \textbf{\bibinfo{volume}{40}},
  \bibinfo{pages}{1181} (\bibinfo{year}{2007}{\natexlab{a}}).

\bibitem[{\citenamefont{Foulkes et~al.}(2001)\citenamefont{Foulkes, Mitas,
  Needs, and Rajagopal}}]{Foulkes:2001p19717}
\bibinfo{author}{\bibfnamefont{W.~M.~C.} \bibnamefont{Foulkes}},
  \bibinfo{author}{\bibfnamefont{L.}~\bibnamefont{Mitas}},
  \bibinfo{author}{\bibfnamefont{R.~J.} \bibnamefont{Needs}}, \bibnamefont{and}
  \bibinfo{author}{\bibfnamefont{G.}~\bibnamefont{Rajagopal}},
  \bibinfo{journal}{Rev. Mod. Phys.} \textbf{\bibinfo{volume}{73}},
  \bibinfo{pages}{33} (\bibinfo{year}{2001}).

\bibitem[{\citenamefont{Reynolds et~al.}(1982)\citenamefont{Reynolds, Ceperley,
  Alder, and Lester}}]{Reynolds:1982en}
\bibinfo{author}{\bibfnamefont{P.~J.} \bibnamefont{Reynolds}},
  \bibinfo{author}{\bibfnamefont{D.~M.} \bibnamefont{Ceperley}},
  \bibinfo{author}{\bibfnamefont{B.~J.} \bibnamefont{Alder}}, \bibnamefont{and}
  \bibinfo{author}{\bibfnamefont{W.~A.} \bibnamefont{Lester}},
  \bibinfo{journal}{J. Chem. Phys.} \textbf{\bibinfo{volume}{77}},
  \bibinfo{pages}{5593} (\bibinfo{year}{1982}).

\bibitem[{\citenamefont{DePasquale et~al.}(1988)\citenamefont{DePasquale,
  Rothstein, and Vrbik}}]{DePasquale:1988bc}
\bibinfo{author}{\bibfnamefont{M.~F.} \bibnamefont{DePasquale}},
  \bibinfo{author}{\bibfnamefont{S.~M.} \bibnamefont{Rothstein}},
  \bibnamefont{and} \bibinfo{author}{\bibfnamefont{J.}~\bibnamefont{Vrbik}},
  \bibinfo{journal}{J. Chem. Phys.} \textbf{\bibinfo{volume}{89}},
  \bibinfo{pages}{3629} (\bibinfo{year}{1988}).

\bibitem[{\citenamefont{Umrigar et~al.}(1993)\citenamefont{Umrigar,
  Nightingale, and Runge}}]{UMRIGAR:1993p25301}
\bibinfo{author}{\bibfnamefont{C.~J.} \bibnamefont{Umrigar}},
  \bibinfo{author}{\bibfnamefont{M.~P.} \bibnamefont{Nightingale}},
  \bibnamefont{and} \bibinfo{author}{\bibfnamefont{K.~J.} \bibnamefont{Runge}},
  \bibinfo{journal}{J. Chem. Phys.} \textbf{\bibinfo{volume}{99}},
  \bibinfo{pages}{2865} (\bibinfo{year}{1993}).

\bibitem[{\citenamefont{Mitas et~al.}(1991)\citenamefont{Mitas, Shirley, and
  Ceperley}}]{Mitas:1991kr}
\bibinfo{author}{\bibfnamefont{L.}~\bibnamefont{Mitas}},
  \bibinfo{author}{\bibfnamefont{E.~L.} \bibnamefont{Shirley}},
  \bibnamefont{and} \bibinfo{author}{\bibfnamefont{D.~M.}
  \bibnamefont{Ceperley}}, \bibinfo{journal}{J. Chem. Phys.}
  \textbf{\bibinfo{volume}{95}}, \bibinfo{pages}{3467} (\bibinfo{year}{1991}).

\bibitem[{\citenamefont{Kalos}(1962)}]{Kalos:1962gq}
\bibinfo{author}{\bibfnamefont{M.}~\bibnamefont{Kalos}},
  \bibinfo{journal}{Phys. Rev.} \textbf{\bibinfo{volume}{128}},
  \bibinfo{pages}{1791} (\bibinfo{year}{1962}).

\bibitem[{\citenamefont{Trivedi and Ceperley}(1990)}]{Trivedi:1990bj}
\bibinfo{author}{\bibfnamefont{N.}~\bibnamefont{Trivedi}} \bibnamefont{and}
  \bibinfo{author}{\bibfnamefont{D.}~\bibnamefont{Ceperley}},
  \bibinfo{journal}{Phys. Rev. B} \textbf{\bibinfo{volume}{41}},
  \bibinfo{pages}{4552} (\bibinfo{year}{1990}).

\bibitem[{\citenamefont{Buonaura and Sorella}(1998)}]{Buonaura:1998p25304}
\bibinfo{author}{\bibfnamefont{M.}~\bibnamefont{Buonaura}} \bibnamefont{and}
  \bibinfo{author}{\bibfnamefont{S.}~\bibnamefont{Sorella}},
  \bibinfo{journal}{Phys. Rev. B} \textbf{\bibinfo{volume}{57}},
  \bibinfo{pages}{11446} (\bibinfo{year}{1998}).

\bibitem[{\citenamefont{Sorella and Capriotti}(2000)}]{Sorella:2000p17651}
\bibinfo{author}{\bibfnamefont{S.}~\bibnamefont{Sorella}} \bibnamefont{and}
  \bibinfo{author}{\bibfnamefont{L.}~\bibnamefont{Capriotti}},
  \bibinfo{journal}{Phys. Rev. B} \textbf{\bibinfo{volume}{61}},
  \bibinfo{pages}{2599} (\bibinfo{year}{2000}).

\bibitem[{\citenamefont{Casula et~al.}(2005)\citenamefont{Casula, Filippi, and
  Sorella}}]{Casula:2005p14138}
\bibinfo{author}{\bibfnamefont{M.}~\bibnamefont{Casula}},
  \bibinfo{author}{\bibfnamefont{C.}~\bibnamefont{Filippi}}, \bibnamefont{and}
  \bibinfo{author}{\bibfnamefont{S.}~\bibnamefont{Sorella}},
  \bibinfo{journal}{Phys. Rev. Lett.} \textbf{\bibinfo{volume}{95}},
  \bibinfo{pages}{100201} (\bibinfo{year}{2005}).

\bibitem[{\citenamefont{Casula et~al.}(2010)\citenamefont{Casula, Moroni,
  Sorella, and Filippi}}]{Casula:2010p14082}
\bibinfo{author}{\bibfnamefont{M.}~\bibnamefont{Casula}},
  \bibinfo{author}{\bibfnamefont{S.}~\bibnamefont{Moroni}},
  \bibinfo{author}{\bibfnamefont{S.}~\bibnamefont{Sorella}}, \bibnamefont{and}
  \bibinfo{author}{\bibfnamefont{C.}~\bibnamefont{Filippi}},
  \bibinfo{journal}{J. Chem. Phys.} \textbf{\bibinfo{volume}{132}},
  \bibinfo{pages}{154113} (\bibinfo{year}{2010}).

\bibitem[{\citenamefont{Al-Saidi et~al.}(2006)\citenamefont{Al-Saidi, Zhang,
  and Krakauer}}]{AlSaidi:2006gj}
\bibinfo{author}{\bibfnamefont{W.~A.} \bibnamefont{Al-Saidi}},
  \bibinfo{author}{\bibfnamefont{S.}~\bibnamefont{Zhang}}, \bibnamefont{and}
  \bibinfo{author}{\bibfnamefont{H.}~\bibnamefont{Krakauer}},
  \bibinfo{journal}{J. Chem. Phys.} \textbf{\bibinfo{volume}{124}},
  \bibinfo{pages}{224101} (\bibinfo{year}{2006}).

\bibitem[{\citenamefont{Zhang and Krakauer}(2003)}]{Zhang:2003ed}
\bibinfo{author}{\bibfnamefont{S.}~\bibnamefont{Zhang}} \bibnamefont{and}
  \bibinfo{author}{\bibfnamefont{H.}~\bibnamefont{Krakauer}},
  \bibinfo{journal}{Phys. Rev. Lett.} \textbf{\bibinfo{volume}{90}},
  \bibinfo{pages}{136401} (\bibinfo{year}{2003}).

\bibitem[{\citenamefont{Baer et~al.}(1998)\citenamefont{Baer, Head-Gordon, and
  Neuhauser}}]{Baer:1998jh}
\bibinfo{author}{\bibfnamefont{R.}~\bibnamefont{Baer}},
  \bibinfo{author}{\bibfnamefont{M.}~\bibnamefont{Head-Gordon}},
  \bibnamefont{and}
  \bibinfo{author}{\bibfnamefont{D.}~\bibnamefont{Neuhauser}},
  \bibinfo{journal}{J. Chem. Phys.} \textbf{\bibinfo{volume}{109}},
  \bibinfo{pages}{6219} (\bibinfo{year}{1998}).

\bibitem[{\citenamefont{Chen and Anderson}(1995)}]{Chen:1995cf}
\bibinfo{author}{\bibfnamefont{B.}~\bibnamefont{Chen}} \bibnamefont{and}
  \bibinfo{author}{\bibfnamefont{J.~B.} \bibnamefont{Anderson}},
  \bibinfo{journal}{J. Chem. Phys.} \textbf{\bibinfo{volume}{102}},
  \bibinfo{pages}{4491} (\bibinfo{year}{1995}).

\bibitem[{\citenamefont{Ceperley and Alder}(1984)}]{Ceperley:1984ha}
\bibinfo{author}{\bibfnamefont{D.~M.} \bibnamefont{Ceperley}} \bibnamefont{and}
  \bibinfo{author}{\bibfnamefont{B.~J.} \bibnamefont{Alder}},
  \bibinfo{journal}{J. Chem. Phys.} \textbf{\bibinfo{volume}{81}},
  \bibinfo{pages}{5833} (\bibinfo{year}{1984}).

\bibitem[{\citenamefont{Bajdich et~al.}(2010)\citenamefont{Bajdich, Tiago,
  Hood, Kent, and Reboredo}}]{Bajdich:2010iu}
\bibinfo{author}{\bibfnamefont{M.}~\bibnamefont{Bajdich}},
  \bibinfo{author}{\bibfnamefont{M.~L.} \bibnamefont{Tiago}},
  \bibinfo{author}{\bibfnamefont{R.~Q.} \bibnamefont{Hood}},
  \bibinfo{author}{\bibfnamefont{P.~R.~C.} \bibnamefont{Kent}},
  \bibnamefont{and} \bibinfo{author}{\bibfnamefont{F.~A.}
  \bibnamefont{Reboredo}}, \bibinfo{journal}{Phys. Rev. Lett.}
  \textbf{\bibinfo{volume}{104}}, \bibinfo{pages}{193001}
  (\bibinfo{year}{2010}).

\bibitem[{\citenamefont{Baroni and Moroni}(1999)}]{Baroni:1999p27757}
\bibinfo{author}{\bibfnamefont{S.}~\bibnamefont{Baroni}} \bibnamefont{and}
  \bibinfo{author}{\bibfnamefont{S.}~\bibnamefont{Moroni}},
  \bibinfo{journal}{Phys. Rev. Lett.} \textbf{\bibinfo{volume}{82}},
  \bibinfo{pages}{4745} (\bibinfo{year}{1999}).

\bibitem[{\citenamefont{Yuen et~al.}(2009)\citenamefont{Yuen, Oblinsky,
  Giacometti, and Rothstein}}]{Yuen:2009gv}
\bibinfo{author}{\bibfnamefont{W.~K.} \bibnamefont{Yuen}},
  \bibinfo{author}{\bibfnamefont{D.~G.} \bibnamefont{Oblinsky}},
  \bibinfo{author}{\bibfnamefont{R.~D.} \bibnamefont{Giacometti}},
  \bibnamefont{and} \bibinfo{author}{\bibfnamefont{S.~M.}
  \bibnamefont{Rothstein}}, \bibinfo{journal}{Int. J. Quantum Chem.}
  \textbf{\bibinfo{volume}{109}}, \bibinfo{pages}{3229} (\bibinfo{year}{2009}).

\bibitem[{\citenamefont{Booth et~al.}(2009)\citenamefont{Booth, Thom, and
  Alavi}}]{Booth:2009p27842}
\bibinfo{author}{\bibfnamefont{G.~H.} \bibnamefont{Booth}},
  \bibinfo{author}{\bibfnamefont{A.~J.~W.} \bibnamefont{Thom}},
  \bibnamefont{and} \bibinfo{author}{\bibfnamefont{A.}~\bibnamefont{Alavi}},
  \bibinfo{journal}{J. Chem. Phys.} \textbf{\bibinfo{volume}{131}},
  \bibinfo{pages}{054106} (\bibinfo{year}{2009}).

\bibitem[{\citenamefont{Coccia and Guidoni}(2012)}]{Coccia:2012kz}
\bibinfo{author}{\bibfnamefont{E.}~\bibnamefont{Coccia}} \bibnamefont{and}
  \bibinfo{author}{\bibfnamefont{L.}~\bibnamefont{Guidoni}},
  \bibinfo{journal}{J. Comput. Chem.} \textbf{\bibinfo{volume}{33}},
  \bibinfo{pages}{2332} (\bibinfo{year}{2012}).

\bibitem[{\citenamefont{Barborini et~al.}(2012)\citenamefont{Barborini,
  Sorella, and Guidoni}}]{Barborini:2012iy}
\bibinfo{author}{\bibfnamefont{M.}~\bibnamefont{Barborini}},
  \bibinfo{author}{\bibfnamefont{S.}~\bibnamefont{Sorella}}, \bibnamefont{and}
  \bibinfo{author}{\bibfnamefont{L.}~\bibnamefont{Guidoni}},
  \bibinfo{journal}{J. Chem. Theory Comput.} \textbf{\bibinfo{volume}{8}},
  \bibinfo{pages}{1260} (\bibinfo{year}{2012}).

\bibitem[{\citenamefont{Filippi et~al.}(2012)\citenamefont{Filippi, Buda,
  Guidoni, and Sinicropi}}]{Filippi:2012hg}
\bibinfo{author}{\bibfnamefont{C.}~\bibnamefont{Filippi}},
  \bibinfo{author}{\bibfnamefont{F.}~\bibnamefont{Buda}},
  \bibinfo{author}{\bibfnamefont{L.}~\bibnamefont{Guidoni}}, \bibnamefont{and}
  \bibinfo{author}{\bibfnamefont{A.}~\bibnamefont{Sinicropi}},
  \bibinfo{journal}{J. Chem. Theory Comput.} \textbf{\bibinfo{volume}{8}},
  \bibinfo{pages}{112} (\bibinfo{year}{2012}).

\bibitem[{\citenamefont{Kolorenc and Mitas}(2011)}]{Kolorenc:2011hv}
\bibinfo{author}{\bibfnamefont{J.}~\bibnamefont{Kolorenc}} \bibnamefont{and}
  \bibinfo{author}{\bibfnamefont{L.}~\bibnamefont{Mitas}},
  \bibinfo{journal}{Rep. Prog. Phys.} \textbf{\bibinfo{volume}{74}},
  \bibinfo{pages}{026502} (\bibinfo{year}{2011}).

\bibitem[{\citenamefont{Maezono et~al.}(2010)\citenamefont{Maezono, Drummond,
  Ma, and Needs}}]{Maezono:2010ic}
\bibinfo{author}{\bibfnamefont{R.}~\bibnamefont{Maezono}},
  \bibinfo{author}{\bibfnamefont{N.~D.} \bibnamefont{Drummond}},
  \bibinfo{author}{\bibfnamefont{A.}~\bibnamefont{Ma}}, \bibnamefont{and}
  \bibinfo{author}{\bibfnamefont{R.~J.} \bibnamefont{Needs}},
  \bibinfo{journal}{Phys. Rev. B} \textbf{\bibinfo{volume}{82}},
  \bibinfo{pages}{184108} (\bibinfo{year}{2010}).

\bibitem[{\citenamefont{Valsson and Filippi}(2010)}]{Valsson:2010p25419}
\bibinfo{author}{\bibfnamefont{O.}~\bibnamefont{Valsson}} \bibnamefont{and}
  \bibinfo{author}{\bibfnamefont{C.}~\bibnamefont{Filippi}},
  \bibinfo{journal}{J. Chem. Theory Comput.} \textbf{\bibinfo{volume}{6}},
  \bibinfo{pages}{1275} (\bibinfo{year}{2010}).

\bibitem[{\citenamefont{Spanu et~al.}(2009)\citenamefont{Spanu, Sorella, and
  Galli}}]{Spanu:2009p12613}
\bibinfo{author}{\bibfnamefont{L.}~\bibnamefont{Spanu}},
  \bibinfo{author}{\bibfnamefont{S.}~\bibnamefont{Sorella}}, \bibnamefont{and}
  \bibinfo{author}{\bibfnamefont{G.}~\bibnamefont{Galli}},
  \bibinfo{journal}{Phys. Rev. Lett.} \textbf{\bibinfo{volume}{103}},
  \bibinfo{pages}{196401} (\bibinfo{year}{2009}).

\bibitem[{\citenamefont{Zimmerman et~al.}(2009)\citenamefont{Zimmerman,
  Toulouse, Zhang, Musgrave, and Umrigar}}]{Zimmerman:2009hh}
\bibinfo{author}{\bibfnamefont{P.~M.} \bibnamefont{Zimmerman}},
  \bibinfo{author}{\bibfnamefont{J.}~\bibnamefont{Toulouse}},
  \bibinfo{author}{\bibfnamefont{Z.}~\bibnamefont{Zhang}},
  \bibinfo{author}{\bibfnamefont{C.~B.} \bibnamefont{Musgrave}},
  \bibnamefont{and} \bibinfo{author}{\bibfnamefont{C.~J.}
  \bibnamefont{Umrigar}}, \bibinfo{journal}{J. Chem. Phys.}
  \textbf{\bibinfo{volume}{131}}, \bibinfo{pages}{124103}
  (\bibinfo{year}{2009}).

\bibitem[{\citenamefont{Sterpone et~al.}(2008)\citenamefont{Sterpone, Spanu,
  Ferraro, Sorella, and Guidoni}}]{Sterpone:2008p12640}
\bibinfo{author}{\bibfnamefont{F.}~\bibnamefont{Sterpone}},
  \bibinfo{author}{\bibfnamefont{L.}~\bibnamefont{Spanu}},
  \bibinfo{author}{\bibfnamefont{L.}~\bibnamefont{Ferraro}},
  \bibinfo{author}{\bibfnamefont{S.}~\bibnamefont{Sorella}}, \bibnamefont{and}
  \bibinfo{author}{\bibfnamefont{L.}~\bibnamefont{Guidoni}},
  \bibinfo{journal}{J. Chem. Theory Comput.} \textbf{\bibinfo{volume}{4}},
  \bibinfo{pages}{1428} (\bibinfo{year}{2008}).

\bibitem[{\citenamefont{Sorella et~al.}(2007)\citenamefont{Sorella, Casula, and
  Rocca}}]{Sorella:2007p12646}
\bibinfo{author}{\bibfnamefont{S.}~\bibnamefont{Sorella}},
  \bibinfo{author}{\bibfnamefont{M.}~\bibnamefont{Casula}}, \bibnamefont{and}
  \bibinfo{author}{\bibfnamefont{D.}~\bibnamefont{Rocca}}, \bibinfo{journal}{J.
  Chem. Phys.} \textbf{\bibinfo{volume}{127}}, \bibinfo{pages}{014105}
  (\bibinfo{year}{2007}).

\bibitem[{\citenamefont{Schautz and Filippi}(2004)}]{Schautz:2004p25285}
\bibinfo{author}{\bibfnamefont{F.}~\bibnamefont{Schautz}} \bibnamefont{and}
  \bibinfo{author}{\bibfnamefont{C.}~\bibnamefont{Filippi}},
  \bibinfo{journal}{J. Chem. Phys.} \textbf{\bibinfo{volume}{120}},
  \bibinfo{pages}{10931} (\bibinfo{year}{2004}).

\bibitem[{\citenamefont{Caffarel et~al.}(1993)\citenamefont{Caffarel, Rerat,
  and Pouchan}}]{Caffarel:1993td}
\bibinfo{author}{\bibfnamefont{M.}~\bibnamefont{Caffarel}},
  \bibinfo{author}{\bibfnamefont{M.}~\bibnamefont{Rerat}}, \bibnamefont{and}
  \bibinfo{author}{\bibfnamefont{C.}~\bibnamefont{Pouchan}},
  \bibinfo{journal}{Phys. Rev. A} \textbf{\bibinfo{volume}{47}},
  \bibinfo{pages}{3704} (\bibinfo{year}{1993}).

\bibitem[{\citenamefont{Bartlett and Musia{\l}}(2007)}]{Bartlett:2007kv}
\bibinfo{author}{\bibfnamefont{R.}~\bibnamefont{Bartlett}} \bibnamefont{and}
  \bibinfo{author}{\bibfnamefont{M.}~\bibnamefont{Musia{\l}}},
  \bibinfo{journal}{Rev. Mod. Phys.} \textbf{\bibinfo{volume}{79}},
  \bibinfo{pages}{291} (\bibinfo{year}{2007}).

\bibitem[{\citenamefont{Sherrill and Schaefer}(1999)}]{Sherrill:1999dw}
\bibinfo{author}{\bibfnamefont{C.~D.} \bibnamefont{Sherrill}} \bibnamefont{and}
  \bibinfo{author}{\bibfnamefont{H.~F.} \bibnamefont{Schaefer},
  \bibfnamefont{III}} (\bibinfo{publisher}{Elsevier}, \bibinfo{year}{1999}),
  pp. \bibinfo{pages}{143--269}.

\bibitem[{\citenamefont{M{\o}ller and Plesset}(1934)}]{Moller:1934kf}
\bibinfo{author}{\bibfnamefont{C.}~\bibnamefont{M{\o}ller}} \bibnamefont{and}
  \bibinfo{author}{\bibfnamefont{M.~S.} \bibnamefont{Plesset}},
  \bibinfo{journal}{Phys. Rev.} \textbf{\bibinfo{volume}{46}},
  \bibinfo{pages}{618} (\bibinfo{year}{1934}).

\bibitem[{\citenamefont{Head-Gordon et~al.}(1988)\citenamefont{Head-Gordon,
  Pople, and Frisch}}]{HeadGordon:1988fd}
\bibinfo{author}{\bibfnamefont{M.}~\bibnamefont{Head-Gordon}},
  \bibinfo{author}{\bibfnamefont{J.~A.} \bibnamefont{Pople}}, \bibnamefont{and}
  \bibinfo{author}{\bibfnamefont{M.~J.} \bibnamefont{Frisch}},
  \bibinfo{journal}{Chem. Phys. Lett.} \textbf{\bibinfo{volume}{153}},
  \bibinfo{pages}{503} (\bibinfo{year}{1988}).

\bibitem[{\citenamefont{Filippi and Umrigar}(2000)}]{Filippi:2000p25406}
\bibinfo{author}{\bibfnamefont{C.}~\bibnamefont{Filippi}} \bibnamefont{and}
  \bibinfo{author}{\bibfnamefont{C.~J.} \bibnamefont{Umrigar}},
  \bibinfo{journal}{Phys. Rev. B} \textbf{\bibinfo{volume}{61}},
  \bibinfo{pages}{R16291} (\bibinfo{year}{2000}).

\bibitem[{\citenamefont{Umrigar}(1989)}]{UMRIGAR:1989tq}
\bibinfo{author}{\bibfnamefont{C.~J.} \bibnamefont{Umrigar}},
  \bibinfo{journal}{Int. J. Quantum Chem.} pp. \bibinfo{pages}{217--230}
  (\bibinfo{year}{1989}).

\bibitem[{\citenamefont{Assaraf and Caffarel}(2000)}]{Assaraf:2000vr}
\bibinfo{author}{\bibfnamefont{R.}~\bibnamefont{Assaraf}} \bibnamefont{and}
  \bibinfo{author}{\bibfnamefont{M.}~\bibnamefont{Caffarel}},
  \bibinfo{journal}{J. Chem. Phys.} \textbf{\bibinfo{volume}{113}},
  \bibinfo{pages}{4028} (\bibinfo{year}{2000}).

\bibitem[{\citenamefont{Assaraf and Caffarel}(2003)}]{Assaraf:2003gq}
\bibinfo{author}{\bibfnamefont{R.}~\bibnamefont{Assaraf}} \bibnamefont{and}
  \bibinfo{author}{\bibfnamefont{M.}~\bibnamefont{Caffarel}},
  \bibinfo{journal}{J. Chem. Phys.} \textbf{\bibinfo{volume}{119}},
  \bibinfo{pages}{10536} (\bibinfo{year}{2003}).

\bibitem[{\citenamefont{Attaccalite and
  Sorella}(2008)}]{Attaccalite:2008p12639}
\bibinfo{author}{\bibfnamefont{C.}~\bibnamefont{Attaccalite}} \bibnamefont{and}
  \bibinfo{author}{\bibfnamefont{S.}~\bibnamefont{Sorella}},
  \bibinfo{journal}{Phys. Rev. Lett.} \textbf{\bibinfo{volume}{100}},
  \bibinfo{pages}{114501} (\bibinfo{year}{2008}).

\bibitem[{\citenamefont{Sorella and Capriotti}(2010)}]{Sorella:2010p23644}
\bibinfo{author}{\bibfnamefont{S.}~\bibnamefont{Sorella}} \bibnamefont{and}
  \bibinfo{author}{\bibfnamefont{L.}~\bibnamefont{Capriotti}},
  \bibinfo{journal}{J. Chem. Phys.} \textbf{\bibinfo{volume}{133}},
  \bibinfo{pages}{234111} (\bibinfo{year}{2010}).

\bibitem[{\citenamefont{Chiesa et~al.}(2005)\citenamefont{Chiesa, Ceperley, and
  Zhang}}]{Chiesa:2005p28489}
\bibinfo{author}{\bibfnamefont{S.}~\bibnamefont{Chiesa}},
  \bibinfo{author}{\bibfnamefont{D.~M.} \bibnamefont{Ceperley}},
  \bibnamefont{and} \bibinfo{author}{\bibfnamefont{S.}~\bibnamefont{Zhang}},
  \bibinfo{journal}{Phys. Rev. Lett.} \textbf{\bibinfo{volume}{94}},
  \bibinfo{pages}{036404} (\bibinfo{year}{2005}).

\bibitem[{\citenamefont{Wagner and Grossman}(2010)}]{Wagner:2010p25393}
\bibinfo{author}{\bibfnamefont{L.~K.} \bibnamefont{Wagner}} \bibnamefont{and}
  \bibinfo{author}{\bibfnamefont{J.~C.} \bibnamefont{Grossman}},
  \bibinfo{journal}{Phys. Rev. Lett.} \textbf{\bibinfo{volume}{104}},
  \bibinfo{pages}{210201} (\bibinfo{year}{2010}).

\bibitem[{\citenamefont{Barborini and Guidoni}(2012)}]{Barborini:2012it}
\bibinfo{author}{\bibfnamefont{M.}~\bibnamefont{Barborini}} \bibnamefont{and}
  \bibinfo{author}{\bibfnamefont{L.}~\bibnamefont{Guidoni}},
  \bibinfo{journal}{J. Chem. Phys.} \textbf{\bibinfo{volume}{137}},
  \bibinfo{pages}{224309} (\bibinfo{year}{2012}).

\bibitem[{\citenamefont{Coccia et~al.}(2013)\citenamefont{Coccia, Varsano, and
  Guidoni}}]{Coccia:2012ex}
\bibinfo{author}{\bibfnamefont{E.}~\bibnamefont{Coccia}},
  \bibinfo{author}{\bibfnamefont{D.}~\bibnamefont{Varsano}}, \bibnamefont{and}
  \bibinfo{author}{\bibfnamefont{L.}~\bibnamefont{Guidoni}},
  \bibinfo{journal}{J. Chem. Theory Comput.} \textbf{\bibinfo{volume}{9}},
  \bibinfo{pages}{8} (\bibinfo{year}{2013}).

\bibitem[{\citenamefont{Coccia et~al.}(2012)\citenamefont{Coccia, Chernomor,
  Barborini, Sorella, and Guidoni}}]{Coccia:2012fi}
\bibinfo{author}{\bibfnamefont{E.}~\bibnamefont{Coccia}},
  \bibinfo{author}{\bibfnamefont{O.}~\bibnamefont{Chernomor}},
  \bibinfo{author}{\bibfnamefont{M.}~\bibnamefont{Barborini}},
  \bibinfo{author}{\bibfnamefont{S.}~\bibnamefont{Sorella}}, \bibnamefont{and}
  \bibinfo{author}{\bibfnamefont{L.}~\bibnamefont{Guidoni}},
  \bibinfo{journal}{J. Chem. Theory Comput.} \textbf{\bibinfo{volume}{8}},
  \bibinfo{pages}{1952} (\bibinfo{year}{2012}).

\bibitem[{\citenamefont{Zen et~al.}(2012)\citenamefont{Zen, Zhelyazov, and
  Guidoni}}]{Zen:2012br}
\bibinfo{author}{\bibfnamefont{A.}~\bibnamefont{Zen}},
  \bibinfo{author}{\bibfnamefont{D.}~\bibnamefont{Zhelyazov}},
  \bibnamefont{and} \bibinfo{author}{\bibfnamefont{L.}~\bibnamefont{Guidoni}},
  \bibinfo{journal}{J. Chem. Theory Comput.} \textbf{\bibinfo{volume}{8}},
  \bibinfo{pages}{4204} (\bibinfo{year}{2012}).

\bibitem[{\citenamefont{Umrigar et~al.}(2007)\citenamefont{Umrigar, Toulouse,
  Filippi, Sorella, and Hennig}}]{Umrigar:2007p12662}
\bibinfo{author}{\bibfnamefont{C.~J.} \bibnamefont{Umrigar}},
  \bibinfo{author}{\bibfnamefont{J.}~\bibnamefont{Toulouse}},
  \bibinfo{author}{\bibfnamefont{C.}~\bibnamefont{Filippi}},
  \bibinfo{author}{\bibfnamefont{S.}~\bibnamefont{Sorella}}, \bibnamefont{and}
  \bibinfo{author}{\bibfnamefont{R.~G.} \bibnamefont{Hennig}},
  \bibinfo{journal}{Phys. Rev. Lett.} \textbf{\bibinfo{volume}{98}},
  \bibinfo{pages}{110201} (\bibinfo{year}{2007}).

\bibitem[{\citenamefont{Clark et~al.}(2011)\citenamefont{Clark, Morales,
  McMinis, Kim, and Scuseria}}]{Clark:2011ie}
\bibinfo{author}{\bibfnamefont{B.~K.} \bibnamefont{Clark}},
  \bibinfo{author}{\bibfnamefont{M.~A.} \bibnamefont{Morales}},
  \bibinfo{author}{\bibfnamefont{J.}~\bibnamefont{McMinis}},
  \bibinfo{author}{\bibfnamefont{J.}~\bibnamefont{Kim}}, \bibnamefont{and}
  \bibinfo{author}{\bibfnamefont{G.~E.} \bibnamefont{Scuseria}},
  \bibinfo{journal}{J. Chem. Phys.} \textbf{\bibinfo{volume}{135}},
  \bibinfo{pages}{244105} (\bibinfo{year}{2011}).

\bibitem[{\citenamefont{Morales et~al.}(2012)\citenamefont{Morales, McMinis,
  Clark, Kim, and Scuseria}}]{Morales:2012kk}
\bibinfo{author}{\bibfnamefont{M.~A.} \bibnamefont{Morales}},
  \bibinfo{author}{\bibfnamefont{J.}~\bibnamefont{McMinis}},
  \bibinfo{author}{\bibfnamefont{B.~K.} \bibnamefont{Clark}},
  \bibinfo{author}{\bibfnamefont{J.}~\bibnamefont{Kim}}, \bibnamefont{and}
  \bibinfo{author}{\bibfnamefont{G.~E.} \bibnamefont{Scuseria}},
  \bibinfo{journal}{J. Chem. Theory Comput.} \textbf{\bibinfo{volume}{8}},
  \bibinfo{pages}{2181} (\bibinfo{year}{2012}).

\bibitem[{\citenamefont{Xu et~al.}(2013)\citenamefont{Xu, Deible, Peterson, and
  Jordan}}]{Xu:2013fc}
\bibinfo{author}{\bibfnamefont{J.}~\bibnamefont{Xu}},
  \bibinfo{author}{\bibfnamefont{M.~J.} \bibnamefont{Deible}},
  \bibinfo{author}{\bibfnamefont{K.~A.} \bibnamefont{Peterson}},
  \bibnamefont{and} \bibinfo{author}{\bibfnamefont{K.~D.}
  \bibnamefont{Jordan}}, \bibinfo{journal}{J. Chem. Theory Comput.}
  \textbf{\bibinfo{volume}{9}}, \bibinfo{pages}{2170} (\bibinfo{year}{2013}).

\bibitem[{\citenamefont{Jastrow}(1955)}]{Jastrow:1955en}
\bibinfo{author}{\bibfnamefont{R.}~\bibnamefont{Jastrow}},
  \bibinfo{journal}{Phys. Rev.} \textbf{\bibinfo{volume}{98}},
  \bibinfo{pages}{1479} (\bibinfo{year}{1955}).

\bibitem[{\citenamefont{Casula and Sorella}(2003)}]{Casula:2003p12694}
\bibinfo{author}{\bibfnamefont{M.}~\bibnamefont{Casula}} \bibnamefont{and}
  \bibinfo{author}{\bibfnamefont{S.}~\bibnamefont{Sorella}},
  \bibinfo{journal}{J. Chem. Phys.} \textbf{\bibinfo{volume}{119}},
  \bibinfo{pages}{6500} (\bibinfo{year}{2003}).

\bibitem[{\citenamefont{Bajdich et~al.}(2008)\citenamefont{Bajdich, Mitas,
  Wagner, and Schmidt}}]{Bajdich:2008p18507}
\bibinfo{author}{\bibfnamefont{M.}~\bibnamefont{Bajdich}},
  \bibinfo{author}{\bibfnamefont{L.}~\bibnamefont{Mitas}},
  \bibinfo{author}{\bibfnamefont{L.~K.} \bibnamefont{Wagner}},
  \bibnamefont{and} \bibinfo{author}{\bibfnamefont{K.~E.}
  \bibnamefont{Schmidt}}, \bibinfo{journal}{Phys. Rev. B}
  \textbf{\bibinfo{volume}{77}}, \bibinfo{pages}{115112}
  (\bibinfo{year}{2008}).

\bibitem[{\citenamefont{Bajdich et~al.}(2006)\citenamefont{Bajdich, Mitas,
  Drobny, Wagner, and Schmidt}}]{Bajdich:2006p18510}
\bibinfo{author}{\bibfnamefont{M.}~\bibnamefont{Bajdich}},
  \bibinfo{author}{\bibfnamefont{L.}~\bibnamefont{Mitas}},
  \bibinfo{author}{\bibfnamefont{G.}~\bibnamefont{Drobny}},
  \bibinfo{author}{\bibfnamefont{L.}~\bibnamefont{Wagner}}, \bibnamefont{and}
  \bibinfo{author}{\bibfnamefont{K.}~\bibnamefont{Schmidt}},
  \bibinfo{journal}{Phys. Rev. Lett.} \textbf{\bibinfo{volume}{96}},
  \bibinfo{pages}{130201} (\bibinfo{year}{2006}).

\bibitem[{\citenamefont{Holzmann et~al.}(2003)\citenamefont{Holzmann, Ceperley,
  Pierleoni, and Esler}}]{Holzmann:2003p28608}
\bibinfo{author}{\bibfnamefont{M.}~\bibnamefont{Holzmann}},
  \bibinfo{author}{\bibfnamefont{D.~M.} \bibnamefont{Ceperley}},
  \bibinfo{author}{\bibfnamefont{C.}~\bibnamefont{Pierleoni}},
  \bibnamefont{and} \bibinfo{author}{\bibfnamefont{K.}~\bibnamefont{Esler}},
  \bibinfo{journal}{Phys. Rev. E} \textbf{\bibinfo{volume}{68}},
  \bibinfo{pages}{046707} (\bibinfo{year}{2003}).

\bibitem[{\citenamefont{Toulouse and Umrigar}(2008)}]{Toulouse:2008p27527}
\bibinfo{author}{\bibfnamefont{J.}~\bibnamefont{Toulouse}} \bibnamefont{and}
  \bibinfo{author}{\bibfnamefont{C.~J.} \bibnamefont{Umrigar}},
  \bibinfo{journal}{J. Chem. Phys.} \textbf{\bibinfo{volume}{128}},
  \bibinfo{pages}{174101} (\bibinfo{year}{2008}).

\bibitem[{\citenamefont{Fracchia et~al.}(2012)\citenamefont{Fracchia, Filippi,
  and Amovilli}}]{Fracchia:2012el}
\bibinfo{author}{\bibfnamefont{F.}~\bibnamefont{Fracchia}},
  \bibinfo{author}{\bibfnamefont{C.}~\bibnamefont{Filippi}}, \bibnamefont{and}
  \bibinfo{author}{\bibfnamefont{C.}~\bibnamefont{Amovilli}},
  \bibinfo{journal}{J. Chem. Theory Comput.} \textbf{\bibinfo{volume}{8}},
  \bibinfo{pages}{1943} (\bibinfo{year}{2012}).

\bibitem[{\citenamefont{Neuscamman}(2012)}]{Neuscamman:2012hm}
\bibinfo{author}{\bibfnamefont{E.}~\bibnamefont{Neuscamman}},
  \bibinfo{journal}{Phys. Rev. Lett.} \textbf{\bibinfo{volume}{109}},
  \bibinfo{pages}{203001} (\bibinfo{year}{2012}).

\bibitem[{\citenamefont{Petruzielo et~al.}(2011)\citenamefont{Petruzielo,
  Toulouse, and Umrigar}}]{Petruzielo:2011p24345}
\bibinfo{author}{\bibfnamefont{F.~R.} \bibnamefont{Petruzielo}},
  \bibinfo{author}{\bibfnamefont{J.}~\bibnamefont{Toulouse}}, \bibnamefont{and}
  \bibinfo{author}{\bibfnamefont{C.~J.} \bibnamefont{Umrigar}},
  \bibinfo{journal}{J. Chem. Phys.} \textbf{\bibinfo{volume}{134}},
  \bibinfo{pages}{064104} (\bibinfo{year}{2011}).

\bibitem[{\citenamefont{Marchi et~al.}(2009)\citenamefont{Marchi, Azadi,
  Casula, and Sorella}}]{Marchi:2009p12614}
\bibinfo{author}{\bibfnamefont{M.}~\bibnamefont{Marchi}},
  \bibinfo{author}{\bibfnamefont{S.}~\bibnamefont{Azadi}},
  \bibinfo{author}{\bibfnamefont{M.}~\bibnamefont{Casula}}, \bibnamefont{and}
  \bibinfo{author}{\bibfnamefont{S.}~\bibnamefont{Sorella}},
  \bibinfo{journal}{J. Chem. Phys.} \textbf{\bibinfo{volume}{131}},
  \bibinfo{pages}{154116} (\bibinfo{year}{2009}).

\bibitem[{\citenamefont{Clough et~al.}(1973)\citenamefont{Clough, Beers, Klein,
  and Rothman}}]{Clough:1973bh}
\bibinfo{author}{\bibfnamefont{S.~A.} \bibnamefont{Clough}},
  \bibinfo{author}{\bibfnamefont{Y.}~\bibnamefont{Beers}},
  \bibinfo{author}{\bibfnamefont{G.~P.} \bibnamefont{Klein}}, \bibnamefont{and}
  \bibinfo{author}{\bibfnamefont{L.~S.} \bibnamefont{Rothman}},
  \bibinfo{journal}{J. Chem. Phys.} \textbf{\bibinfo{volume}{59}},
  \bibinfo{pages}{2254} (\bibinfo{year}{1973}).

\bibitem[{\citenamefont{Verhoeven and Dymanus}(1970)}]{Verhoeven:1970jd}
\bibinfo{author}{\bibfnamefont{J.}~\bibnamefont{Verhoeven}} \bibnamefont{and}
  \bibinfo{author}{\bibfnamefont{A.}~\bibnamefont{Dymanus}},
  \bibinfo{journal}{J. Chem. Phys.} \textbf{\bibinfo{volume}{52}},
  \bibinfo{pages}{3222} (\bibinfo{year}{1970}).

\bibitem[{\citenamefont{Benedict et~al.}(1956)\citenamefont{Benedict, Gailar,
  and Plyler}}]{Benedict:1956id}
\bibinfo{author}{\bibfnamefont{W.~S.} \bibnamefont{Benedict}},
  \bibinfo{author}{\bibfnamefont{N.}~\bibnamefont{Gailar}}, \bibnamefont{and}
  \bibinfo{author}{\bibfnamefont{E.~K.} \bibnamefont{Plyler}},
  \bibinfo{journal}{J. Chem. Phys.} \textbf{\bibinfo{volume}{24}},
  \bibinfo{pages}{1139} (\bibinfo{year}{1956}).

\bibitem[{\citenamefont{Feller et~al.}(1987)\citenamefont{Feller, Boyle, and
  Davidson}}]{Feller:1987dm}
\bibinfo{author}{\bibfnamefont{D.}~\bibnamefont{Feller}},
  \bibinfo{author}{\bibfnamefont{C.~M.} \bibnamefont{Boyle}}, \bibnamefont{and}
  \bibinfo{author}{\bibfnamefont{E.~R.} \bibnamefont{Davidson}},
  \bibinfo{journal}{J. Chem. Phys.} \textbf{\bibinfo{volume}{86}},
  \bibinfo{pages}{3424} (\bibinfo{year}{1987}).

\bibitem[{\citenamefont{Partridge and Schwenke}(1997)}]{Partridge:1997p26051}
\bibinfo{author}{\bibfnamefont{H.}~\bibnamefont{Partridge}} \bibnamefont{and}
  \bibinfo{author}{\bibfnamefont{D.}~\bibnamefont{Schwenke}},
  \bibinfo{journal}{J. Chem. Phys.} \textbf{\bibinfo{volume}{106}},
  \bibinfo{pages}{4618} (\bibinfo{year}{1997}).

\bibitem[{\citenamefont{Schwenke and Partridge}(2000)}]{Schwenke:2000hf}
\bibinfo{author}{\bibfnamefont{D.~W.} \bibnamefont{Schwenke}} \bibnamefont{and}
  \bibinfo{author}{\bibfnamefont{H.}~\bibnamefont{Partridge}},
  \bibinfo{journal}{J. Chem. Phys.} \textbf{\bibinfo{volume}{113}},
  \bibinfo{pages}{6592} (\bibinfo{year}{2000}).

\bibitem[{\citenamefont{Lodi et~al.}(2008)\citenamefont{Lodi, Tolchenov,
  Tennyson, Lynas-Gray, Shirin, Zobov, Polyansky, Csaszar, van Stralen, and
  Visscher}}]{Lodi:2008ic}
\bibinfo{author}{\bibfnamefont{L.}~\bibnamefont{Lodi}},
  \bibinfo{author}{\bibfnamefont{R.~N.} \bibnamefont{Tolchenov}},
  \bibinfo{author}{\bibfnamefont{J.}~\bibnamefont{Tennyson}},
  \bibinfo{author}{\bibfnamefont{A.~E.} \bibnamefont{Lynas-Gray}},
  \bibinfo{author}{\bibfnamefont{S.~V.} \bibnamefont{Shirin}},
  \bibinfo{author}{\bibfnamefont{N.~F.} \bibnamefont{Zobov}},
  \bibinfo{author}{\bibfnamefont{O.~L.} \bibnamefont{Polyansky}},
  \bibinfo{author}{\bibfnamefont{A.~G.} \bibnamefont{Csaszar}},
  \bibinfo{author}{\bibfnamefont{J.~N.~P.} \bibnamefont{van Stralen}},
  \bibnamefont{and} \bibinfo{author}{\bibfnamefont{L.}~\bibnamefont{Visscher}},
  \bibinfo{journal}{J. Chem. Phys.} \textbf{\bibinfo{volume}{128}},
  \bibinfo{pages}{044304} (\bibinfo{year}{2008}).

\bibitem[{\citenamefont{Csaszar et~al.}(2005)\citenamefont{Csaszar, Czako,
  Furtenbacher, Tennyson, Szalay, Shirin, Zobov, and
  Polyansky}}]{Csaszar:2005dl}
\bibinfo{author}{\bibfnamefont{A.~G.} \bibnamefont{Csaszar}},
  \bibinfo{author}{\bibfnamefont{G.}~\bibnamefont{Czako}},
  \bibinfo{author}{\bibfnamefont{T.}~\bibnamefont{Furtenbacher}},
  \bibinfo{author}{\bibfnamefont{J.}~\bibnamefont{Tennyson}},
  \bibinfo{author}{\bibfnamefont{V.}~\bibnamefont{Szalay}},
  \bibinfo{author}{\bibfnamefont{S.~V.} \bibnamefont{Shirin}},
  \bibinfo{author}{\bibfnamefont{N.~F.} \bibnamefont{Zobov}}, \bibnamefont{and}
  \bibinfo{author}{\bibfnamefont{O.~L.} \bibnamefont{Polyansky}},
  \bibinfo{journal}{J. Chem. Phys.} \textbf{\bibinfo{volume}{122}},
  \bibinfo{pages}{214305} (\bibinfo{year}{2005}).

\bibitem[{\citenamefont{Kim et~al.}(1995)\citenamefont{Kim, Lee, Lee, Mhin, and
  Kim}}]{KIM:1995p23441}
\bibinfo{author}{\bibfnamefont{J.~S.} \bibnamefont{Kim}},
  \bibinfo{author}{\bibfnamefont{J.~Y.} \bibnamefont{Lee}},
  \bibinfo{author}{\bibfnamefont{S.}~\bibnamefont{Lee}},
  \bibinfo{author}{\bibfnamefont{B.~J.} \bibnamefont{Mhin}}, \bibnamefont{and}
  \bibinfo{author}{\bibfnamefont{K.~S.} \bibnamefont{Kim}},
  \bibinfo{journal}{J. Chem. Phys.} \textbf{\bibinfo{volume}{102}},
  \bibinfo{pages}{310} (\bibinfo{year}{1995}).

\bibitem[{\citenamefont{Feller and Peterson}(2009)}]{Feller:2009p23440}
\bibinfo{author}{\bibfnamefont{D.}~\bibnamefont{Feller}} \bibnamefont{and}
  \bibinfo{author}{\bibfnamefont{K.~A.} \bibnamefont{Peterson}},
  \bibinfo{journal}{J. Chem. Phys.} \textbf{\bibinfo{volume}{131}},
  \bibinfo{pages}{154306} (\bibinfo{year}{2009}).

\bibitem[{\citenamefont{Kato}(1957)}]{Kato:1957jo}
\bibinfo{author}{\bibfnamefont{T.}~\bibnamefont{Kato}},
  \bibinfo{journal}{Commun. on Pure Appl. Math.} \textbf{\bibinfo{volume}{10}},
  \bibinfo{pages}{151} (\bibinfo{year}{1957}).

\bibitem[{\citenamefont{Foster and Weinhold}(1980)}]{NYO}
\bibinfo{author}{\bibfnamefont{J.~P.} \bibnamefont{Foster}} \bibnamefont{and}
  \bibinfo{author}{\bibfnamefont{F.}~\bibnamefont{Weinhold}},
  \bibinfo{journal}{J. Am. Chem. Soc.} \textbf{\bibinfo{volume}{102}},
  \bibinfo{pages}{7211} (\bibinfo{year}{1980}).

\bibitem[{\citenamefont{Hurley et~al.}(1953)\citenamefont{Hurley,
  Lennard-Jones, and Pople}}]{Hurley:1953fo}
\bibinfo{author}{\bibfnamefont{A.~C.} \bibnamefont{Hurley}},
  \bibinfo{author}{\bibfnamefont{J.}~\bibnamefont{Lennard-Jones}},
  \bibnamefont{and} \bibinfo{author}{\bibfnamefont{J.~A.} \bibnamefont{Pople}},
  \bibinfo{journal}{Proc. R. Soc. London, Ser. A}
  \textbf{\bibinfo{volume}{220}}, \bibinfo{pages}{446} (\bibinfo{year}{1953}).

\bibitem[{\citenamefont{Coleman}(1972)}]{Coleman:1972ds}
\bibinfo{author}{\bibfnamefont{A.~J.} \bibnamefont{Coleman}},
  \bibinfo{journal}{J. Math. Phys.} \textbf{\bibinfo{volume}{13}},
  \bibinfo{pages}{214} (\bibinfo{year}{1972}).

\bibitem[{\citenamefont{Casula et~al.}(2004)\citenamefont{Casula, Attaccalite,
  and Sorella}}]{Casula:2004p12689}
\bibinfo{author}{\bibfnamefont{M.}~\bibnamefont{Casula}},
  \bibinfo{author}{\bibfnamefont{C.}~\bibnamefont{Attaccalite}},
  \bibnamefont{and} \bibinfo{author}{\bibfnamefont{S.}~\bibnamefont{Sorella}},
  \bibinfo{journal}{J. Chem. Phys.} \textbf{\bibinfo{volume}{121}},
  \bibinfo{pages}{7110} (\bibinfo{year}{2004}).

\bibitem[{\citenamefont{Filippi and Umrigar}(1996)}]{Filippi:1996eb}
\bibinfo{author}{\bibfnamefont{C.}~\bibnamefont{Filippi}} \bibnamefont{and}
  \bibinfo{author}{\bibfnamefont{C.~J.} \bibnamefont{Umrigar}},
  \bibinfo{journal}{J. Chem. Phys.} \textbf{\bibinfo{volume}{105}},
  \bibinfo{pages}{213} (\bibinfo{year}{1996}).

\bibitem[{\citenamefont{Trail}(2008)}]{Trail:2008ft}
\bibinfo{author}{\bibfnamefont{J.~R.} \bibnamefont{Trail}},
  \bibinfo{journal}{Phys. Rev. E} \textbf{\bibinfo{volume}{77}},
  \bibinfo{pages}{016703} (\bibinfo{year}{2008}).

\bibitem[{\citenamefont{Kunsch}(1989)}]{KUNSCH:1989ws}
\bibinfo{author}{\bibfnamefont{H.~R.} \bibnamefont{Kunsch}},
  \bibinfo{journal}{Ann. Stat.} \textbf{\bibinfo{volume}{17}},
  \bibinfo{pages}{1217} (\bibinfo{year}{1989}).

\bibitem[{\citenamefont{Wolff}(2004)}]{Wolff:2004cu}
\bibinfo{author}{\bibfnamefont{U.}~\bibnamefont{Wolff}},
  \bibinfo{journal}{Comput. Phys. Commun.} \textbf{\bibinfo{volume}{156}},
  \bibinfo{pages}{143} (\bibinfo{year}{2004}).

\bibitem[{\citenamefont{Sorella}(2005)}]{Sorella:2005p14143}
\bibinfo{author}{\bibfnamefont{S.}~\bibnamefont{Sorella}},
  \bibinfo{journal}{Phys. Rev. B} \textbf{\bibinfo{volume}{71}},
  \bibinfo{pages}{241103} (\bibinfo{year}{2005}).

\bibitem[{\citenamefont{Toulouse and Umrigar}(2007)}]{Toulouse:2007p27522}
\bibinfo{author}{\bibfnamefont{J.}~\bibnamefont{Toulouse}} \bibnamefont{and}
  \bibinfo{author}{\bibfnamefont{C.~J.} \bibnamefont{Umrigar}},
  \bibinfo{journal}{J. Chem. Phys.} \textbf{\bibinfo{volume}{126}},
  \bibinfo{pages}{084102} (\bibinfo{year}{2007}).

\bibitem[{\citenamefont{Sorella}(2013)}]{Sorella:2013bz}
\bibinfo{author}{\bibfnamefont{S.}~\bibnamefont{Sorella}}
  (\bibinfo{publisher}{Springer Berlin Heidelberg}, \bibinfo{address}{Berlin,
  Heidelberg}, \bibinfo{year}{2013}), pp. \bibinfo{pages}{207--236}.

\bibitem[{\citenamefont{Mazzola et~al.}(2012)\citenamefont{Mazzola, Zen, and
  Sorella}}]{Mazzola:2012ch}
\bibinfo{author}{\bibfnamefont{G.}~\bibnamefont{Mazzola}},
  \bibinfo{author}{\bibfnamefont{A.}~\bibnamefont{Zen}}, \bibnamefont{and}
  \bibinfo{author}{\bibfnamefont{S.}~\bibnamefont{Sorella}},
  \bibinfo{journal}{J. Chem. Phys.} \textbf{\bibinfo{volume}{137}},
  \bibinfo{pages}{134112} (\bibinfo{year}{2012}).

\bibitem[{\citenamefont{Assaraf
  et~al.}(2007{\natexlab{b}})\citenamefont{Assaraf, Caffarel, and
  Scemama}}]{Assaraf:2007p19843}
\bibinfo{author}{\bibfnamefont{R.}~\bibnamefont{Assaraf}},
  \bibinfo{author}{\bibfnamefont{M.}~\bibnamefont{Caffarel}}, \bibnamefont{and}
  \bibinfo{author}{\bibfnamefont{A.}~\bibnamefont{Scemama}},
  \bibinfo{journal}{Phy. Rev. E} \textbf{\bibinfo{volume}{75}},
  \bibinfo{pages}{035701} (\bibinfo{year}{2007}{\natexlab{b}}).

\bibitem[{\citenamefont{Sorella}()}]{TurboRVB}
\bibinfo{author}{\bibfnamefont{S.}~\bibnamefont{Sorella}},
  \emph{\bibinfo{title}{{\em TurboRVB} quantum monte carlo package (accessed
  date 1 november 2012)}},
  \urlprefix\url{http://people.sissa.it/~sorella/web/index.html}.

\bibitem[{\citenamefont{Burkatzki et~al.}(2007)\citenamefont{Burkatzki,
  Filippi, and Dolg}}]{Burkatzki:2007p25447}
\bibinfo{author}{\bibfnamefont{M.}~\bibnamefont{Burkatzki}},
  \bibinfo{author}{\bibfnamefont{C.}~\bibnamefont{Filippi}}, \bibnamefont{and}
  \bibinfo{author}{\bibfnamefont{M.}~\bibnamefont{Dolg}}, \bibinfo{journal}{J.
  Chem. Phys.} \textbf{\bibinfo{volume}{126}}, \bibinfo{pages}{234105}
  (\bibinfo{year}{2007}).

\bibitem[{\citenamefont{Trail and Needs}(2005)}]{Trail:2005iw}
\bibinfo{author}{\bibfnamefont{J.}~\bibnamefont{Trail}} \bibnamefont{and}
  \bibinfo{author}{\bibfnamefont{R.}~\bibnamefont{Needs}}, \bibinfo{journal}{J.
  Chem. Phys.} \textbf{\bibinfo{volume}{122}}, \bibinfo{pages}{174109}
  (\bibinfo{year}{2005}).

\bibitem[{\citenamefont{Azadi et~al.}(2010)\citenamefont{Azadi, Cavazzoni, and
  Sorella}}]{Azadi:2010p14081}
\bibinfo{author}{\bibfnamefont{S.}~\bibnamefont{Azadi}},
  \bibinfo{author}{\bibfnamefont{C.}~\bibnamefont{Cavazzoni}},
  \bibnamefont{and} \bibinfo{author}{\bibfnamefont{S.}~\bibnamefont{Sorella}},
  \bibinfo{journal}{Phys. Rev. B} \textbf{\bibinfo{volume}{82}},
  \bibinfo{pages}{125112} (\bibinfo{year}{2010}).

\bibitem[{\citenamefont{Dunning}(1989)}]{DUNNING:1989uk}
\bibinfo{author}{\bibfnamefont{T.~H.} \bibnamefont{Dunning}},
  \bibinfo{journal}{J. Chem. Phys.} \textbf{\bibinfo{volume}{90}},
  \bibinfo{pages}{1007} (\bibinfo{year}{1989}).

\bibitem[{\citenamefont{Kendall et~al.}(1992)\citenamefont{Kendall, Dunning,
  and Harrison}}]{KENDALL:1992vx}
\bibinfo{author}{\bibfnamefont{R.}~\bibnamefont{Kendall}},
  \bibinfo{author}{\bibfnamefont{T.}~\bibnamefont{Dunning}}, \bibnamefont{and}
  \bibinfo{author}{\bibfnamefont{R.}~\bibnamefont{Harrison}},
  \bibinfo{journal}{J. Chem. Phys.} \textbf{\bibinfo{volume}{96}},
  \bibinfo{pages}{6796} (\bibinfo{year}{1992}).

\bibitem[{NIS()}]{NISTweb}
\emph{\bibinfo{title}{Nist chemistry webbook (accessed date march 2013).}},
  \urlprefix\url{http://webbook.nist.gov/chemistry}.

\bibitem[{ccc()}]{cccbdb}
\emph{\bibinfo{title}{Computational chemistry comparison and benchmark database
  (accessed date march 2013).}}, \urlprefix\url{http://cccbdb.nist.gov/}.

\bibitem[{\citenamefont{Chakravorty et~al.}(1993)\citenamefont{Chakravorty,
  Gwaltney, Davidson, Parpia, and p~Fischer}}]{Chakravorty:1993gg}
\bibinfo{author}{\bibfnamefont{S.}~\bibnamefont{Chakravorty}},
  \bibinfo{author}{\bibfnamefont{S.}~\bibnamefont{Gwaltney}},
  \bibinfo{author}{\bibfnamefont{E.}~\bibnamefont{Davidson}},
  \bibinfo{author}{\bibfnamefont{F.}~\bibnamefont{Parpia}}, \bibnamefont{and}
  \bibinfo{author}{\bibfnamefont{C.}~\bibnamefont{p~Fischer}},
  \bibinfo{journal}{Phys Rev A} \textbf{\bibinfo{volume}{47}},
  \bibinfo{pages}{3649} (\bibinfo{year}{1993}).

\bibitem[{\citenamefont{Barone}(2005)}]{Barone:2005p24347}
\bibinfo{author}{\bibfnamefont{V.}~\bibnamefont{Barone}}, \bibinfo{journal}{J.
  Chem. Phys.} \textbf{\bibinfo{volume}{122}}, \bibinfo{pages}{014108}
  (\bibinfo{year}{2005}).

\bibitem[{\citenamefont{Brown et~al.}(2007)\citenamefont{Brown, Trail, Rios,
  and Needs}}]{Brown:2007gh}
\bibinfo{author}{\bibfnamefont{M.~D.} \bibnamefont{Brown}},
  \bibinfo{author}{\bibfnamefont{J.~R.} \bibnamefont{Trail}},
  \bibinfo{author}{\bibfnamefont{P.~L.} \bibnamefont{Rios}}, \bibnamefont{and}
  \bibinfo{author}{\bibfnamefont{R.~J.} \bibnamefont{Needs}},
  \bibinfo{journal}{J. Chem. Phys.} \textbf{\bibinfo{volume}{126}},
  \bibinfo{pages}{224110} (\bibinfo{year}{2007}).

\bibitem[{\citenamefont{L{\'o}pez~R{\'\i}os
  et~al.}(2012)\citenamefont{L{\'o}pez~R{\'\i}os, Seth, Drummond, and
  Needs}}]{LopezRios:2012cg}
\bibinfo{author}{\bibfnamefont{P.}~\bibnamefont{L{\'o}pez~R{\'\i}os}},
  \bibinfo{author}{\bibfnamefont{P.}~\bibnamefont{Seth}},
  \bibinfo{author}{\bibfnamefont{N.~D.} \bibnamefont{Drummond}},
  \bibnamefont{and} \bibinfo{author}{\bibfnamefont{R.~J.} \bibnamefont{Needs}},
  \bibinfo{journal}{Phys. Rev. E} \textbf{\bibinfo{volume}{86}},
  \bibinfo{pages}{036703} (\bibinfo{year}{2012}).

\bibitem[{\citenamefont{Franks}(1972)}]{Franks:1972wj}
\bibinfo{editor}{\bibfnamefont{F.}~\bibnamefont{Franks}}, ed.,
  \emph{\bibinfo{title}{{Water, a comprehensive treatise}}}
  (\bibinfo{publisher}{New York: Plenum Press}, \bibinfo{year}{1972}).

\bibitem[{\citenamefont{Kukolich}(1969)}]{Kukolich:1969es}
\bibinfo{author}{\bibfnamefont{S.~G.} \bibnamefont{Kukolich}},
  \bibinfo{journal}{J. Chem. Phys.} \textbf{\bibinfo{volume}{50}},
  \bibinfo{pages}{3751} (\bibinfo{year}{1969}).

\end{thebibliography}

\end{document}